\documentclass{ws-rmp1}
\usepackage{amsfonts,amssymb}
\usepackage{stmaryrd}
\usepackage{amstext}

\newcommand{\beq}{\begin{equation}}
\newcommand{\eeq}{\end{equation}}
\newcommand{\beqa}{\begin{eqnarray}}
\newcommand{\eeqa}{\end{eqnarray}}
\newcommand{\nn}{\nonumber \\}

\newcommand{\prlabel}[1]{\addtocounter{theorem}{-1}
  \refstepcounter{theorem}\label{#1}}
\newcommand{\rmlabel}[1]{\addtocounter{remark}{-1}
  \refstepcounter{remark}\label{#1}}

\newcommand{\mbf}[1]{\ensuremath{\mathchoice
                    {\mbox{\boldmath$\displaystyle\mathbf{\mathit{#1}}$}}
                    {\mbox{\boldmath$\textstyle\mathbf{\mathit{#1}}$}}
                    {\mbox{\boldmath$\scriptstyle\mathbf{\mathit{#1}}$}}
                    {\mbox{\boldmath$\scriptscriptstyle\mathbf{\mathit{#1}}$}}}}
\newcommand{\Mbf}[1]{\ensuremath{\mathchoice
                    {\mbox{\boldmath$\displaystyle\mathbf{#1}$}}
                    {\mbox{\boldmath$\textstyle\mathbf{#1}$}}
                    {\mbox{\boldmath$\scriptstyle\mathbf{#1}$}}
                    {\mbox{\boldmath$\scriptscriptstyle\mathbf{#1}$}}}}

\def \hev {\theta}

\DeclareSymbolFont{ltrs}     {OT1}{pzc}{m}{it}
\DeclareSymbolFont{ltrsa}     {OMS}{cmsy}{m}{n}
\DeclareSymbolFont{ltrsb}     {OMS}{cmsy}{b}{n}

\DeclareSymbolFont{ltrsA}{U}{txmia}{m}{it}
\DeclareSymbolFontAlphabet{\mfrak}{ltrsA}

\def \czeta {\zeta}
\DeclareMathSymbol{\mmzeta}{\mathord}{ltrsa}{"5A}
\DeclareMathSymbol{\bmzeta}{\mathbf}{ltrsb}{"5A}
\def \mzeta {\text{\scriptsize $\mmzeta$}}
\def \bzeta {\text{\scriptsize $\bmzeta$}}
\def \cmu {\alpha}
\def \cnu {\beta}

\def \aa {A}
\def \bb {B}

\def \nnu {\nu}
\def \mtrx {{\raisebox{-3pt}{\hspace{-2.5pt}}}}
\DeclareMathSymbol{\cnfalg}{\mathord}{ltrs}{"63}
\DeclareMathSymbol{\confgr}{\mathord}{ltrsa}{"43}
\DeclareMathSymbol{\compgr}{\mathord}{ltrsa}{"4B}
\DeclareMathSymbol{\Wf}{\mathord}{ltrs}{"57}
\DeclareMathSymbol{\lgf}{\mathord}{ltrs}{"6C}
\def \Spin {\mathit{Spin}}
\def \spin {\mathit{spin}}
\def \SO {\mathit{SO}}
\def \so {\mathit{so}}
\def \spr {\cdot}
\def \SCT {C}

\def \trn {t}
\newcommand{\OMG}[1]{\Omega}
\def \MINK {M}

\newcounter{tmpc}

\newenvironment{mlist}{%
\setcounter{tmpc}{0}
\begin{list}{{\rm (}\textit{\alph{tmpc}}{\rm )}}{\usecounter{tmpc}
\setlength{\leftmargin}{20pt}
\setlength{\rightmargin}{0cm}
\setlength{\itemsep}{2.5pt}
\setlength{\topsep}{5pt}
\setlength{\labelsep}{1pt}
\setlength{\labelwidth}{18pt}
\setlength{\listparindent}{18pt}}
}{\end{list}}

\newenvironment{plist}{%
\setcounter{tmpc}{0}
\begin{list}{{\rm (\alph{tmpc})}}{\usecounter{tmpc}
\setlength{\leftmargin}{16pt}
\setlength{\rightmargin}{0cm}
\setlength{\itemsep}{2.5pt}
\setlength{\topsep}{5pt}
\setlength{\labelsep}{2pt}
\setlength{\labelwidth}{13pt}
\setlength{\listparindent}{18pt}}
}{\end{list}}

\newcommand{\gvspc}[1]{\raisebox{#1}{$\,$}$\!$}
\newcommand{\mgvspc}[1]{\raisebox{#1}{$\,$}\!}

\def \l| {\! \left| \,}
\def \r| {\right|}
\def \La {
\left\langle \!\!{\,}^{\mathop{}\limits_{}}_{\mathop{}\limits^{}}\right.}
\def \Ra {
\left. \!\!{\,}^{\mathop{}\limits_{}}_{\mathop{}\limits^{}}\right\rangle}
\def \Vl {
\left. \!\!{\,}^{\mathop{}\limits_{}}_{\mathop{}\limits^{}} \right|}

\newcommand{\vrestr}[2]{\!\left.\raisebox{#1}{$\,$}\!\right|_{\,
\raisebox{1pt}{\small \(#2\)}}}

\newcommand{\Brk}[1]{[ #1 ]}

\newcommand{\Brkl}[1]{[ #1}
\newcommand{\Brkr}[1]{#1 ]}

\newcommand{\Bbrk}[1]{\llbracket #1 \rrbracket}

\newcommand{\Aabrk}[1]{\llbracket #1 \rrbracket}
\newcommand{\Abrk}[1]{[ #1 ]}

\newcommand{\Aalbrk}[1]{\llbracket #1}
\newcommand{\Aarbrk}[1]{#1 \rrbracket}
\newcommand{\Albrk}[1]{[ #1}
\newcommand{\Arbrk}[1]{#1 ]}

\newcommand{\frc}[2]{%
\text{\raisebox{3.5pt}{\(#1\)}\hspace{-1pt}$/$\hspace{-1.5pt}\raisebox{-4pt}{\(#2\)}}}
\newcommand{\lfrc}[2]{%
\text{{\footnotesize \raisebox{3.5pt}{\(#1\)}\hspace{0pt}$/$\hspace{-1.5pt}\raisebox{-4pt}{\(#2\)}}}}

\newcommand{\lscw}[3]{\text{{\small \({#1}^{\text{{\tiny ${#2}$}}}_{\text{{\tiny ${#3}$}}}\)}}}

\def \R {{\mathbb R}}
\def \C {{\mathbb C}}
\def \Z {{\mathbb Z}}
\def \N {{\mathbb N}}

\def \Sr {{\mathbb S}}

\def \M {\overline{M}}

\def \HS {\mathcal{H}}
\def \DOM {\mathcal{D}}

\def \lvac {\left\langle 0 \! \left| \right. \right.  \!}
\def \rvac {\!\! \left. \left. \right| \! 0 \right\rangle}

\def \Cnj {+}
\def \Cnj {\star}
\newcommand{\cnj}[1]{{#1}^{\hspace{1pt} \Cnj}}
\newcommand{\cnjj}[1]{{#1}^{\hspace{1pt} \Cnj \hspace{1pt} \Cnj}}

\def \Lin {V}

\newcommand{\hr}[2]{h^{\hspace{-0.5pt}\left( \hspace{-0.5pt} #1 \hspace{-0.5pt} \right)}_{#2}}

\def \di {\partial}

\def \ID {{\mathbb I}}

\def \mgrt {\gg}

\def \Su {\mathop{\sum}\limits}

\newcommand{\LVEC}[1]{{\mathop{#1}\limits^{\leftarrow}}}

\def \dirz {z \hspace{-5pt} \text{\small $/$}}
\def \dirw {w \hspace{-5pt} \text{\small $/$}}
\def \diru {u \hspace{-5pt} \text{\small $/$}}
\def \dirv {v \hspace{-5pt} \text{\small $/$}}
\def \dirdi {\di \hspace{-5.5pt} /}
\def \pfun {\wp}
\def \zfun {\mathfrak{Z}}
\def \hcom {\mathfrak{H}}
\def \fundom {\mathfrak{D}}
\def \sphi {{\phi\raisebox{7.5pt}{\hspace{-4pt}\scriptsize $*$}}}
\def \schi {{\chi\raisebox{7.5pt}{\hspace{-4pt}\scriptsize $*$}}}

\newcommand{\txfrac}[2]{\frac{\raisebox{1pt}{$#1$}}{\raisebox{-3pt}{$#2$}}}
\newcommand{\Txfrac}[2]{\frac{\raisebox{2pt}{$#1$}}{\raisebox{-5pt}{$#2$}}}
\newcommand{\smtxfrac}[2]{\text{\scriptsize \(\frac{\raisebox{0pt}{$#1$}}{\raisebox{0pt}{$#2$}}\)}}
\newcommand{\nfrc}[2]{%
\text{\raisebox{1pt}{\(#1\)}\hspace{0pt}$/$\hspace{-1.5pt}\raisebox{-4pt}{\(#2\)}}}

\def \podr {\hspace{-2pt}}
\DeclareMathSymbol{\DLT}{\mathord}{ltrs}{"44}

\DeclareMathSymbol{\Vol}{\mathord}{ltrs}{"56}
\newcommand{\VOL}[1]{\Vol_{\hspace{-0.5pt}#1}}
\def \MED {\text{\small $\mathcal{E}$}}

\begin{document}

\markboth{N.M. Nikolov, I.T. Todorov}
{Elliptic Thermal
Functions and Modular Forms in a GCI QFT}

%
\catchline{}{}{}{}{}
%

\title{ELLIPTIC THERMAL
	CORRELATION FUNCTIONS AND MODULAR FORMS
	IN A GLOBALLY CONFORMAL INVARIANT QFT\footnote{%
	to appear in Rev. Math. Phys.}}

\author{NIKOLAY M. NIKOLOV\footnote{mitov@inrne.bas.bg, nikolov@theorie.physik.uni-goe.de} 
\quad and \quad IVAN T. TODOROV\footnote{todorov@inrne.bas.bg, itodorov@theorie.physik.uni-goe.de}}

\address{Institute for Nuclear Research and Nuclear Energy,  \\
Tsarigradsko Chaussee 72,
BG-1784 Sofia, Bulgaria \\
and Institut f\"ur Theoretische Physik, Universit\"at G\"ottingen, \\
Friedrich--Hund--Platz 1, D--37077 G\"ottingen, Germany}

\maketitle

\begin{history}
\today
\end{history}

\begin{abstract}
Global conformal invariance (GCI) of quantum field theory (QFT)
in two and higher space--time dimensions implies the Huygens' principle,
and hence, rationality of correlation functions of observable
fields~\cite{NT01}.
The conformal Hamiltonian $H$ has discrete spectrum
assumed here to be finitely degenerate.
We then prove that
thermal expectation values of field products on compactified Minkowski space
can be represented as finite linear combinations of basic (doubly periodic)
\textit{elliptic functions} in the conformal time variables
(of periods $1$ and $\tau$) whose coefficients are, in general,
formal power series in  \(q^{\frac{1}{2}} = e^{i \pi \tau}\) involving
spherical functions of the ``space--like'' fields' arguments.
As a corollary, if the resulting expansions converge to meromorphic
functions, then the finite temperature correlation functions are elliptic.
Thermal $2$--point functions of free fields are computed and shown to display
these features.
We also study modular transformation properties of Gibbs energy mean
values with respect to the (complex) inverse temperature $\tau$
(\(\mathit{Im} \, \tau = \txfrac{\beta}{2\pi} > 0\)).
The results are used to obtain the thermodynamic limit
of thermal energy densities and correlation functions.
\end{abstract}




\tableofcontents

\markboth{N.M. Nikolov, I.T. Todorov}
{Elliptic Thermal
Functions and Modular Forms in a GCI QFT}


\section{Introduction}\label{Sec.1}

The {\it modular group}
\(\mathit{SL} \left( 2,\Z \right) \, ( \, =: \Gamma \left( 1 \right))\)
arises as the symmetry
group of an oriented 2-dimensional lattice.
Usually~--~including our case~--~this is the period lattice
of an elliptic function.
The factor group \(\mathit{PSL} \left( 2,\Z \right)=
\nfrc{\mathit{SL} \left( 2,\Z \right)}{\Z_2}\) of $\mathit{SL} \left( 2,\Z \right)$
with respect to its 2-element centre \(\Z_2 \equiv \nfrc{\Z}{2\Z}\)
acts faithfully by fractional linear transformations on the upper half-plane
\begin{equation}\label{n1}
\hcom :=
\left\{ \tau \in \C : \mathit{Im} \, \tau > 0 \right\}
, \hspace{8pt}
g \left( \tau \right) =
\frac{a \, \tau \! + \! b}{c \, \tau \! + \! d}
\hspace{8pt} \mathrm{for} \hspace{8pt}
g = \left(\hspace{-2pt}\begin{array}{cc}
a & b \\ c & d \end{array}\hspace{-1pt}\right)
\! \in
\mathit{SL} \left( 2, \Z \right)
.
\end{equation}
The modular inversion, the involutive $S$--transformation of
$\hcom$,
\begin{equation}\label{eq1.1}
S \, = \, \left( \begin{array}{cr} 0 & -1 \\ 1 & 0 \end{array} \right)
: \tau \to -\,\frac{1}{\tau}
\quad (\mathit{Im} \, \tau > 0)
\, , \qquad
\end{equation}
which relates high and low temperature behaviour, is the oldest and
best studied example of a duality transformation \cite{KW41}
(for a recent reference in the context of elliptic functions that
provides a historical review going back to 19th century
work~--~see~\cite{DK02}).
It naturally appears in the study of finite temperature correlation
functions in a conformally invariant field theory (CFT).
The case of 2-dimensional (2D) CFT has been thoroughly studied
from the point of view of vertex algebras in \cite{Zh96}.
The present paper builds on the observation that this
analysis can be extended in a straightforward manner to the
recently developed GCI~QFT (see~\cite{NT01} \cite{NST02} \cite{NST03}).

\subsection{Conformal invariance in QFT}\label{sbsc1.1}

The conformal group $\confgr$ acts on Minkowski space $M$ by local diffeomorphisms 
which preserve the conformal class of the metric form, - i.e. multiply it 
by a positive factor. Unlike the Poincar\'e group, $\confgr$ acts, in general, by 
nonlinear transformations on $M$ which may have singularities on a cone 
(or on a hyperplane). Furthermore, $\confgr$ has an infinite sheeted universal 
cover. In view of these peculiarities, there exist different notions of 
conformal invariance in QFT. In order to make clear the concept of GCI 
QFT we shall briefly discuss these notions in the framework of axiomatic 
QFT~(\cite{SW}).

The weakest condition of conformal invariance in QFT is the {\it infinitesimal 
conformal invariance} of the Wightman functions. This yields 
a system of first order differential equations for each Wightman function 
(displayed in Sect.~2.2). According to the Bargman-Hall-Wightman theorem
(see \cite{Jost 65} Sect.~IV.4--5) the Wightman functions are boundary values of 
analytic functions, holomorphic in a complex domain, the so called 
{\it symmetrized extended tube}, which contains all non-coinciding euclidean 
arguments. As a result, the same (complexified) system of differential 
equations is satisfied by these analytic functions and hence by their 
(real analytic) euclidean restrictions that define the so called 
{\it non-coinciding Schwinger} (or {\it Euclidean Green}) functions which are, 
therefore, invariant under (Euclidean) infinitesimal conformal 
transformations.

We thus see that the conditions of infinitesimal conformal invariance for 
each element in the hierarchy of functions~--~Wightman distributions, their 
analytic continuations, and the Schwinger functions~--~are equivalent. The 
global, i.e. group, version of conformal invariance is more subtle.

We recall that there exist {\it conformal compactifications} of both Minkowski and Euclidean 
space, such that the local actions of the corresponding conformal groups can be extended 
to everywhere defined ones.
The conformally compactified Euclidean space is just the (simply connected) 
sphere $\Sr^D$. It follows that the infinitesimal Euclidean conformal 
invariance is equivalent to invariance under the Euclidean conformal 
group. Compactified Minkowski space $\M$, on the other hand, is 
isomorphic to $\nfrc{\Sr^1 \times \Sr^{D-1}}{\Z_2}$ so that it has an infinite sheeted 
universal cover, $\widetilde{M}$. One can only integrate, in general, the conditions 
of infinitesimal invariance to invariance on this infinite sheeted cover,
the finite conformal transformations on $\widetilde{M}$ becoming multivalued if 
projected on $\M$. Assuming euclidean conformal invariance L\"uscher 
and Mack \cite{LM} have thus established invariance of the QFT continued to 
$\widetilde{M}$ under the infinite sheeted covering of the Minkowski space conformal 
group and called it ``global conformal invariance'' {\it on the universal covering space}.
The above analysis shows that it is, in fact, equivalent to infinitesimal
conformal invariance in Minkowski space.

By contrast, the GCI condition introduced in \cite{NT01}\footnote{A special case of such a 
condition~--~in the context of (generalized)free fields~--~has been displayed earlier in \cite{HSS}; 
see also \cite{GO} where a condition of this type is discussed in the framework of 2D CFT
and \cite{H} for a retrospective view on the subject.}, invariance under finite conformal 
transformations $g$ {\it in Minkowski space} $M$ (whenever both $x$ and $gx$ are in $M$) is 
stronger since it allows to continue the Wightman functions to invariant distributions on $\M$.
Combined with locality GCI on $M$ implies the {\it Huygens' principle}~--~the
vanishing of field (anti)commutators for non-isotropic separations°--°since it
allows to transform space--like into time like intervals.
Under Wightman axioms the Huygens principle is equivalent to
{\it strong locality} (i.e. the algebraic condition~(\ref{eqn2.2})). 
(Note that only in even dimensional space-time the canonical free massless fields and the stress-energy 
tensor satisfy the Huygens principle.)
Strong locality and energy 
positivity imply the rationality of Wightman functions (cf. Theorem~3.3 
below), thus excluding non-integer anomalous field dimensions. In the case 
of 2D CFT the GCI incorporates the notion of chiral algebra which has served as 
a starting point for developing the important mathematical concept of a 
{\it vertex algebra} - see \cite{Bo86,Bo97}, \cite{Ka96}, \cite{FBZ01} and further references in the 
latter two books. Moreover, it also includes non-chiral 2D fields with rational 
correlation functions. (The simplest example is given by the energy 
density of conformal weight $(\txfrac{1}{2}, \txfrac{1}{2})$ in the vacuum sector of the 
critical Ising model.) The {\it primary fields}, which may well have 
anomalous dimensions and non-trivial braiding properties (as the 
magnetization field in the Ising model), appear in this framework as 
intertwiners between the vacuum and other positive energy representations 
of the GCI algebra. It is then expected that only the 4-dimensional 
counterpart of such intertwining primary fields may display anomalous 
time-like braiding relations of the type discussed in \cite{S01}.

\subsection{Why thermal correlation functions should be elliptic in the conformal time 
differences?}\label{sbsc1.2}

The main idea is simple to explain.
A GCI~QFT lives on compactified Min{\-}kow{\-}ski space $\M$
of dimension $D$ which has a natural complex vector
parametrization:
\begin{equation}\label{eq1.2}
\M
\hspace{-1pt} = \hspace{-1pt}
\Sr^1 \hspace{-1pt} \times \Sr^{D-1}
\hspace{-2pt}\raisebox{-1pt}{$\left/\raisebox{9pt}{}\right.$}
\raisebox{-4pt}{$\Z_2$}
\hspace{-1pt} = \hspace{-1pt}
\left\{ z_{\cmu}
\hspace{-1pt} = \hspace{-1pt}
e^{2\pi i \hspace{1pt} \czeta} \hspace{1pt} u_{\cmu} :
\czeta \in \R,\hspace{1pt}
u^{\, 2}
\hspace{-1pt} := \hspace{-1pt}
\mbf{u}^{\, 2} \hspace{-2pt} + u_D^2
\hspace{-1pt} = \hspace{-1pt} 1,\hspace{1pt}
u \in \R^D \right\}
\hspace{-2pt},
\end{equation}
$\czeta$ being the \textit{conformal time} variable.
The coordinates $z$ in Eq.~(\ref{eq1.2}) are obtained by a complex
conformal transformation introduced in Sect.~\ref{ssec2.1} (Eq.~(\ref{t2.2}))
of the Cartesian coordinates of Minkowski space
generalizing the Cayley transform
(inverse stereographic projection) of the chiral (i.~e. 1-dimensional) case.
They have been first introduced for \(D=4\) in \cite{T86} using the Cayley
(\(u(2) \to U(2)\)) compactification map, and were generalized
to arbitrary $D$ in~\cite{NT02}; the reader will find a geometric
introduction to this and more general systems of charts in~\cite{N03}.
The use of Euclidean metric in (\ref{eq1.2}) does not mean, of course,
that we are working within the Euclidean picture of QFT at this point.
We recall that the Euclidean rotation group $SO(D)$ is a subgroup of 
the Minkowski space conformal group $SO(D,2)$ and that (\ref{eq1.2}) 
(involving the Euclidean unit sphere $\Sr^{D-1}$) does indeed represent 
compactifed Minkowski space. (This is made clear in Sect.~2.1 by 
exhibiting its relation to the Dirac projective quadric.)

Transforming the fields in these coordinates we obtain an equivalent
representation of the GCI local fields on $\M$ called
(\textit{analytic} or) $z$--\textit{picture}.
Since the transformation is conformal the vacuum correlation (Wightman)
functions do not change their form.
For example, the $z$--picture scalar field $\phi \left( z \right)$ of (integer)
dimension $d$ has rational correlation functions like
\begin{equation}\label{eq1.3}
\lvac \phi \left( z_1 \right) \phi \left( z_2 \right) \rvac
\, = \, \left( z_{12}^{\, 2} \right)^{-d}
\, , \quad
z_{12} \, = \, z_1 - z_2
\, , \qquad
\end{equation}
invariant under $D$--dimensional inhomogeneous complex rotation group.
Let us note that we will treat the fields as formal power series in $z$ and 
$\Txfrac{1}{z^{\, 2}}$ which is shown in \cite{N03} to be completely equivalent to 
the Wightman approach (with GCI) which treats local fields as operator valued 
distributions.

The conformal Hamiltonian $H$, with respect to which we will consider the
thermal correlation functions,
gives rise to a multiplication of $z$ by a phase factor (and hence, to
a translation of $\czeta$ in Eq.~(\ref{eq1.2})).
This suggests introducing
a \textit{real compact picture field} $\phi \left( \czeta,\, u \right)$
related to $\phi \left( z \right)$ by
\begin{equation}\label{eq1.4}
\phi \left( \czeta,\, u \right) \, = \,
e^{2\pi i\hspace{1pt}d\hspace{1pt}\czeta} \hspace{1pt} \phi
\left( e^{2\pi i\hspace{1pt}\czeta}\hspace{1pt} u \right)
\, . \
\end{equation}
Then $H$ acts on it by (infinitesimal) translation in $\czeta\,$:
\begin{equation}\label{eq1.5n}
e^{2 \pi i \hspace{1pt} t \hspace{1pt} H}
\, \phi \left( \czeta,\, u \right) \,
e^{-2 \pi i \hspace{1pt} t \hspace{1pt} H}
\, = \, \phi \left( \czeta+ t ,\, u \right)
\, . \qquad
\end{equation}
Since $H$ has an integer or half-integer spectrum in the
vacuum sector state space it follows that:
\begin{equation}\label{eq1.6n}
\phi \left( \czeta+1,\, u \right) =
\left( -1 \right)^{2 \, d}
\phi\left( \czeta,\, u \right)
\, , \
\end{equation}
i.~e.,
\textit{the conformal time evolution is periodic or anti-periodic
with period~$1$} in the compact picture,
so that the vacuum and the thermal correlation
functions will be also (anti-) periodic.

The second period $\tau$ comes from the statistical quantum physics:
there it is pure imaginary and proportional to the inverse absolute temperature.
More precisely, for any (real) Bose field $\phi$ with an invariant dense domain
$\DOM{}$ (common for all fields' and actually coinciding with the finite energy
space spanned by eigenvectors of $H$~--~see Proposition~\ref{prp:2.2}),
we are going to construct the {\it partition function}
\begin{equation}\label{eq1.7n}
Z \left( \tau \right) \, = \, \mathit{tr}_{\DOM{}} \,
\left( q^H \right)
\, , \quad
q \, = \, e^{2 \pi i \, \tau}
\, , \quad
\mathit{Im} \, \tau > 0
\quad (|q|<1)
\end{equation}
and the {\it Gibbs correlation functions}
\begin{equation}\label{eq1.8n}
\La \phi \left( \czeta_1,\, u_1 \right)
\dots \phi \left( \czeta_n,\, u_n \right) \Ra_q
\, := \,
\frac{\textstyle 1}{\textstyle Z \left( \tau \right)} \, \mathit{tr}_{\DOM{}}
\left\{ \phi \left( \czeta_1,\, u_1 \right)
\dots \phi\left( \czeta_n,\, u_n \right) \, q^H	\right\}
\end{equation}
as meromorphic functions in $\tau$, $\czeta_k$ and $u_k$ (\(k=1,\dots,n\))
in a suitable domain of $\C^{nD+1}$.
Sure, the existence of Gibbs equilibrium states with the above properties
requires additional assumptions extending the notion of classical phase
space volume.
Such extra assumptions are needed in any axiomatic treatment of
thermodynamic properties of local QFT\footnote{%
For a general discussion of this point within Haag's operator algebra
approach~--~see Sect.~V.5 of \cite{Haag}, where Buchholz nuclearity condition
\cite{Bu86} \cite{Bu90} is advocated and reviewed.}.
Our study of GCI~QFT is facilitated by the fact that the conformal
Hamiltonian $H$ has a (bounded below) discrete spectrum
\(\left\{\raisebox{10pt}{\hspace{-2pt}}\right.
\txfrac{n}{2} : n = 0,1,\dots
\left.\raisebox{10pt}{\hspace{-2pt}}\right\}\).
The partition function~(\ref{eq1.7n}) exists for any inverse temperature
\(\beta = 2\pi \, \mathit{Im} \, \tau \, ( \, > 0)\) iff the growth of dimension
\(d \! \left(\raisebox{10pt}{\hspace{-2pt}}\right.
\txfrac{n}{2}
\left.\raisebox{10pt}{\hspace{-2pt}}\right)\)
of the $n$th eigenspace of $H$ is slower than any exponential
$e^{\varepsilon n}$ (\(\varepsilon > 0\)).
Moreover, in the GCI~QFT it is sufficient to assume that $H$ has
just finitely degenerate spectrum to ensure the existence of
thermal correlation functions~(\ref{eq1.8n})
(and the partition function~(\ref{eq1.7n})) as \textit{formal power series} in
\(q^{\frac{1}{2}} = e^{i \pi \tau}\) with coefficients
which are {\it symmetric rational functions} in
\(\left( e^{\pi i \, \czeta_1},u_1 \right)\), $\dots$,
\(\left( e^{\pi i \, \czeta_n},u_n \right)\), as it is shown in
Sect.~\ref{sec.2n}.
This allows then to extend the heuristic argument, given in \cite{Zh96},
which makes it plausible that the
{\it Kubo-Martin-Schwinger (KMS) property \cite{HHW67}}\footnote{%
For a later discussion combining nuclearity, KMS and Lorentz
invariance~--~see~\cite{BB}.}
\begin{equation}\label{addXX}
\La \phi \left( \czeta_1,\, u_1 \right)
\dots \phi \left( \czeta_n,\, u_n \right) \Ra_q
\, = \,
\La
\phi \left( \czeta_2,\, u_2 \right)
\dots \phi \left( \czeta_n,\, u_n \right)
\phi \left( \czeta_1+\tau,\, u_1 \right)
\Ra_q
\, \
\end{equation}
implies that the functions (\ref{eq1.8n}) are doubly periodic meromorphic
functions with periods $1$ and $\tau$ in
\(\czeta_{jk} = \czeta_j - \czeta_k\); in other words,
they are {\it elliptic} functions.
In Sect.~\ref{msec:3} we give a rigorous interpretation of this argument
thus proving that the finite temperature correlation functions~(\ref{eq1.8n}) has
the form of \textit{finite linear combinations}
of basic (series of) \textit{elliptic functions} in the conformal time variables
whose coefficients are, in general, formal power series in
\(q^{\frac{1}{2}} = e^{i \pi \tau}\) involving spherical functions of the
angular fields' arguments $u_k$ (Theorem~\ref{adprp1}).\footnote{%
We note that our results are valid in any space--time dimension $D$
which, in particular, for the \(D=1\) case,
corresponding to the chiral projection of the 2-dimensional CFT, implies
that under the assumptions of convergence of all the traces of
products of fields' modes (including the partition function~(\ref{eq1.7n})),
the finite temperature correlation functions are
convergent to elliptic functions (since then there are no additional
angular variables).}
Let us stress that our main result, Theorem~\ref{adprp1}, takes into account the 
most general {\it purely algebraic} properties of the theory only. 
As noted above additional hypotheses of topological character are necessary 
in order to guarantee the existence of the thermal expectation values as 
meromorphic functions. 
In this case our analysis tells us that these meromorphic functions 
are automatically elliptic (Corollary~3.6). 
We shall demonstrate that this is indeed the case for conformally invariant
free fields by computing explicitly their Gibbs 2--point functions.

\subsection{Basic (anti)periodic functions. Content of the paper}\label{sbsc1.3}

An elliptic function is characterized by its poles
and their residues
(in the fundamental domain).
The poles of the thermal correlation functions should be the same
as the poles of the operator product expansions (OPE): they only appear at
mutually isotropic field arguments.
In the compact picture the light cone equation factorizes:
\begin{equation}\label{eqn?}
0 \, = \,
z_{12}^{\, 2} \, = \, \left(
e^{2\pi i\hspace{1pt}\czeta_1}\hspace{1pt} u_1 -
e^{2\pi i\hspace{1pt}\czeta_2}\hspace{1pt} u_2 \right)^2
\, \equiv \,
-4 \, e^{2\pi i\hspace{1pt}\left( \czeta_1+\czeta_2 \right)} \,
\sin \,\pi \czeta_+ \,
\sin \, \pi \czeta_- ;
\end{equation}
here we have introduced the variables
\begin{equation}\label{eq1.6}
\czeta_{\pm} \, = \, \czeta_{12} \pm \alpha
\, , \quad
\mathrm{for} \quad
u_1 \spr u_2 \, = \, \cos \, 2\pi\alpha
\, . \qquad
\end{equation}
Therefore, the basic elliptic functions occurring in the theory depend on
the variables of type (\ref{eq1.6}) and have poles on the
lattice spanned by the periods $1$ and~$\tau\,$.

Taking into account the fermionic case the above statements
are modified, the periodicity in both periods $1$ and $\tau$
being replaced by antiperiodicity.
We are thus led to the set
\(\left\{\raisebox{9pt}{\hspace{-2pt}}\right.
p_k^{\kappa,\lambda} \left( \zeta,\, \tau \right)
: k=1,2,\dots,\, \kappa,\lambda = 0,1
\left.\raisebox{9pt}{\hspace{-2pt}}\right\} \) of basic (elliptic)
functions, uniquely characterized by the conditions:
\begin{mlist}
\item[(\textit{i})\hspace{8pt}]
\(p_k^{\kappa,\lambda} \left( \zeta, \tau \right)\) are meromorphic functions
in \(\left( \zeta, \tau \right) \in \C \times \hcom\)
with exactly one pole at \(\zeta = 0\)
of order $k$ and residue $1$
in the domain
\(\left\{\raisebox{9pt}{\hspace{-3pt}}\right.
\alpha\tau+\beta :\) \(\alpha, \beta \in\)
\(\left[ 0,\, 1 \right)
\left.\raisebox{9pt}{\hspace{-3pt}}\right\} \subset \C\)
for all \(\tau \in \hcom\) and \(k = 1,2,\dots\);\gvspc{10pt}
\item[(\textit{ii})\hspace{5pt}]
\(p_{k+1}^{\kappa,\lambda} \left( \zeta, \tau \right) =
- \Txfrac{1}{k} \,
\Txfrac{\di}{\di \zeta} \,
p_k^{\kappa,\lambda} \left( \zeta, \tau \right)
\) \ for \ \(k = 1,2,\dots\);
\item[(\textit{iii})\hspace{2pt}]
\(p_k^{\kappa,\lambda} \left( \zeta+1, \tau \right) = \left( -1 \right)^{\lambda}
p_k^{\kappa,\lambda} \left( \zeta, \tau \right)\) \ for \ \(k = 1,2,\dots\);
\item[(\textit{iv})\hspace{3pt}]
\(p_k^{\kappa,\lambda} \left( \zeta+\tau, \tau \right) =
\left( -1 \right)^{\kappa} p_k^{\kappa,\lambda} \left( \zeta, \tau \right)\)
\ for \ \(k+\kappa+\lambda > 1\);\gvspc{10pt}
\item[(\textit{v})\hspace{6pt}]
\(p_k^{\kappa,\lambda} \left( -\zeta, \tau \right)
= \left( -1 \right)^k \, p_k^{\kappa,\lambda} \left( \zeta, \tau \right)\)
\ for \ \(k = 1,2,\dots\).\gvspc{10pt}
\end{mlist}
Note that for \(k=1\) and \(\kappa=\lambda=0\)
at most one of conditions (\textit{iii}) and (\textit{iv}) can be satisfied
and we have chosen the first one.
This is a natural choice since the periodicity with
period $1$ in the conformal time is coupled to the periodicity in the angle
$\alpha$. It leads to a difference
between our \(p_1^{00} =: p_1\) and
\(p_2^{00} =: p_2\)--functions, and the
Weierstrass $\zfun$-- and $\pfun$--functions
(Eqs.~(\ref{fn_W1}) and~(\ref{fn_W2})), respectively,
by linear functions in $\zeta$ (see Eqs.~(\ref{eqnA.19n}) and~(\ref{eqnA.21a});
the Weierstrass functions have the advantage that they
have simple modular transformation laws).
In Appendix~A
(see Proposition~\ref{prp:A.2})
we allow for a more general
$U(1)$ character replacing $(-1)^{\kappa}$ in condition
(\textit{iv}):
\(p_k^{\kappa,\lambda} \left( \zeta+\tau
,\tau,\mu \right) =
\left( -1 \right)^{\kappa} \, e^{-2\pi i \hspace{1pt} \mu}
\, p_k^{\kappa,\lambda} \left( \zeta,\tau,\mu \right)
\), where the parameter $\mu$ can be interpreted as
{\it chemical potential} in physical applications
(and \(p_k^{\kappa,\lambda} \left( \zeta,\tau,0 \right)
= p_k^{\kappa,\lambda} \left( \zeta,\tau \right)\)).

The $n$--point correlation functions (\ref{eq1.8n})
are elliptic in \(\czeta_{jk} = \czeta_j -\czeta_k\) with poles
at \(\pm\alpha_{jk} + m+n\tau\) (\(n,\, m \in \Z\)),
where \(\cos \, 2\pi\alpha_{jk} = u_j \spr u_k\,\).
One cannot expect, however, that they are homogeneous
under \textit{modular transformations}
\begin{equation}\label{eq1.11n1}
g \left( \czeta,\, \tau \right) \, = \,
\left( \frac{\czeta_{jk}}{c \, \tau + d},\,
\frac{a \, \tau + b}{c \, \tau + d} \right)
\quad \mathrm{for} \quad
g \, = \, \left(\hspace{-2pt}\begin{array}{cc}
a & b \\ c & d \end{array}\hspace{-1pt}\right)
\, \in \, \mathit{SL} \left( 2,\, \Z \right)
\end{equation}
since $\alpha_{jk}$, playing the role of spherical
distances for \(D > 2\), are not invariant under rescaling
\(\alpha_{jk} \mapsto
\Txfrac{\alpha_{jk}}{c \, \tau + d}\,\).
One can hope to recover modular covariance for
their (always well defined,
in Wightman theories~--~see \cite{Bo64})
1-dimensional restrictions corresponding to
\begin{equation}\label{eq1.12n2}
u_1 \, = \, u_2 \, = \, \dots \, = \, u_n
\, , \quad
\alpha_{jk} \, = \, 0
\, . \
\end{equation}
It turns out that the restricted 2-point function of a
\(d = 1\) free massless scalar field for \(D = 4\)
indeed transforms homogeneously (of degree $2$) under
the modular transformations~(\ref{eq1.11n1}).
The corresponding
energy mean value in an equilibrium state,
\begin{equation}\label{eq1.13n2}
\La \hspace{-1pt} H \hspace{-1pt} \Ra_q \, = \, \frac{1}{Z \left( \tau \right)}
\, \mathit{tr}_{\DOM{}} \left( H \, q^H \right)
\, \
\end{equation}
is a modular form of weight $4$
(and level \(\Gamma \left( 1 \right) \hspace{1pt} ( \, \equiv
\mathit{SL} \left( 2,\, \Z \right))\))~--~after
shifting the vacuum energy (Sect.~\ref{Ssec.4.1}).

The paper is organized as follows.

In Sect. 2.1 we give a concise review of the basic properties of the conformal 
group and its Lie algebra, and introduce the basic complex parametrization 
of Minkowski space which we use throughout this paper. It allows us to 
formulate in Sect. 2.2 an algebraic counterpart of the Wightman axioms in 
what we call the {\it analytic ($z$) picture}. We sum up the implications of 
these axioms in Sect. 3.1 where we also give an introduction to the purely 
algebraic approach to GCI QFT in terms of formal power series. In Sect. 3.2 
we obtain the general form of the thermal correlation functions. 
In Sect.~\ref{SEC:3} we calculate the finite temperature correlation
functions in the (generalized) free field GCI models starting
with their relation to the Wightman functions found in Sect.~\ref{Sec:3}.
The cases of physical free fields in \(D = 4\) dimensions:
the massless scalar, Weyl and electromagnetic fields are considered in
Sects.~\ref{sec:5}, \ref{Ssec.5.1} and~\ref{Ssec.5.2}, respectively.
We have also studied examples of subcanonical free fields
(for \(D=4\) and \(D=6\)).
The ``thermodynamic limit'' in which the compactification radius
$R$ goes to infinity
(so that $\M$ is restricted to $M$ and time is no longer cyclic)
is considered in Sect.~\ref{Sect.7n} where it is shown that
the thermal correlation functions have Minkowski space limits.
The results are summed up and discussed in Sect.~\ref{sec:7}.
The reader will find our conventions about elliptic
functions and modular forms, used in the text, in
Appendix~A.

\section{Globally Conformal Invariant QFT on
Compactified Min\-kow\-ski Space}\label{sec.2n}

In the GCI~QFT the natural choice of the
\textit{conformal group} $\confgr$
is the connected spinor group
\(\Spin \left( D,2 \right) \, ( \, \cong \confgr)\).
Then the complexified conformal group will be
\(\confgr_{\C} \cong \Spin_{\C} \left( D+2 \right)\).
The conformal Lie algebra will be denoted by
\(\cnfalg \, ( \, \cong \spin (D,2) \cong \so (D,2))\)
and its complexification, by~$\cnfalg_{\C}$.
We begin this section with recalling some basic facts about the conformal
group and its action on compactified Minkowski space.

\subsection{Affine
coordinate systems on compactified Minkowski space}\label{ssec2.1}

Let $M$ be the $D$-dimensional Minkowski space, with coordinates
\(x = (x^0,\)
\(\mbf{x} = ( x^1, \dots,\) \(x^{D-1} ) ) \in \R^D\)
and Poincar\'e invariant interval
\(x_{12}^{\, 2} = \mbf{x}_{12}^{\, 2} -
\left( x_{12}^0 \right)^2\),
\(x_{12} = x_1-x_2\),
\(\mbf{x}_{12}^{\, 2} = \Su_{j \, = \, 1}^{D-1}
\left( x_{12}^j \right)^2\).
The group of conformal transformations of $M$
is defined as the group of
(local) diffeomorphisms of $M$ (\(\ni x \mapsto y\))
preserving the conformal class of the infinitesimal metric
\(dx^{\, 2} \, ( \, \equiv dx^{\mu} dx_{\mu})\),
i.e., mapping $dx^{\, 2}$ on
\(dy^{\, 2} =\) \(\omega^{-2} \! \left( x \right)\) \(dx^{\, 2}\)
(\(\omega \left( x \right) \neq 0\)).
It is finite
(\(\txfrac{\left( D+1 \right) \left( D+2 \right)}{2}\))
dimensional
for \(D \geqslant 3\), due to the Liouville theorem,
and is generated by:
\begin{itemize}
\item
the Poincar\'e translations
\(e^{i \, a \spr P} \left( x \right) \, ( \, \equiv
e^{i \, a^{\mu} P_{\mu}} \left( x \right) \, ) \, = x + a\)
(for \(x,a \in M\)),
\item
the Lorentz transformations
\(e^{t \Omega_{\mu\nu}}\),
\(0 \leqslant \mu < \nu \leqslant D-1\)
(\(\Omega_{\nu\mu} = -\Omega_{\mu\nu}\)),
\item
the dilations \(x \mapsto \rho x\), \(\rho > 0\)
\item
and the \textit{special conformal transformations}
\(e^{i \, a \spr K} \left( x \right) = \Txfrac{x+x^{\, 2} \, a}{
1 + 2 \, a \spr x + a^{\, 2} \, x^{\, 2}}\)
(which has obvious singularities).
\end{itemize}
The corresponding Lie algebra is isomorphic to the Lie algebra
of the pseudo--orthogonal group \(\SO (D,2)\) .
Recalling this isomorphism we introduce
the basis of
infinitesimal (pseudo) rotations
\(\Omega_{ab} \, ( \, = -\Omega_{ba})\),
where the indices $a$ and $b$
take values \(-1,0,\dots,D\),
the underlying orthonormal basis
\(\vec{e}_a \in \R^{D,2}\) (\(a =\) $-1,$ $0,$ $1,$ $\dots,$ \(D\))
satisfying \(\vec{e}_{\alpha}^{\, 2} =\) \(1 =\)
\(- \vec{e}_{-1}^{\, 2} =\) \(- \vec{e}_0^{\, 2}\),
\(\alpha = 1,\dots,D\)
(cf. \cite{NT01} Appendix~A).
The generators $\Omega_{ab}$ are characterized by the following
nontrivial commutation relations
\begin{eqnarray}\label{n2.9n}
&
\left[ \hspace{1pt} \Omega_{a\cmu} \, , \, \Omega_{b\cmu} \, \right]
\, = \, \Omega_{ab} \, ( \, = \,
\left[ \hspace{1pt} \Omega_{\cmu a} \, , \, \Omega_{\cmu b} \, \right] )
\quad \text{for} \ \cmu = 1,\dots,D,
& \nonumber \\ &
\left[ \hspace{1pt} \Omega_{\kappa a} \, , \, \Omega_{\kappa b} \, \right] \, = \,
- \Omega_{ab} \quad \text{for} \ \kappa = -1,0
\, ; &
\end{eqnarray}
$iP_{\mu}$, $iK_{\mu}$ and the dilations are expressed in terms of them as:
\beq\label{e2.3n}
i \hspace{1pt} P_{\mu} = -\Omega_{-1\mu} - \Omega_{\mu D}
, \ \
i \hspace{1pt} K_{\mu} = -\Omega_{-1\mu} + \Omega_{\mu D}
, \ \
\rho^{- \Omega_{-1D}} \left(  x\right) = \rho \, x
\ \ (\rho > 0)
\, ;
\eeq
the Lorentz generators \(\Omega_{\mu\nu}\)
correspond to \(0 \leqslant \mu,\nu \leqslant D-1\).
In fact, the group $\SO (D,2)$ itself has
an action on $M$ by (rational) conformal transformations.
It is straightforward to derive this action using the
Klein--Dirac construction of compactified Minkowski space
$\M$~(\ref{eq1.2}), realized as the projective quadric of $\R^{D,2}$
(\cite{D36}, see also \cite{NT01} Appendix~A for
a survey
adapted to our present purposes and notation).
The Minkowski space $M$ is mapped into a dense subset (identified with $M$)
of $\M$ thus providing an affine chart in $\M$.

Other affine charts in $\M$ can be obtained
by applying conformal transformations.
In particular, the following chart in the \textit{complex}
compactified Minkowski space $\M_{\C}$
plays a crucial role in the GCI~QFT.

Let \(M_{\C} := M + i M\) be the complexified Minkowski
space, with coordinates \(\mzeta = \left(\mzeta^0, \bzeta \right) =
x+ iy \in \C^D\),
\(\mzeta_{12}^{\, 2} \equiv \left( \mzeta_1 - \mzeta_2 \right)^2
= \bzeta_{12}^{\, 2} - \left( \mzeta_{12}^0 \right)^2\)
being the Poincar\'e invariant interval
and let $E_{\C}$ be the complex Euclidean $D$--dimensional space
with coordinates $z=\left(\mbf{z},\, z_D\right) \in \C^D$
and Euclidean invariant interval
\(z_{12}^{\, 2} = \mbf{z}_{12}^{\, 2} + \left( z_{12}^D \right)^2\),
\(z_{12} = z_1 - z_2\).
The rational complex coordinate transformation (see \cite{T86},
\cite{NT02},
\cite{N03}):
\begin{equation}\label{t2.2}
g_c : M_{\C} \left( \ni \mzeta \right) \to E_{\C} \left( \ni z\right)
\hspace{-1pt} , \hspace{8pt}
\mbf{z} \hspace{-1pt} = \hspace{-1pt}
\frac{\bzeta}{\omega \! \left( \mzeta \right)}
\hspace{1pt} , \hspace{8pt}
z_D \hspace{-1pt} = \hspace{-1pt}
\frac{1 \hspace{-1pt} - \hspace{-1pt} \mzeta^{\, 2}}{
2\, \omega \! \left( \mzeta \right)}
\hspace{1pt} , \hspace{8pt}
\omega \! \left( \mzeta  \right) \hspace{-1pt} = \hspace{-1pt}
\frac{1 \hspace{-1pt} + \hspace{-1pt} \mzeta^2}{2} - i \mzeta^0
\end{equation}
is a complex conformal map (with singularities) such that
\begin{eqnarray}\label{t2.3}
&&
z_{12}^{\,2 } = \frac{\mzeta_{12}^{\, 2}}{
\omega \left( \mzeta_1 \right)\omega \left( \mzeta_2 \right)}
\, , \quad
d z^{\, 2} \left(= d \mbf{z}^{\, 2} + d z_D^{\, 2} \right)
= \frac{d\mzeta^{\, 2}}{
\omega \left( \mzeta \right)^2}
\, . \quad
\end{eqnarray}
The transformation (\ref{t2.2}) is regular
on the real Minkowski space $M$
and
on the \textit{forward tube domain}
$\mathfrak{T}_+$ $=$
\(\left\{\raisebox{8pt}{\hspace{-3pt}}\right. \mzeta =\)
\(x + i \, y :\) \(y^0 > \left| \mbf{y} \right|
\left.\raisebox{8pt}{\hspace{-3pt}}\right\}\),
and maps them on precompact subsets of $E_{\C}$.
The closure $\overline{g_c(M)}$ of the image
of the real Minkowski space $M$ has the form~(\ref{eq1.2})
(thus being identified with $\M$)
and the image $T_+$ of $\mathfrak{T}_+$ under $g_c$ is
\begin{equation}\label{t2.4}
T_+ \hspace{-1pt} := \hspace{-1pt}
\left\{\raisebox{12pt}{\hspace{-3pt}}\right. z \in \mathbb{C}^D
\hspace{-2pt} : \hspace{2pt}
\left| \hspace{1pt} z^{\, 2} \right| \hspace{-1pt} < \hspace{-1pt} 1,\
z \cdot \overline{z} =
\left| \hspace{1pt} z^1 \right|^2 \hspace{-2pt} + \dots +
\left| \hspace{1pt} z^D	 \right|^2
\hspace{-2pt} < \hspace{-1pt} \frac{1}{2}
\left( 1 + \left| \hspace{1pt} z^{\, 2} \right|^2 \right)
\left.\raisebox{12pt}{\hspace{-3pt}}\right\}
.
\end{equation}
The \textit{conjugation} \(* : \M_{\C} \to \M_{\C}\) which
leaves invariant the real space $\M$ is represented in the $z$--coordinates~as:
\begin{equation}\label{eqn2.15}
z \, \mapsto \, z^* \, := \, \frac{\overline{z}}{\overline{z}^{\ 2}}
\, \equiv \, j_W \left( R_D \left( \overline{z} \right) \right)
\, , \
\end{equation}
where
\begin{equation}\label{eqn2.16}
R_{\cmu} \left( z \right) \, := \,
\left( z^1,\, \dots,\, -z^{\cmu},\, \dots,\, z^D \right)
\, , \quad
j_W \left( z \right) \, := \, \frac{R_D \left( z \right)}{z^{\, 2}}
\, \
\end{equation}
and $j_W$ is a $z$--picture analogue of the \textit{Weyl reflection}.

Let us introduce
the complex Lie algebra generators \(T_{\cmu}\)
and \(\SCT_{\cmu}\) for \(\cmu = 1,\, \dots,\, D\) of
the $z$--\textit{translations}
\(e^{w \spr T} \left( z \right) = z + w\)
and
the $z$--\textit{special conformal transformations}
\(e^{w \spr \SCT} \left( z \right) =
\Txfrac{z+z^2 \, w}{1 + 2 \, w \spr z + w^{\, 2} \, z^{\, 2}}\)
(\(w,\, z \in \C^D\))
which
are conjugated by $g_c$ to the analogous generators
$iP_{\mu}$ and $iK_{\mu}$.
This new basis of generators is expressed in terms of
$\Omega_{ab}$~as:
\begin{equation}\label{n2.10n}
T_{\cmu} = i \Omega_{0\cmu} - \Omega_{-1\cmu}
, \ \,
\SCT_{\cmu} = - i \Omega_{0\cmu} - \Omega_{-1\cmu}
\ \, \text{for} \ \cmu = 1,\dots,D
.
\end{equation}
The set of infinitesimal $z$--rotations is again a subset
$\{\Omega_{\alpha\beta}\}$ of $\{\Omega_{ab}\}$ corresponding to
\(1 \leqslant \alpha < \beta \leqslant D\)
and the conformal generator which gives rise to the dilation
(or, in fact, phase) transformation
of the $z$--coordinates
(to be interpreted as a conformal time translation)
is the conformal Hamiltonian
\beq\label{conf_ham}
H = i \Omega_{-10}
\, , \quad
e^{i \, t \, H} \left( z \right) = e^{i \, t } \, z
\, \
\eeq
(\(\left[ T_{\cmu}, C_{\cnu} \right] =
2 \left( \delta_{\cmu\cnu} H - \Omega_{\cmu\cnu} \right)\),
\(\left[ H, C_{\cmu} \right] = - C_{\cmu}\),
\(\left[ H, T_{\cmu} \right] = T_{\cmu}\)).
The relations~(\ref{n2.10n})--(\ref{conf_ham}) can be easily obtained
in the projective realization of $\M$ where the transformation $g_c$
is represented by a rotation of an angle $\txfrac{\pi}{2}$ in the plane
$\left( i \vec{e}_0, \vec{e}_D \right)$ (\(\in \C^{D+2}\),
see for more detail \cite{N03} Appendix~A).
Note that there is an involutive antilinear automorphism
\(\Cnj :\cnfalg_{\C} \to \cnfalg_{\C}\) leaving invariant
the real algebra $\cnfalg$, i.e.,
\beqa\label{lie-cnj}
&
\cnjj{\Omega} = \Omega
, \quad
\cnj{\left( \lambda \Omega + \lambda' \Omega' \right)} =
\overline{\lambda} \, \cnj{\Omega} + \overline{\lambda'} \, \cnj{\Omega'}
, \quad
\cnj{\left[ \Omega, \Omega' \right]} =
\left[ \cnj{\Omega}, \cnj{\Omega'} \right]
,
& \nn &
\Omega_{ab}^{\Cnj} = \Omega_{ab} \quad \Longrightarrow \quad
P_{\mu}^{\Cnj} = -P_{\mu}
, \quad
K_{\mu}^{\Cnj} = -K_{\mu}
, \quad
H^{\Cnj} = -H
, \quad
T_{\cmu}^{\Cnj} = \SCT_{\cmu}
\, & \qquad
\eeqa
for
\(\lambda,\lambda' \in \C\),
\(\Omega,\Omega' \in \cnfalg_{\C}\),
\(\mu = 0,\dots,D-1\) and \(\cmu = 1,\dots,D\).
In fact, the real generators
$T_{\cmu}$, $\SCT_{\cmu}$, $\Omega_{\cmu\cnu}$, $H$ (\(\cmu,\cnu = 1,\dots,D\))
span an \textit{Euclidean real form} (\(\cong \spin \left( D+1,1 \right)\))
of the complex conformal algebra.

From a group theoretic point of view the compactified Minkowski space $\M$
is a homogeneous space of the conformal group $\confgr$
characterized by the stabilizers of the points.
For the tip $p_{\infty}$ of the light cone at infinity
\(K_{\infty} := \M \backslash M\)
(recall that the isotropy relation extends to a conformally
invariant relation on $\M$) the stabilizer is exactly the
\textit{Weyl group}: the Poinar\'e group with dilations.
In more detail, the Lie algebra of the stabilizer of $p_{\infty}$
is spanned by the generators \(\{iP_{\mu},\Omega_{\mu\nu},\Omega_{-1D}\}\),
while the Lie algebra of the stabilizer of the origin $p_0$
(corresponding to \(x = 0 \in M\)) is spanned by
\(\{iK_{\mu},\Omega_{\mu\nu},\Omega_{-1D}\}\).
Thus every chart in $\M$ as well as in $\M_{\C}$ can be
uniquely characterized, as a vector space, by a pair $(p,q)$ of
mutually nonisotropic points: the origin $p$ of
the chart and the tip $q$ of the its light cone complement.
For the Minkowski space chart the stabilizer of the pair
\((p,q) \equiv (p_0,p_{\infty})\) is the Cartesian product
of the (one--parameter) dilation and the Lorentz subgroups
with Lie algebra spanned by \(\{\Omega_{-1D},\Omega_{\mu\nu}\}\).
The $z$--chart introduced above is characterized by the
pair of mutually conjugate points \((p,q) = (ie_0,-ie_0)\)
(\(e_0 := (1,\Mbf{0}) \in M\), so that \(p \in \mathfrak{T}_+ \subset M_{\C}\))
and the stabilizer $\compgr_{\C}$ of this pair is a
$\Cnj$--invariant subgroup of $\confgr_{\C}$.
The real part of $\compgr_{\C}$ coincides with the
\textit{maximal compact subgroup}
$\compgr$ which is generated by two mutually commuting subgroups:
the $U (1)$--group \(\{e^{\hspace{1pt}2\pi i\hspace{1pt}t\hspace{1pt}H}\}\)
and the \(\Spin \left( D \right)\) group acting on $z$ via real (Euclidean)
rotations (\(\compgr \cong
\nfrc{U \left( 1 \right) \times \Spin \left( D \right)}{\Z_2}\)).
Since the points $p$ and $q$ are mutually conjugate,
$\compgr$ is also the real part of the stabilizer
of the origin in the $z$--chart.
In fact, $T_+$~(\ref{t2.4}) is isomorphic to the homogeneous space
$\nfrc{\confgr}{\compgr}$ of $\confgr$ (cf.~\cite{U63}).
Note that the complex Lie algebra of the stabilizer
of \(z=0\) is spanned by the generators
\(\{\SCT_{\cmu},H,\Omega_{\cmu\cnu}\}\).

\begin{remark}
\rmlabel{rm:2.1new}
In the familiar realization of $\M$ as the Dirac projective quadric~\cite{D36},
\[
\M = Q \left/\right.\! \R^*
, \quad
Q =
\left\{\raisebox{10pt}{\hspace{-2pt}}\right.
\vec{\xi} \!\in \R^{D, 2}
\backslash \{0\}
:
\vec{\xi}{\hspace{-1pt}}^{\, 2} := \mbf{\xi}^{\, 2} +
\xi_D^2 - \xi_0^2 - \xi_{-1}^2
\hspace{1pt}
(= \xi^{a} \eta_{ab} \xi^{b}) = 0
\left.\raisebox{10pt}{\hspace{-2pt}}\right\}
.
\]
the Minkowski space coordinates $x$ and the complex coordinates $z$ of (\ref{eq1.2}) are expressed by
\[
x^{\mu} \, = \, \Txfrac{\xi^{\mu}}{\xi^{D} + \xi^{-1}}
, \quad
z^{\cmu} \, = \, \Txfrac{\xi^{\cmu}}{\xi^{-1} - i \xi^{0}}
.
\]
\end{remark}

\begin{remark}
\rmlabel{rm:2.1}
Only Lorentz types of signatures
\(\left( D-1,1 \right)\) or \(\left( 1, D-1 \right)\)
possess the remarkable property
that there exist affine charts covering the corresponding
conformally compactified \textit{real} space (\cite{N03} Proposition~A.1).
Moreover, every such chart is characterized by the condition that
the tip $q$ of the light cone complement belongs to the union
\(T_+ \cup T_-\) (\(T_- := \left( T_+ \right)^*\) is the image in
the $z$--coordinates of the \textit{backward tube} $\mathfrak{T}_-$).
If \(q \in T_{\pm}\) then $T_{\mp}$ is also covered by the chart.
\end{remark}

\subsection{Wightman axioms for
conformal field theory in the analytic picture}\label{ssec2.2}

We proceed with a brief survey of the axiomatic QFT with GCI.
First, one assumes the existence of a vector bundle over
the complex compactified Minkowski space $\M_{\C}$ called the \textit{field bundle}.
It is endowed with an action of the conformal group
$\confgr_{\C}$ via (bundle) automorphisms.
Thus for every point \(p \in \M_{\C}\) its stabilizer $\confgr_p$
will act by a representation $\pi_p$ on the (finite dimensional) fibre
$F_p$ over $p$.
Then if $p$ is the origin of some affine chart in $\M_{\C}$,
e.g., the $z$--chart,
we can \textit{trivialize} the bundle over the chart using
the coordinate translations
\(\trn_w \left( z \right) \, ( \, \equiv e^{w \spr T} \left( z \right) ) \, = z+w\)
so that the action of $\confgr_{\C}$ in this trivialization will take the form
\beq\label{fib_act}
\left( z = \{z^{\alpha}\},\, \phi = \{\phi_{\aa}\} \right)
\, \mathop{\longmapsto}\limits_{g} \,
\left( g \left( z \right),\,
\pi_z \! \left( g \right) \phi =
\{\pi_z \! \left( g \right)\mtrx^{\bb}_{\aa} \phi_{\bb}\} \right)
\, \in \, \C^D \times F
\eeq
where $\{\phi_{\aa}\}$
are some (spin--tensor) coordinates in the fibre \(F := F_0\) over the origin \(z=0\)
and the matrix valued function
\(\pi_z \! \left( g \right) =
\{\pi_z \! \left( g \right)\mtrx^{\bb}_{\aa}\}\)
(\(g \in \confgr_{\C}\)), regular in the domain of $g$ and called \textit{cocyle},
is characterized by the properties
\beq\label{eqn2.4}
\pi_z \! \left( g_1 g_2 \right) =
\pi_{g_2 \left( z \right)} \! \left( g_1 \right)
\pi_z \! \left( g_2 \right)
, \quad
\pi_z \! \left( \trn_w \right) = \ID_F
\ \, (\Leftrightarrow \ \,
\pi_z \! \left( g \right) =
\pi_0 \! \left( \trn_{g \left( z \right)}^{-1} \ g \ \trn_z \right))
. \ \
\eeq
The fibre $F$ is the space of (classical) \textit{field values}
and the coordinates $\phi_{\aa}$ correspond to the
collection of local fields in the theory.
An example of a field bundle is the
\textit{electro-magnetic} field
defined as
the bundle of $2$--forms
\(F_{\alpha\beta} \, dz^{\alpha} \! \wedge dz^{\beta}\)
$=$
\(F^{\MINK}_{\mu\nu} \, dx^{\mu} \! \wedge dx^{\nu}\)
over $\M$.

The axiomatic assumptions of the GCI~QFT
are the Wightman axioms~\cite{SW} and the condition
of GCI for the correlation functions~\cite{NT01}.
As proven in \cite{NT01}, Theorem~3.1, GCI
is equivalent to the rationality of
the (analytically continued) Wightman functions.
Thus the vacuum $n$--point correlation functions
in the theory can be considered as meromorphic sections
of the $n$th tensor power (over $\M_{\C}^{\times \hspace{1pt} n}$)
of the field bundle and hence, for every affine chart in $\M_{\C}$
we obtain a system of rational correlation functions over the chart.
This provides the general scheme for the passage from the GCI~QFT
over Minkowski space to the theory over
a complex affine chart which contains
the forward ``tube'' $T_+$~(\ref{t2.4})~--~see \cite{N03} Sect.~9.

The (analytic) $z$-\textit{picture} of a GCI~QFT is
equivalent to the theory of vertex algebras
(\cite{Bo86} \cite{Ka96} \cite{Bo97} \cite{FBZ01}) in
higher dimensions (see \cite{N03}).
We proceed to formulate the analogue of Wightman axioms~\cite{SW}
in this picture.

The quantum fields $\phi_{\aa} \left( z \right)$
(\(\aa = 1,\dots, I\) for \(I = \dim F\))
will be treated as formal power
series in $z$ and $\Txfrac{1}{z^2}$.
This is possible because of the analytic properties of the fields in a GCI Wightman QFT 
(\cite{N03}~Theorem~9.1).
Using \textit{harmonic} polynomials one can uniquely
separate the integer powers of the interval $z^{\, 2}$
due to the following (known) fact:
\textit{for every polynomial $p \left( z \right)$ there exist unique
polynomials $h \left( z \right)$ and $q \left( z \right)$ such that
$h \left( z \right)$ is harmonic and}
\(p \left( z \right) = h \left( z \right) + z^{\, 2} \, q \left( z \right)\).
Thus, if we fix a basis
\(\left\{\raisebox{9pt}{\hspace{-2pt}}\right.
\hr{m}{\sigma} \left( z \right)
\left.\raisebox{9pt}{\hspace{-2pt}}\right\}\) of
\textit{homogeneous} harmonic polynomials of degree $m$, for every
\(m = 0,\, 1,\, \dots\), we can write our fields
$\phi_{\aa} \left( z \right)$
in a unique way
as formal series in the monomials
\(\left( z^{\, 2} \right)^n \, \hr{m}{\sigma} \left( z \right)\)
for \(n \in \Z\), \(m = 0,\, 1,\, \dots\) and
the index $\sigma$ takes values in a finite set~$I_m$.
In such a way we end up with the following axiom:

\begin{plist}
\item[\textit{Fields} (\textit{F}).]
The fields are represented by (nonzero) formal series
\begin{equation}\label{eqn2.1}
\phi_{\aa} \left( z \right) =
\Su_{n \, \in \, \Z} \,
\Su_{m \, = \, 0}^{\infty} \,
\Su_{\sigma} \,
\phi_{\,\aa\,\left\{ n,\, m,\, \sigma \right\}} \,
\left( z^{\, 2} \right)^n \, \hr{m}{\sigma}
\left( z \right)
\, , \
\end{equation}
with coefficients
$\phi_{\,\aa\,\left\{ n,\, m,\, \sigma \right\}}$ which are
operators acting on a common invariant dense domain $\DOM{}$ of the
Hilbert space $\HS$ of physical states.
We require that for every vector state \(\Psi \in \DOM{}\)
there exists a constant \(N_{\Psi} \in \N\) such that
\(\phi_{\,\aa\,\left\{ n,\, m,\, \sigma \right\}} \, \Psi = 0\)
for all \(n \leqslant -N_{\Psi}\), \(m = 0,\, 1,\, \dots\)
and all possible values of $\sigma$, or equivalently,
\(\left( z^{\, 2} \right)^{N_{\Psi}}
\phi_{\aa} \left( z \right)\Psi\) is a formal power series
with no negative powers.
(This requirement is related to the \textit{energy positivity},
stated below in the axiom (\textit{SC}).)
\end{plist}

As the properties of the field $\phi_{\aa} \left( z \right)$
do not dependent of the choice on
\(\left\{\raisebox{9pt}{\hspace{-2pt}}\right.
\hr{m}{\sigma} \left( z \right)
\left.\raisebox{9pt}{\hspace{-2pt}}\right\}\)
we may also write it in a \textit{basis independent form}:
\beq\label{eq2.14nn}
\phi_{\aa} \left( z \right) \, = \,
\Su_{n \in \Z} \, \Su_{m \, = \, 0}^{\infty}
\phi_{\,\aa\, \left\{ n,m \right\}} \! \left( z \right) \, \left( z^{\, 2} \right)^n
\eeq
where $\phi_{\,\aa\,\left\{ n,m \right\}} \left( z \right)$ ($=$
$\Su_{\sigma} \, \phi_{\,\aa\, \left\{ n,m,\sigma \right\}} \, \hr{m}{\sigma} \left( z \right)$)
are \textit{operator valued homogeneous harmonic polynomials}.
We shall use this more concise presentation in studding
examples of free fields (Sect.~\ref{SEC:3}).
Using an (arbitrary) basis
\(\left\{\raisebox{9pt}{\hspace{-2pt}}\right.
\hr{m}{\sigma} \left( z \right)
\left.\raisebox{9pt}{\hspace{-2pt}}\right\}\),
on the other hand, makes more transparent the algebraic manipulations
of formal power series in this and next sections.

The next axiom introduces the conformal symmetry of the theory.

\begin{plist}
\item[\textit{Covariance} (\textit{C}).]
There exists a unitary representation, $U \! \left( g \right)$ of the real conformal
group \(\confgr\) on the Hilbert
space $\HS$ such that the hermitian generators of the conformal Lie algebra
$\cnfalg$ leave invariant the fields' domain \(\DOM{}\).
We also require the existence of a rational matrix--valued function
\({\left\{\raisebox{9pt}{\hspace{-3pt}}\right. \pi_z \left( g \right)_{\aa}^{\bb}
\left.\raisebox{9pt}{\hspace{-2pt}}\right\}}_{\aa,\, \bb = 1,\, \dots,\, I}\)
depending on \(z \in \C^D\) and \(g \in \confgr\),
regular for $z$ in the domain of \(g\) on $\C^D$, and such that it satisfies
the properties~(\ref{eqn2.4}).
Then the fields $\phi_{\aa} \left( z \right)$ are assumed to satisfy infinitesimal
conformal covariance, formally written as:
\begin{eqnarray}\label{eqn2.6}
&&
\frac{d}{d \hspace{0.5pt} t}
\left(\raisebox{10pt}{\hspace{-2pt}}\right.
U \left( e^{\hspace{0.5pt} t \hspace{0.5pt} \Omega} \right) \,
\phi_{\aa} \left( z \right) \,
U \left( e^{\hspace{0.5pt} t \hspace{0.5pt} \Omega} \right)^{\hspace{-2pt}-1}
\left.\raisebox{10pt}{\hspace{-2pt}}\right)
\vrestr{10pt}{ t = 0}
\, = \nonumber \\ && = \,
\frac{d}{d \hspace{0.5pt} t }
\left(\raisebox{10pt}{\hspace{-2pt}}\right.
\left[\raisebox{9pt}{\hspace{-2pt}}\right.
\pi_z
\left(\raisebox{9pt}{\hspace{-2pt}}\right.
e^{\hspace{0.5pt} t \hspace{0.5pt} \Omega}
\left.\raisebox{9pt}{\hspace{-2pt}}\right)^{\hspace{-2pt}-1}
\left.\raisebox{9pt}{\hspace{-2pt}}\right]{\hspace{-2pt}}_{\aa}^{\bb} \
\phi_{\bb}
\left(\raisebox{9pt}{\hspace{-2pt}}\right.
e^{\hspace{0.5pt} t \hspace{0.5pt} \Omega} \left( z \right)
\left.\raisebox{9pt}{\hspace{-2pt}}\right) \,
\left.\raisebox{10pt}{\hspace{-2pt}}\right)
\vrestr{10pt}{ t = 0}
\, \
\end{eqnarray}
for \(\Omega \in \spin \left( D,2 \right)\).
\end{plist}

It is simpler to write down the field covariance law if we further
assume that our fields are transforming under an
\textit{elementary induced} representation of the conformal group~$\confgr$.
This means that the cocycle \(\pi_z \left( g \right)\)
is trivial at \(z=0\)
for \(g = e^{a \spr \SCT}\) and it is thus determined by a
representation of the maximal compact subgroup
$\compgr$
of $\confgr$:
\begin{equation}\label{eqn2.10}
\pi_z
\left(\raisebox{9pt}{\hspace{-2pt}}\right.
e^{i \hspace{0.5pt} t \hspace{0.5pt} H}
\left.\raisebox{9pt}{\hspace{-2pt}}\right)\mtrx^{\bb}_{\aa}
\, = \,
e^{i \hspace{0.5pt} t \hspace{0.5pt} d_{\aa}}
\, \delta^{\bb}_{\aa}
\, , \quad
\pi_z
\left(\raisebox{9pt}{\hspace{-2pt}}\right.
e^{
t \hspace{0.5pt} \Omega_{\cmu\cnu}}
\left.\raisebox{9pt}{\hspace{-2pt}}\right)\mtrx^{\bb}_{\aa}
\, = \,
\left[\raisebox{9pt}{\hspace{-2pt}}\right.
e^{
t \hspace{1pt}
\pi_0 \left( \Omega_{\cmu\cnu} \right)}
\left.\raisebox{9pt}{\hspace{-2pt}}\right]{\hspace{-1pt}}\mtrx^{\bb}_{\aa}
\, , \
\end{equation}
where $d_{\aa}$ are positive numbers called \textit{dimensions} of the
corresponding fields \(\phi_{\aa}\).
Under this additional assumption we can present Eq.~(\ref{eqn2.6})
in a more concrete form:
\begin{eqnarray}\label{eqn2.11}
\left[ \hspace{1pt} T_{\cmu} \, , \,
\phi_{\aa} \left( z \right) \, \right]
\, = && \podr
\di_{z^{\cmu}} \, \phi_{\aa} \left( z \right)
\, , \quad
\\ \label{eqn2.12}
\left[ \hspace{1pt} H \, , \,
\phi_{\aa} \left( z \right) \, \right]
\, = && \podr
z \spr \di_{z} \, \phi_{\aa} \left( z \right) \, + \,
d_{\aa} \, \phi_{\aa} \left( z \right)
\, , \quad
\mgvspc{9pt}
\\ \label{eqn2.13}
\left[ \hspace{1pt} \Omega_{\cmu\cnu} \, , \,
\phi_{\aa} \left( z \right) \, \right]
\, = && \podr
\left( z^{\cmu} \, \di_{z^{\cnu}} - z^{\cnu} \, \di_{z^{\cmu}} \right)
\phi_{\aa} \left( z \right)
\, + \,
\pi_0 \left( \Omega_{\cmu\cnu} \right)\mtrx^{\bb}_{\aa}
\phi_{\bb} \left( z \right)
\, , \quad \mgvspc{8pt}
\\ \label{eqn2.14}
\left[ \hspace{1pt} \SCT_{\cmu} \, , \,
\phi_{\aa} \left( z \right) \, \right]
\, = && \podr
\left( z^{\, 2} \, \di_{z^{\cmu}} -
2 \, z^{\cmu} \, z \spr \di_z \right)
\phi_{\aa} \left( z \right) \, - \,
2 \, z^{\cmu} \, d_{\aa} \, \phi_{\aa} \left( z \right) \, + \,
\nonumber \\ && \podr
\mgvspc{11pt}
\, + \,
2 \, z^{\cnu} \,
\pi_0 \left( \Omega_{\cnu\cmu} \right)\mtrx^{\bb}_{\aa} \,
\phi_{\bb} \left( z \right)
\, , \qquad \mgvspc{8pt}
\end{eqnarray}
where \(\di_{z^{\cmu}}\) stands for the partial (formal) derivative
\(\Txfrac{\di}{\di z^{\cmu}}\).

We further assume that the hermitian conjugate
$\phi_{\aa} \! \left( z \right)^*$ of each $\phi_{\aa} \! \left( z \right)$
belongs to the linear span of
the set~\(\left\{ \phi_{\aa} \right\}\).

\begin{plist}
\item[\textit{Field conjugation law} (*).]
For every \(\Psi_1,\, \Psi_2 \in \DOM{}\)
and for any field $\phi_{\aa}$ there exists a \textit{conjugate} field $\sphi_{\aa}$
such that:
\beq\label{eqn2.17}
\La \Psi_1 \Vl \phi_{\aa} \left( z \right) \, \Psi_2 \Ra
\, = \,
\La \pi_{R_D \left( \overline{z} \right)}
\hspace{-2pt} \left( j_W^{-1} \right)\mtrx^{\bb}_{\aa} \,
\sphi_{\bb} \left(\raisebox{9pt}{\hspace{-2pt}}\right.
z^* \left.\raisebox{9pt}{\hspace{-2pt}}\right) \Psi_1
\Vl \Psi_2 \Ra
\, , \
\eeq
where $j_W$ is defined by (\ref{eqn2.15}) and (\ref{eqn2.16}).
The exact meaning of Eq.~(\ref{eqn2.17}) is provided by the fact
that both sides are finite series, i.~e. polynomials in $z$ and
$\frc{1}{z^{\, 2}}$
(see \cite{N03}~Remark~8.1).
The correspondence \(\phi_{\aa} \mapsto \sphi_{\aa}\) gives rise to
an antilinear involution in the standard fibre $F$ of the field bundle,
invariant under the action of $\confgr$.
\end{plist}

The next axiom states energy positivity and determines the physical vacuum.

\begin{plist}
\item[\textit{Spectral condition} (\textit{SC}).]
The conformal time generator $H$ is represented on $\HS$ by
a positive operator.
There is only one norm 1 conformally invariant vector \(\rvac \in \HS\)
(up to phase factor) and it is contained in the fields' domain $\DOM{}$.
\end{plist}

We shall now formulate
a strong form of the locality axiom
also called \textit{Huygens' principle}
stating that the fields are independent for nonisotropic separations.

\begin{plist}
\item[\textit{Strong Locality} or \textit{Huygens' principle} (\textit{SL}).]
Every field \(\phi_{\aa}  \hspace{-2.5pt} \left( \hspace{-0.6pt} z \hspace{-0.6pt} \right)\)
is assumed to have a fixed
\textit{statistical} parity
\(p_{\aa} =0, 1\)
and there exist positive integers \(M_{\aa\bb}\) such that:
\begin{equation}\label{eqn2.2}
\left( z_{12}^{\, 2} \right)^{M_{\aa\bb}}
\left(\raisebox{9pt}{\hspace{-2pt}}\right.
\phi_{\aa} \left( z_1 \right)\, \phi_{\bb} \left( z_2 \right)
- \left( -1 \right)^{p_{\aa}p_{\bb}}
\phi_{\bb} \left( z_2 \right)\, \phi_{\aa} \left( z_1 \right)
\left.\raisebox{9pt}{\hspace{-2pt}}\right)
\, = \, 0
\,  \
\end{equation}
where \(z_{12} := z_1 - z_2\).
\end{plist}

\begin{remark}
\rmlabel{rmt1}
When we deal with formal power series it is more convenient
to use weaker (infinitesimal) conformal invariance but a stronger locality axiom. 
Indeed, for rational functions, GCI follows from infinitesimal conformal invariance.
Thus within the Wightman framework, the two pairs of axioms:
(1)~ordinary locality and GCI, (2)~strong locality and infinitesimal
conformal invariance, are completely equivalent.
\end{remark}

\begin{plist}
\item[\textit{Completeness}.]
The set of vectors $\rvac$,
\(\phi_{\,\aa_1\,\left\{ n_1,\, m_1,\, \sigma_1 \right\}}
\dots \phi_{\,\aa_k\,\left\{ n_k,\, m_k,\, \sigma_k \right\}} \, \rvac\),
for all \(k \in N\)and all possible values of the indices of the $\phi$'s,
spans the fields' domain~$\DOM{}$.
\end{plist}

This completes our analogue of Wightman axioms in the $z$-picture.
Theorems~9.2 and 9.3 of \cite{N03} allow one to state
the following general result:

{\samepage

\begin{theorem}
\prlabel{thr:2.1}
There is a one--to--one correspondence between the finite systems of
Wightman fields with GCI correlation functions \cite{NT01} and the
systems of formal series satisfying the above axioms.
\end{theorem}

}

Using the fact that the cocycle $\pi_z \! \left( g \right)$
is meromorphic (even rational) in $g$ and $z$ we can continue it
to \(\confgr_{\C}\) and
write down the explicit connection between the Wightman fields
$\phi_{\aa}^{\MINK} \! \left( x \right)$ on the Minkowski space
and the analytic picture fields $\phi_{\aa} \! \left( z \right)$:
\begin{equation}\label{WF}
\phi_{\aa}^{\MINK} \left( x \right) \, = \,
\pi_{z} \left(\raisebox{9pt}{\hspace{-2pt}}\right. g_c^{-1}
\left.\raisebox{9pt}{\hspace{-2pt}}\right)\mtrx^{\bb}_{\aa} \
\phi_{\bb}
\left(\raisebox{9pt}{\hspace{-2pt}}\right. g_c \left( x \right)
\left.\raisebox{9pt}{\hspace{-2pt}}\right)
\quad
(z = g_c \left( x \right))
\end{equation}
where $g_c$ is the transformation~(\ref{t2.2})
viewed as an element of $\confgr_{\C}$.
Eq.~(\ref{WF}) is the precise expression of the fact that
the fields $\phi_{\aa}^{\MINK} \! \left( x \right)$
and $\phi_{\aa} \! \left( z \right)$ are different coordinate expressions
of the same (generalized, operator--valued) section
of the field bundle over $\M$.
For example, for the electromagnetic field Eq.~(\ref{WF}) is equivalent~to:
\beq\label{WF_EM}
F_{\mu\nu}^{\MINK} \! \left( x \right) \, dx^{\mu} \! \wedge dx^{\nu}
\, = \,
F_{\cmu\cnu} \! \left( z \right) \, dz^{\cmu} \! \wedge dz^{\cnu}
.
\eeq
The rigorous meaning of Eqs.~(\ref{WF}) and (\ref{WF_EM}) includes,
on one hand, the extension of the operator valued functions
$\phi_{\aa}^{\MINK} \! \left( x \right)$ to a larger class of test
functions which correspond to coordinate expressions of
arbitrary smooth sections over $\M$.
This can be done using the GCI condition (see \cite{NT01} Proposition 2.1).
On the other hand, using the positivity of the scalar product of the
Hilbert state space one can easily extend our formal field
series~(\ref{eqn2.6}) to generalized operator valued functions over $\M$
(in the parametrization~(\ref{eq1.2})).

\medskip

We now proceed to introduce the real
\textit{compact picture representation} which is more convenient in
studying finite temperature correlation functions.

For a local field $\phi \left( z \right)$ of dimension $d$
we set $\phi \left( \czeta,\, u \right)$ to be a formal
Fourier series in \(e^{2\pi i \, \czeta} \in \Sr^1\) and
\(u \in \Sr^{D-1}\) defined as:
\begin{eqnarray}\label{eqn2.20a}
\phi \left( \czeta,\, u \right) \, = && \podr
e^{2\pi i d \, \czeta} \, \phi \left( e^{2\pi i \, \czeta} u \right) \, = \,
\Su_{\nnu \, \in \, d \, + \, \Z} \,
\Su_{m \, = \, 0}^{\infty} \,
\phi_{-\nnu m} \! \left( u \right) \, e^{2\pi i \, \nnu \, \czeta}
\, , \
\nonumber \\ \,
\phi_{-\nnu m} \! \left( u \right)
= && \podr
\Su_{\sigma} \,
\phi_{-\nnu,\, m,\, \sigma} \ \hr{m}{\sigma}
\left( u \right)
\, . \
\end{eqnarray}
(The space of harmonic polynomials, \(\mathit{Span}_{\C} \hspace{1pt}
\left\{\raisebox{9pt}{\hspace{-3pt}}\right. \hr{m}{\sigma}
\left( u \right) \left.\raisebox{9pt}{\hspace{-3pt}}\right\}_{m,\, \sigma}\),
is identified with the algebra of complex polynomials restricted to the sphere
$\Sr^{D-1}$.)
Then the connection with the previous analytic picture modes is:
\beq\label{eqn2.21a}
\phi_{\left\{ n,\, m,\, \sigma \right\}}
\, = \,
\phi_{\nnu,\, m,\, \sigma}
\quad \text{for} \quad
\nnu \, = \, -d-2n-m
\quad
(n \, = \, -\frac{\nnu +m+d}{2})
\, . \
\eeq
Note that the index $n$ in the analytic picture modes
\(\phi_{\left\{ n,m,\sigma \right\}}\) is always integer
while in the compact picture modes,
\(\phi_{\nnu,m,\sigma}\), it is integer or half--integer depending on $d$
(which is reflected in the first sum in~(\ref{eqn2.20a})).
In accord with the commutation relation ~(\ref{eqn2.12})
we obtain
\begin{eqnarray}\label{eqn2.22a}
\left[ H,\, \phi_{\nnu,\, m} \! \left( u \right) \right] =
-\nnu \, \phi_{\nnu,\, m} \! \left( u \right)
\, , && \podr \quad
e^{2 \pi i \hspace{1pt} t \, H} \, \phi_{\nnu,\, m} \! \left( u \right) \, e^{-2 \pi i \hspace{1pt} t \, H}
\, = \,
e^{-2\pi i \hspace{1pt} t \, \nnu} \,
\phi_{\nnu,\, m} \! \left( u \right)
\nn
\left[ H,\, \phi_{\left\{ n,\, m \right\}} \! \left( z \right) \right] = && \podr
\left( d+2n+m \right) \phi_{\left\{ n,\, m \right\}} \! \left( z \right)
\end{eqnarray}
It follows that $2\pi i H$ acts as a translation generator in $\czeta$
in accord with Eq.~(\ref{eq1.5n}).

\medskip

As a realization of the above axioms we will consider the
case of a neutral scalar field
\(\phi \left( z \right) \equiv
\phi^{\left( d \right)} \left( z \right)\)
of dimension~$d$.
Its 2-point function is proportional to the unique scalar
conformal invariant function of dimension~$d$,
\begin{equation}\label{eqn5.1}
\lvac \phi \left( z_1 \right)
\phi \left( z_2 \right)
\rvac \, = \, \frac{
1}{
\left(\raisebox{9pt}{\hspace{-3pt}}\right.
z_{12}^{\, 2} \left.\raisebox{9pt}{\hspace{-3pt}}\right)^d}
\, \quad (z_{12} = z_1 - z_2)
\end{equation}
viewed as a Taylor series in the second argument, $z_2$,
with coefficients~--~rational functions in~$z_1$
(see Theorem~\ref{thm:2.3} below).
The field cocycle is \(\pi_z \left( g \right) = \omega \left( g,\, z \right)^{-d}\),
where \(\omega \left( g,\, z \right)\)
is a quadratic polynomial in $z$ (\(\omega \! \left( g, z \right) \! = \!
\left(\raisebox{10pt}{\hspace{-4pt}}\right.
\mathit{det} \! \left(\raisebox{9pt}{\hspace{-4pt}}\right.
\Txfrac{\partial g \left( z \right)^{\cmu}}{\partial z^{\beta}}
\left.\raisebox{9pt}{\hspace{-4pt}}\right)
\left.\raisebox{10pt}{\hspace{-5pt}}\right)^{\hspace{-2pt}
\text{\scriptsize $-$}
\frac{\raisebox{0pt}{\footnotesize $1$}}{\raisebox{0pt}{\footnotesize $D$}}}\hspace{-5pt}\),
see \cite{NT01} Eq. (A.5)).
(Note that the Minkowski space transform of the correlation function~(\ref{eqn5.1})
is proportional to
\(\Txfrac{1}{x_{12}^{\, 2} \hspace{-1pt} + \hspace{-1pt} i 0
x_{12}^0}\)~--~cf. with Eq.~(\ref{ea5.17}).)
The hermiticity of the field is expressed by
\begin{equation}\label{eqn5.2}
\phi \hspace{-1pt} \left( z \right)^* \, = \,
\frac{1}{\left( \overline{z}^{\, 2} \right)^d} \
\phi \hspace{-1pt} \left( z^* \right)
\quad \
(z^* \, = \, \frac{\overline{z}}{\overline{z}^{\, 2}}
\, , \ \,
\phi \hspace{-1pt} \left( z \right) \equiv \sphi \hspace{-1pt} \left( z \right)
)
\, \
\end{equation}
since \(\omega \left( j_W,\, z \right) \, = \, z^{\, 2}\).
This conjugation law simplifies in the compact picture,
since we are using real coordinates \(\left( \czeta,\, u \right)\);
the hermiticity condition for the field modes reads:
\begin{equation}\label{eqn2.24n}
\phi_{\nnu,\, m,\, \sigma}^* \, = \, \phi_{-\nnu,\, m,\, \sigma}
\, . \
\end{equation}

\section{GCI Correlation Functions as Meromorphic Functions}\label{msec:3}

\subsection{Rationality of the vacuum correlation functions}\label{mssec:3.1p}

Theorem~\ref{thr:2.1}
implies the rationality of the Wightman
functions as well as the analyticity properties of the fields.
Since these facts are of major importance we shall prove them
independently.
We begin with stating some basic properties of the formal series
which arise in the analytic picture of GCI~QFT.

We introduce, following~\cite{N03},
the space $\Lin \Bbrk{z, \frc{1}{z^{\, 2}}}$ of formal series
\beq\label{fs}
v \! \left( z \right) =
\Su_{n \, \in \, \Z} \,
\Su_{m \, = \, 0}^{\infty} \,
\Su_{\sigma} \,
v_{\left\{ n,\, m,\, \sigma \right\}} \,
\left( z^{\, 2} \right)^n \, \hr{m}{\sigma}
\left( z \right)
\eeq
with coefficients $v_{\left\{ n,\, m,\, \sigma \right\}}$ belonging
to a complex vector space $\Lin$.
The space of finite series of $\Lin \Bbrk{z, \frc{1}{z^{\, 2}}}$
will be denoted by $\Lin \Brk{z, \frc{1}{z^{\, 2}}}$.
Obviously, $\C \Brk{z, \frc{1}{z^{\, 2}}}$ is a complex algebra
and $\Lin \Brk{z, \frc{1}{z^{\, 2}}}$ is a module over this algebra.
Nevertheless, the product between a series
\(f \! \left( z \right) \in \C \Bbrk{z, \frc{1}{z^{\, 2}}}\)
and a series
\(v \! \left( z \right) \in \Lin \Bbrk{z, \frc{1}{z^{\, 2}}}\)
is not defined in general but it is not difficult to define the product
\(f \! \left( z \right) v \! \left( z \right)\) if
\(f \! \left( z \right) \in \C \Brk{z, \frc{1}{z^{\, 2}}}\)
(thus turning $\Lin \Bbrk{z, \frc{1}{z^{\, 2}}}$ into a
$\C \Brk{z, \frc{1}{z^{\, 2}}}$--module).
We emphasize that all the products between formal series
throughout this paper will be
treated in \textit{purely algebraic} sense, i.e., every
coefficient of the product series should be obtained
by a finite number operations of summation and multiplication
on the coefficients of the initial series.

On the other hand, the space of Taylor series
$\Lin \Bbrk{z}$ in $z$, with coefficients in $\Lin$,
is naturally identified with the subspace of
$\Lin \Bbrk{z, \frc{1}{z^{\, 2}}}$ of formal series~(\ref{fs})
whose sum in $n$ runs from $0$ to $\infty$.
Evidently, $\Lin \Bbrk{z}$ is a module over the algebra
$\C \Bbrk{z}$ (i.e., the product
\(f \! \left( z \right) v \! \left( z \right)\) is well defined
for \(f \! \left( z \right) \in \C \Bbrk{z}\) and
\(v \! \left( z \right) \in \Lin \Bbrk{z}\)).
Similarly, $\Lin \Brk{z}$ (the space of polynomials in $z$
with coefficients in $\Lin$) is a subspace of
$\Lin \Brk{z, \frc{1}{z^{\, 2}}}$
and it is a module over the polynomial algebra $\C \Brk{z}$.
There is a larger space of formal series of \(\C \Bbrk{z, \frc{1}{z^{\, 2}}}\)
than $\C \Bbrk{z}$ which still possesses a complex algebra structure:
this is the space $\C \Bbrk{z}_{\lscw{z}{\, 2}{}}$ of those series
\(f \! \left( z \right) \in \C \Bbrk{z, \frc{1}{z^{\, 2}}}\)
whose sum in $n$ in~(\ref{fs}) is bounded below.
The more general spaces $\Lin \Bbrk{z}_{\lscw{z}{\, 2}{}}$ are defined in a
similar way and we have in fact a shorter equivalent definition
\beq\label{loc_def}
v \! \left( z \right) \in \Lin \Bbrk{z}_{\lscw{z}{\, 2}{}}
\quad \Leftrightarrow \quad
\left( z^{\, 2} \right)^N \, v \! \left( z \right) \in \Lin \Bbrk{z}
\ \, \text{for} \ \, N \mgrt 0
\, .
\eeq

\begin{remark}
\rmlabel{rm:3.1}
The notation $\C \Bbrk{z}_{\lscw{z}{\, 2}{}}$ comes from
commutative algebra: for a commutative ring $R$ and \(f \in R\)
the \textit{localized} ring $R_f$ is defined as the ring of
``ratios'' $\txfrac{a}{f^n}$ for \(a \in R\) and \(n = 0,1,\dots\)
(more precisely, this is the quotient ring
\(\nfrc{R \Brk{t}}{\left( ft-1 \right)}\) of the ring
$R \Brk{t}$ of polynomials in the one--dimensional variable $t$ over
ideal generated by \(ft-1\)~--~see~\cite{AM}).
In a similar way, if $\Lin$ is a module over the ring $R$ then the localized
space $\Lin_f$ is defined as a module over the localized ring $R_f$.
Note that \(\C \Brk{z, \frc{1}{z^{\, 2}}} \equiv
\C \Brk{z}_{\lscw{z}{\, 2}{}}\).
\end{remark}

\begin{proposition}
\prlabel{pr:3.1}
Let $\Lin$ be a complex vector space.
\begin{plist}
\item
The space $\C \Bbrk{z}_{\lscw{z}{\, 2}{}}$ is a complex
algebra containing $\C \Bbrk{z}$ as a subalgebra and
$\Lin \Bbrk{z}_{\lscw{z}{\, 2}{}}$ is a
module over this algebra that extends the
module structure of $\Lin \Bbrk{z}$ over the algebra $\C \Bbrk{z}$.
\item
There are no zero divisors in the
$\C \Bbrk{z}_{\lscw{z}{\, 2}{}}$--module
$\Lin \Bbrk{z}_{\lscw{z}{\, 2}{}}$, i.e.,
if \(f \! \left( z \right) \in \C \Bbrk{z}_{\lscw{z}{\, 2}{}}\)
and \(v \! \left( z \right) \in \Lin \Bbrk{z}_{\lscw{z}{\, 2}{}}\)
are such that \(f \! \left( z \right) v \! \left( z \right) = 0\)
then either \(f = 0\) or \(v = 0\).
\item
If $w$ is another $D$--dimensional formal variable and
$$
\Lin \Bbrk{z}_{\lscw{z}{\, 2}{}} \Bbrk{w}_{\lscw{w}{\, 2}{}}
\, := \,
\left(\raisebox{10pt}{\hspace{-2pt}}\right.
\Lin \Bbrk{z}_{\lscw{z}{\, 2}{}}
\left.\raisebox{10pt}{\hspace{-2pt}}\right)
\Bbrk{w}_{\lscw{w}{\, 2}{}}
$$
then the polynomial $\left( z \hspace{-1pt} - \hspace{-1pt} w \right)^2$
is invertible in
$\Lin \Bbrk{z}_{\lscw{z}{\, 2}{}} \Bbrk{w}_{\lscw{w}{\, 2}{}}$
and its inverse, denoted by
\(\iota_{z,w} \Txfrac{1}{\left( z \hspace{-1pt} - \hspace{-1pt} w \right)^{\, 2}}\),
is the Taylor series of
$\Txfrac{1}{\left( z \hspace{-1pt} - \hspace{-1pt} w \right)^{\, 2}}$ in $w$
with coefficients belonging to $\C \Brk{z, \frc{1}{z^{\, 2}}}$
(i.e.,
\(\iota_{z,w}
\Txfrac{1}{\left( z \hspace{-1pt} - \hspace{-1pt} w \right)^{\, 2}} \! := \!
\Su_{n \, = \, 0}^{\infty} \hspace{-1pt}
\Txfrac{\left( -1 \right)^n}{n!}
\di_{z^{\alpha_1}} \! \dots \di_{z^{\alpha_n}} \hspace{-3pt}
\left( \hspace{-1pt} \Txfrac{1}{z^{{\, 2}}} \hspace{-1pt} \right)\)
$\hspace{-1pt} w^{\alpha_1}$ $\! \dots$ $w^{\alpha_n}$).
\end{plist}
\end{proposition}

The \textit{proof} of
Proposition~\ref{pr:3.1}
is quite simple.
We just remark that the product of
\(f \! \left( z \right) \in \C \Bbrk{z}_{\lscw{z}{\, 2}{}}\)
and \(v \! \left( z \right) \in \Lin \Bbrk{z}_{\lscw{z}{\, 2}{}}\)
can be defined as
$\left( z^{\, 2} \right)^{-N_1-N_2}$
\(\left( \left( z^{\, 2} \right)^{N_1} f \! \left( z \right) \right)\)
\(\left( \left( z^{\, 2} \right)^{N_2} v \! \left( z \right) \right)\)
for \(N_1,N_2 \mgrt 0\), according to Eq.~(\ref{loc_def}),
and does not depend on $N_1$ and $N_2$.
Then condition (\textit{b}) follows from the absence of zero
divisors in the $\C \Bbrk{z}$--module $\Lin \Bbrk{z}$.

\medskip

Having several $D$--dimensional variables \(z_1,\dots,z_n\)
one can inductively define
\beq\label{svar}
\Lin \Bbrk{z_1, \lfrc{1}{z^{\, 2}_1}
\hspace{1pt} ; \dots ; \hspace{1pt} z_n, \lfrc{1}{z^{\, 2}_n}}
\, := \,
\left( \Lin \Bbrk{z_1, \lfrc{1}{z^{\, 2}_1}
\hspace{1pt} ; \dots ; \hspace{1pt} z_{n-1}, \lfrc{1}{z^{\, 2}_{n-1}}} \right)
\Bbrk{z_n, \lfrc{1}{z^{\, 2}_n}}
.
\eeq
A different order of \(z_1,\dots z_n\) in the right hand side of~(\ref{svar})
will correspond to another way of summations in the formal series.
Nevertheless, the order of $z_k$ in the \textit{successive localizations}
\(\Lin \Bbrk{z_1}_{\lscw{z}{\, 2}{1}} \dots \Bbrk{z_n}_{\lscw{z}{\, 2}{n}}\)
is important.
For example,
\(\iota_{z,w} \Txfrac{1}{\left( z \hspace{-1pt} - \hspace{-1pt} w \right)^{\, 2}}\)
and
\(\iota_{w,z} \Txfrac{1}{\left( z \hspace{-1pt} - \hspace{-1pt} w \right)^{\, 2}}\)
are different formal series of
\(\C \Bbrk{z, \lfrc{1}{z^{\, 2}} \hspace{1pt} ; \hspace{1pt}
w, \lfrc{1}{w^{\, 2}}}\).
From
Proposition~\ref{pr:3.1}~(\textit{c})
it follows that
\beq\label{eq3.5-n}
\left( z \hspace{-1pt} - \hspace{-1pt} w \right)^{\, 2}
\left(
\iota_{z,w} \Txfrac{1}{\left( z \hspace{-1pt} - \hspace{-1pt} w \right)^{\, 2}}
-
\iota_{w,z} \Txfrac{1}{\left( z \hspace{-1pt} - \hspace{-1pt} w \right)^{\, 2}}
\right) \, = \, 0
,\,
\eeq
so we see that in the
\(\C \Brk{z, \lfrc{1}{z^{\, 2}} \hspace{1pt} ; \hspace{1pt}
w, \lfrc{1}{w^{\, 2}}}\)--module
\(\C \Bbrk{z, \lfrc{1}{z^{\, 2}} \hspace{1pt} ; \hspace{1pt}
w, \lfrc{1}{w^{\, 2}}}\) there are \textit{zero divisors}
(the same is true in any \(\C \Brk{z, \lfrc{1}{z^{\, 2}}}\)--module
\(\Lin \Bbrk{z, \lfrc{1}{z^{\, 2}}}\)).

The spaces of successive localizations play important role in
the analytic picture GCI~QFT since, according to Axiom~(\textit{F}),
we have
\beq\label{suc_phi}
\phi_{\aa_1} \! \left( z_1 \right) \dots
\phi_{\aa_n} \! \left( z_n \right) \Psi  \, \in \,
\DOM{} \Bbrk{z_1}_{\lscw{z}{\, 2}{1}} \dots \Bbrk{z_n}_{\lscw{z}{\, 2}{n}}
\eeq
for any state vector $\Psi$ belonging to the fields' domain $\DOM{}$.
They are convenient, on one hand, since these spaces have no zero divisors
and, on the other, since products of the type
\(\mathop{\prod}\limits_{1 \, \leqslant \, k \, < \, l \, \leqslant \, n}
\, z_{kl}^{\, 2}\)
are invertible in the algebra
\(\C \Bbrk{z_1}_{\lscw{z}{\, 2}{1}} \dots \Bbrk{z_n}_{\lscw{z}{\, 2}{n}}\).

We will further assume in this section the
validity of the postulates of Sect.~\ref{ssec2.2}.

\begin{proposition}
\prlabel{prp:2.2}
\begin{plist}
\item
The formal series \(\phi_{\aa} \left( z \right) \rvac\) is nonzero and
does not contain negative powers of $z^{\, 2}$, i.e.
\(\phi_{\aa} \left( z \right) \rvac \in \DOM{} \Bbrk{z}\),
for \(\aa = 1,\, \dots,\, I\).
\item
The field dimensions $d_{\aa}$ are positive integers or half integers.
They are related to the statistical parities
\(p_{\aa} \, ( \, = 0,1 )\)
by the equality \(\left( -1 \right)^{2d_{\aa}} = \left( -1 \right)^{p_{\aa}}\).
\item
The operator $H$ has a pure discrete spectrum containing
only integer or half integer nonnegative numbers.
The fields' domain $\DOM{}$ coincides with the linear span
of all eigenvectors of $H$, i.~e. this is the
finite energy space.
Moreover, the domain $\DOM{}$ is invariant under the maximal
compact subgroup
\(\compgr\)
of the conformal group
and decomposes under its action into a direct sum of
finite dimensional irreducible representations.
\end{plist}
\end{proposition}

\begin{proof} (\textit{a})
Take the formal series \(F \left( z,w \right) =
e^{w \spr T} \phi_{\aa} \left( z \right) \rvac\).
The commutation relation~(\ref{eqn2.11}) implies that
\(\di_{z^{\cmu}} F \left( z, w \right) = \di_{w^{\cmu}} F \left( z,w \right)\).
Suppose, on the other hand, in accord with the existence of harmonic
decomposition and Axiom~(\textit{F}) that
\(F  \hspace{-2pt} \left( \hspace{-0.4pt} z,w \hspace{-0.4pt} \right)
\hspace{-0.9pt} = \hspace{-0.9pt} \left( z^{\, 2} \right)^{-N} h \left( z,w \right)
+ \left( z^{\, 2} \right)^{-N+1} g \left( z,w \right)\), where
$h \left( z,w \right)$ and $g \left( z,w \right)$ are formal series with no
negative powers of $z$ and $w$ and $h$ is in addition nonzero and
harmonic with respect to~$z$.
If we now assume that \(N > 0\) it will turn out that
\(\di_{z^{\cmu}} F \left( z,w \right) =
\left( z^{\, 2} \right)^{-N-1} h_1 \left( z,w \right)
+ \left( z^{\, 2} \right)^{-N} g_1 \left( z,w \right)\)
where \(h_1 \left( z,w \right)\) and \(g_1 \left( z,w \right)\)
are series with similar properties.
Then the equation
\(\di_{z^{\cmu}} F \left( z, w \right) = \di_{w^{\cmu}} F \left( z,w \right)\)
will imply an equality of type
\(h_2 \left( z,w \right) = z^{\, 2} g_2 \left( z,w \right)\)
for the series \(h_2 \left( z,w \right)\) and \(g_2 \left( z,w \right)\)
with no negative powers, $h_2$ being nonzero and harmonic with respect to~$z$.
But this would contradict the uniqueness of
the harmonic decomposition.

The series \(\phi_{\aa} \left( z \right) \rvac\) is nonzero: otherwise
Axioms~(\textit{F}) and~(\textit{SL}) imply that
\linebreak
\(\left( \prod_{k < l} z_{kl}^{\, 2} \right)^N
\phi_{\aa} \left( z_1 \right) \phi_{\aa_2} \left( z_2 \right) \dots
\phi_{\aa_n} \left( z_n \right) \rvac = 0\), for large \(N \in \N\),
and we can cancel the
polynomial prefactor due to the fact that in the
\(\C \Bbrk{z_1}_{\lscw{z}{\, 2}{1}} \dots \Bbrk{z_n}_{\lscw{z}{\, 2}{n}}\)--module
\(\DOM{} \Bbrk{z_1}_{\lscw{z}{\, 2}{1}} \dots \Bbrk{z_n}_{\lscw{z}{\, 2}{n}}\)
there are no zero divisors;
then the completeness axiom would imply that the field
\(\phi_{\aa} \left( z \right)\) is zero.

\vspace{0.1in} \noindent (\textit{b})
The positivity of $d_{\aa}$ is a consequence,
on the one hand, of Eq.~(\ref{eqn2.22a}) implying
\(H \phi_{\aa \, \left\{ 0, 0 \right\}} \rvac =
d_{\aa} \phi_{\aa \, \left\{ 0, 0 \right\}} \rvac\)
($\phi_{\aa \, \left\{ 0,0 \right\}}$ $=$
$\phi_{\aa \, \left\{ 0,0 \right\}} \left( z \right)$
is the mode multiplying the unique, up to proportionality,
harmonic polynomial \(\hr{0}{1} \left( z \right) = 1\) of degree $0$)
and on the other, of the positivity of $H$ (axiom~(SC)).
Note also that, according to condition~(\textit{a}),
the vector $\phi_{\aa \, \left\{ 0, 0 \right\}} \rvac$
$\equiv$ $\phi_{\aa} \left( z \right) \rvac \vrestr{10pt}{z=0}$
is nonzero and noncollinear with $\rvac$ since
otherwise, it will follow that
\(\phi_{\aa \, \left\{ n, m, \sigma \right\}} \rvac = 0\) for \(n > 0\)
(and hence, \(\phi_{\aa} \left( z \right) \rvac = 0\)),
or \(\phi_{\aa} \left( z \right) \sim \ID\), respectively
(because
\(\phi_{\aa} \left( z \right) \rvac\)
$=$ $e^{\hspace{1pt} z \spr T}$
\(\phi_{\aa \, \left\{ 0, 0 \right\}} \rvac\)).
The second statement follows from Proposition~7.1 of~\cite{N03}
and the assumed rationality of the field cocycle (see the covariance
axiom).

\vspace{0.1in} \noindent (\textit{c})
The set of vectors in the axiom of completeness is actually a set
of eigenvectors of $H$ with integer or half--integer eigenvalues.
From the commutation relation~(\ref{eqn2.13}) it also follows that
every vector of this system is contained in a finite dimensional
subrepresentation of the Lie algebra of the maximal compact subgroup.
\end{proof}

\begin{remark}
\rmlabel{rm:3.2}
The vector \(\Phi_{\aa} = \phi_{\aa} \left( z \right) \rvac
\left.\raisebox{9pt}{\hspace{-2pt}}\right|_{z \, = \, 0}\)\gvspc{-5pt}
uniquely characterizes the field \(\phi_{\aa} \left( z \right)\)
and we have
\(\phi_{\aa} \left( z \right) \rvac = e^{z \spr T} \, \Phi_{\aa}\).
Moreover, for every \(v \in \DOM\) there exists a unique translation
covariant local field $Y(v,z)$ such that
\(Y \left( v,z \right) \rvac =
e^{z \spr T} \, v\) (see [N03], Sect. 3).
This is a compact formulation of the {\it state--field correspondence}
in the vertex algebra approach.
\end{remark}

\begin{theorem}
\prlabel{thm:2.3}
Every scalar product
\(\La \Psi_1 \Vl
\phi_{\aa_1} \left( z_1 \right) \dots \phi_{\aa_n} \left( z_n \right) \, \Psi_2 \Ra\)
(for arbitrary $\Psi_1$, $\Psi_2$ $\in$ $\DOM{}$),
regarded as a power series,
is absolutely convergent in the domain
\begin{eqnarray}\label{eqn2.20n}
\mathbb{U}_n^{<} :=
\left\{\raisebox{9pt}{\hspace{-2pt}}\right. && \podr
\left( z_1, \dots , z_n \right) \in \C^{D \hspace{0.5pt} n}
\, : \
z_k = e^{\hspace{0.5pt} 2 \hspace{0.5pt} \pi \hspace{0.5pt} i
\hspace{0.5pt} \czeta_k} \, u_k,\ \, \czeta_k \in \C \ \, \mathit{and} \ \,
u_k \in \Sr^{D-1} \! \subset \! \R^D
\nonumber \\ && \podr
(u_k^{\ 2} = 1)
\ \ \mathit{for} \ \ k = 1, \dots, n;\ \
\mathit{Im} \, \czeta_{kl} < 0
\ \ \mathit{for} \ \ 1 \leqslant k < l \leqslant n
\left.\raisebox{9pt}{\hspace{-2pt}}\right\}
\, \
\end{eqnarray}
and its limit is
a rational function of the form
\(
\left(\raisebox{10pt}{\hspace{-2pt}}\right.
\mathop{\prod}\limits_{k \, = \, 1}^n \, z_k^{\, 2}
\left.\raisebox{10pt}{\hspace{-2pt}}\right)^{-N} \!
\left(\raisebox{10pt}{\hspace{-2pt}}\right.
\mathop{\prod}\limits_{1 \, \leqslant \, k \, < \, l \, \leqslant \, n}
\, z_{kl}^{\, 2}
\left.\raisebox{10pt}{\hspace{-2pt}}\right)^{-N}\) \(
P \left( z_1, \dots, z_n \right)
\),
where $P \left( z_1, \dots, z_n \right)$ is a polynomial with coefficients
depending on a finite number of modes' scalar products
\(\La \Psi_1 \Vl
\phi_{\aa_1 \left\{ k_1,m_1,\sigma_1 \right\}} \dots
\phi_{\aa_n \left\{ k_n,m_1,\sigma_1 \right\}} \Psi_2\Ra\)\gvspc{9pt}.
The Wightman functions
\(\lvac
\phi_{\aa_1} \left( z_1 \right) \dots \phi_{\aa_n} \left( z_n \right)
\rvac\)\gvspc{9pt}
(viewed as rational functions)
are in addition globally conformal invariant:
\begin{eqnarray}\label{eqn2.GCI}
&
\lvac
\phi_{\aa_1} \left( z_1 \right) \dots \phi_{\aa_n} \left( z_n \right)
\rvac
\, = & \nonumber \\ & = \,
\lvac
\left[\raisebox{9pt}{\hspace{-2pt}}\right.
\pi_z \! \left( g \right)^{-1}
\left.\raisebox{9pt}{\hspace{-2pt}}\right]{\hspace{-2pt}}_{\aa_1}^{\bb_1} \, \dots \,
\left[\raisebox{9pt}{\hspace{-2pt}}\right.
\pi_z \! \left( g \right)^{-1}
\left.\raisebox{9pt}{\hspace{-2pt}}\right]{\hspace{-2pt}}_{\aa_n}^{\bb_n} \,
\phi_{\bb_1} \left( g \left( z_1 \right) \right) \dots
\phi_{\bb_n} \left( g \left( z_n \right) \right) \,
\rvac
\, , &
\end{eqnarray}
and $\Z_2$--symmetric in the sense
\begin{equation}\label{eqn2.Z2}
\hspace{0pt}
\lvac
\phi_{\aa_{\sigma \left( 1 \right)}} \!
\left(\raisebox{9pt}{\hspace{-3pt}}\right.
z_{\sigma \left( 1 \right)}
\left.\raisebox{9pt}{\hspace{-4pt}}\right) \dots
\phi_{\aa_{\sigma \left( n \right)}} \!
\left(\raisebox{9pt}{\hspace{-3pt}}\right.
z_{\sigma \left( n \right)}
\left.\raisebox{9pt}{\hspace{-4pt}}\right) \rvac
=
\left( -1 \right)^{\varepsilon \left( \sigma \right)} \, \lvac
\phi_{\aa_1} \! \left( z_1 \right) \dots \phi_{\aa_n} \! \left( z_n \right) \rvac
,
\end{equation}
for every permutation $\sigma$;
\(\left( -1 \right)^{\varepsilon \left( \sigma \right)}\) is the corresponding
statistical factor:
\(\varepsilon \left( \sigma \right) =
\sum \, p_{a_{\sigma \left( i \right)}}
p_{a_{\sigma \left( j \right)}}
\, (\text{\textit{mod}} \, 2)\) the sum being over all
pairs of indices \(i<j\) such that
\(\sigma \left( i \right) > \sigma \left( j \right)\).
\end{theorem}

\begin{proof}
Set
\begin{equation}\label{eqn2.23n}
\rho_n \, := \,
\left(\raisebox{10pt}{\hspace{-2pt}}\right.
\mathop{\prod}\limits_{k \, = \, 1}^n \, z_k^{\, 2}
\left.\raisebox{10pt}{\hspace{-2pt}}\right)
\left(\raisebox{10pt}{\hspace{-2pt}}\right.
\mathop{\prod}\limits_{1 \, \leqslant \, k \, < \, l \, \leqslant \, n}
\, z_{kl}^{\, 2}
\left.\raisebox{10pt}{\hspace{-2pt}}\right)
\, . \
\end{equation}
From Axioms (\textit{F}) and (\textit{SL}) it follows that
for sufficiently large \(N \in \N\) the formal series
\(
\rho_n^N \,
\phi_{\aa_1} \left( z_1 \right) \dots \phi_{\aa_n} \left( z_n \right) \, \Psi
\)
is $\Z_2$--sym\-met\-ric
Taylor series in the
$z$'s with coefficients in $\DOM{}$.
Hence,
\(
P ( z_1,\) $\dots,$ \(z_n ) :=\)
$\rho_n^N$
\(\La \Psi_1 \Vl \phi_{\aa_1} \left( z_1 \right)\)
$\dots$ \(\phi_{\aa_n} \left( z_n \right) \, \Psi_2 \Ra\)
is a complex Taylor series, and
if \(H \Psi_1 = \epsilon_1 \Psi_1\) and \(H \Psi_2 = \epsilon_2 \Psi_2\)
it satisfies, in addition, the Euler equation
\begin{equation}\label{eqn2.24nn}
\hspace{0pt}
\mathop{\sum}\limits_{k \, = \, 1}^n
z_k \spr \di_{z_k} P \left( z_1, \dots, z_n \right)
\hspace{-1pt} = \hspace{-1pt}
\left(\raisebox{10pt}{\hspace{-2pt}}\right.
2
n^{2N} \hspace{-3pt} \left( n \! - \! 1 \right)^N \!
+ \epsilon_1 - \epsilon_2 -
\! \Su_{k \, = \, 1}^n  d_k
\left.\raisebox{10pt}{\hspace{-2pt}}\right)
P \left( z_1, \dots, z_n \right)
\,
\end{equation}
as a consequence of the commutation relations~(\ref{eqn2.12}).
Therefore, \(P \left( z_1, \dots , z_n \right)\) is a polynomial
and it is clear that its coefficients are linear combinations of
scalar products of the type
\(\La \Psi_1 \Vl
\phi_{\aa_1 \left\{ k_1,m_1,\sigma_1 \right\}} \dots
\phi_{\aa_n \left\{ k_n,m_1,\sigma_1 \right\}} \Psi_2\Ra\)\gvspc{9pt}.
Because of
Proposition~\ref{prp:2.2}~(b)
we find that the
series
\(\rho_n^N \,
\La \Psi_1 \Vl
\phi_{\aa_1} \left( z_1 \right) \dots \phi_{\aa_n} \left( z_n \right) \, \Psi_2 \Ra\)
is a polynomial for all \(\Psi_1,\, \Psi_2 \in \DOM\)
for \(N \in \N\) sufficiently large.

We now divide by $\rho_n^N$ in the space
\(\C \Bbrk{z_1}_{\lscw{z}{\, 2}{1}} \dots \Bbrk{z_n}_{\lscw{z}{\, 2}{n}}\)
which contains
\(\La \Psi_1 \Vl\) \(\phi_{\aa_1} \left( z_1 \right)\)
$\dots$ \(\phi_{\aa_n} \left( z_n \right) \, \Psi_2 \Ra\),
because of the property~(\ref{suc_phi}).
(The inverse series of $\rho_n^N$ is obtained in
\(\C \Bbrk{z_1}_{\lscw{z}{\, 2}{1}} \dots \Bbrk{z_n}_{\lscw{z}{\, 2}{n}}\)
by expanding every factor \(\left( z_{kl}^{\, 2} \right)^{-N}
= \left(\raisebox{9pt}{\hspace{-3pt}}\right.
\left( z_k - z_l \right)^{2} \left.\raisebox{9pt}{\hspace{-2pt}}\right)^{-N}\),
for \(1 \leqslant k < l \leqslant n\), in Taylor series in $z_l$ around
\(z_l = 0\)~--~see Proposition~\ref{pr:3.1}~(\textit{c})).

Thus the domain of absolute convergence of the formal series
\(\La \Psi_1 \Vl\)
\(\phi_{\aa_1} \left( z_1 \right)\)
\(\dots\) \(\phi_{\aa_n} \left( z_n \right)\) \(\Psi_2 \Ra\)
coincides with the domain of absolute convergence of the above expanded
``propagators'' $\left( z_{kl}^{\, 2} \right)^{-N}$,
which contains $\mathbb{U}_n^<$.

The covariance law~(\ref{eqn2.GCI}) and the
$\Z_2$--symmetry~(\ref{eqn2.Z2}) follow from the
uniqueness of the analytic continuation.
\end{proof}

{\samepage

\begin{remark}
\prlabel{rmr:2.1}
The domain $T_+$ can be characterized as the connected component of \(z = 0\)
of the subset
\(\left\{\raisebox{10pt}{\hspace{-3pt}}\right.
z \in \C : \Txfrac{\left( z - z^* \right)^{2}}{\left( z^* \right)^2}
\equiv 1 - 2 \, z \spr \overline{z} + \left| z^2 \right|^2
\neq 0
\left.\raisebox{10pt}{\hspace{-3pt}}\right\}\).
Note that due to the axiom (*)
the product of $\phi_{\aa} \left( z \right)$
with its conjugate \(\phi_{\aa} \left( z \right)^*\)
will be singular for \(\left( z - z^* \right)^{2} = 0\).
\end{remark}

}

\subsection{Ellipticity of the finite temperature correlation functions}\label{mssec:3.2p}

To study the thermal correlation functions it is
convenient to use the compact picture representation
of the GCI fields.
Therefore, we begin with stating the basic properties of
the compact picture formal series which are analogous to those of
the formal power series of
the previous subsection.

Denote by $\Lin \Aabrk{e^{\pm \pi i \, \czeta},u}$ the space of
infinite formal Fourier series
\beq\label{cps}
v \! \left( \czeta, u \right) =
\Su_{\nnu \, \in \, \smtxfrac{1}{2} \Z} \
\Su_{m \, = \, 0}^{\infty} \,
\Su_{\sigma} \,
v_{\nnu,\, m,\, \sigma} \
e^{-2\pi i \, \nnu \czeta} \, \hr{m}{\sigma}
\left( u \right)
\eeq
with coefficients in $\Lin$. It is a module over the algebra
$\C \Abrk{e^{\pm \pi i \, \czeta},u}$ of Fourier polynomials
(the space of finite complex series of type (\ref{cps})).
When $\nnu$ in Eq.~(\ref{cps}) runs over $\Z$ the resulting space of series
will be denoted by
$\Lin \Bbrk{e^{\pm 2 \pi i \, \czeta},u}$.
In the product
\(f \! \left( \czeta,u \right) v \! \left( \czeta,u \right)\)
of
\(f \! \left( \czeta,u \right) \in \C \Abrk{e^{\pm \pi i \, \czeta},u}\)
and
\(v \! \left( \czeta,u \right) \in \Lin \Aabrk{e^{\pm \pi i \, \czeta},u}\)
the harmonic polynomials
$\hr{m}{\sigma} \left( u \right)$
are treated as spanning the algebra $\C \Bbrk{u}$ of
polynomial functions over the unit sphere \(\Sr^{D-1} \, ( \, \ni u)\).
The space of $n$--point formal series, denoted by
\(\Lin \Aalbrk{e^{\pm \pi i \, \czeta_1},u_1} ;\) $\dots;$
\(\Aarbrk{e^{\pm \pi i \, \czeta_n},u_n}\), is a
module over the algebra
\(\C \Albrk{e^{\pm \pi i \, \czeta_1},u_1} ;\) $\dots;$
\(\Arbrk{e^{\pm \pi i \, \czeta_n},u_n}\).
Following the line of arguments of the previous subsection
we need a compact picture analogue of the localized space
$\Lin \Bbrk{z}_{\lscw{z}{\, 2}{}}$.
Observing that the basic terms in~(\ref{cps}) can be represented as:
\beq\label{repr}
e^{-2\pi i \, \nnu \czeta} \, \hr{m}{\sigma} \! \left( u \right)
\, = \,
\left( z^{\, 2} \right)^{n}
\hr{m}{\sigma} \! \left( z \right)
\quad \text{for} \quad
z = e^{2\pi i \, \czeta} u
\quad \text{and} \quad
\nnu = -\frac{n+m}{2}
\eeq
we are led to introduce
the space $\Lin \Aabrk{e^{\pm \pi i \, \czeta},u}_+$
defined as containing those series~(\ref{cps}) for which
there exists $N \in \N$ such that
\(v_{\nnu,m,\sigma} = 0\) if \(\nnu+m > N\)
(thus excluding arbitrary large powers of $z^{\, 2}$
in the $z$--picture).
It then follows from~(\ref{repr}) that
\(v \! \left( \czeta,u \right) \in\)
\(\Lin \Aabrk{e^{\pm \pi i \, \czeta},u}_+\)
iff
$v \! \left( \czeta,u \right)$ $=$
$e^{-\pi i \, N \czeta}$
\(v' \! \left(\raisebox{9pt}{\hspace{-2pt}}\right.
e^{2\pi i \, \czeta} u
\left.\raisebox{9pt}{\hspace{-2pt}}\right)\)
for some
\(N \in \N\)
and
\(v' \! \left( z \right) \in \Lin \Bbrk{z}\).
We conclude, using
Proposition~\ref{pr:3.1},
that \(\C \Aabrk{e^{\pm \pi i \, \czeta},u}_+\)
is a complex algebra and
\(\Lin \Aabrk{e^{\pm \pi i \, \czeta},u}_+\)
is a module over this algebra with no zero divisors.

Recall that \(4\sin \pi \czeta_+ \sin \pi\czeta_-\), defined by
Eqs.~(\ref{eqn?}) and (\ref{eq1.6}), the compact
picture analogue of the interval $z_{12}^{\, 2}$,
is a Fourier polynomial belonging to
\(\C \Albrk{e^{\pm 2 \pi i \, \czeta_1},u_1} ;\)
\(\Arbrk{e^{\pm 2 \pi i \, \czeta_2},u_2}\)
(i.e., the space of series containing just \textit{integer} powers of
$e^{\pm 2 \pi i \, \czeta_k}$).
We shall now introduce an \textit{elliptic} version of the
\textit{compact picture  interval}:
\beqa\label{Th}
&& \hspace{-30pt}
\Theta \! \left( \czeta_1,u_1;\, \czeta_2,u_2;\, \tau \right) \, := \,
e^{-\smtxfrac{\pi i \tau}{2}} \,
\vartheta_{11} \! \left( \czeta_+,\, \tau \right)
\vartheta_{11} \! \left( \czeta_-,\, \tau \right)
\, = \,
\nn && \hspace{-30pt} \quad = \,
4\sin \pi\czeta_+ \sin \pi\czeta_-
- 4 \left(
\sin 3\pi\czeta_+ \sin \pi\czeta_- \! + \! \sin \pi\czeta_+ \sin 3\pi\czeta_-
\right) e^{2\pi i \tau} + \dots
,\ \
\eeqa
where $\vartheta_{11} \left( \zeta,\tau \right)$ are
the Jacobi $\vartheta$--function~(\ref{eqn_new}).
Having $n$ compact picture points
$\left( \czeta_1,u_1 \right)$ $\dots$ $\left( \czeta_n,u_n \right)$
$(\in \R \times \Sr^{D-1})$
we introduce the shorthand notation:
\beq\label{jk}
\Theta_{jk} \, := \, \Theta \! \left( \czeta_j,u_j;\, \czeta_k,u_k;\, \tau \right)
\, , \quad
\OMG{n} \! \left( \czeta_1,u_1;\, \dots;\, \czeta_n,u_n \right) \, := \,
\mathop{\prod}\limits_{1 \, \leqslant \, j \, < \, k \, \leqslant \, n}
\Theta_{jk}
.\,
\eeq
Each term of the series~(\ref{Th}) and (\ref{jk}) is a
function of the coordinate differences \(\czeta_{jk} = \czeta_j - \czeta_k\);
hence, we can write
\(\Theta \left( \czeta_{12};\, u_1,u_2 \right)\) and
$\OMG{n} \! (\czeta_{12},$ $\dots,$ $\czeta_{n-1 \, n};$ $u_1,$ $\dots,$ $u_n)$
treating them, however, as series in different spaces.

\begin{proposition}
\nopagebreak
\prlabel{pr:3.4}
\begin{plist}
\item
$\Theta_{12}$ is a formal series belonging to
\(\C \Albrk{e^{\pm2\pi i \, \czeta_1},u_1} ;\)
\(\Arbrk{e^{\pm2\pi i \, \czeta_2},u_2} \Bbrk{q}\)
(\(q = e^{2\pi i \tau}\)), i.e.,
$\Theta_{12}$ is a Taylor series in $q$ with coefficients which are
Fourier polynomials belonging to
\(\C \Albrk{e^{\pm2\pi i \, \czeta_1},u_1} ;\)
\(\Arbrk{e^{\pm2\pi i \, \czeta_2},u_2}\).
Moreover, $\Theta_{12}$ is symmetric,
\(\Theta_{12} = \Theta_{21}\),
and it is divisible by
\(\sin \pi\czeta_+ \sin \pi\czeta_-\):
\beqa\label{div}
\hspace{-30pt}
\Theta_{12} \, = && \podr
4 \sin \pi\czeta_+ \sin \pi\czeta_- \,
\left( 1 + q \, \Theta'_{12} \right)
\, = \,
\nn \hspace{-30pt} = && \podr
4\sin \pi\czeta_+ \sin \pi\czeta_-
\left\{\raisebox{10pt}{\hspace{-3pt}}\right. 1
- 2 \left( 1 + 2 \cos 2\pi\czeta_{12} \cos 2\pi\alpha \right)
q
+ \dots
\left.\raisebox{10pt}{\hspace{-3pt}}\right\}
,\ \
\eeqa
where $\Theta'_{12}$ is again a series belonging to
\(\C \Albrk{e^{\pm2\pi i \, \czeta_1},u_1} ;\)
\(\Arbrk{e^{\pm2\pi i \, \czeta_2},u_2} \Bbrk{q}\).
\item
The formal series
$\Theta_{12}$ has inverse:
\beq\label{inv_th}
\Theta_{12}^{-1} \in \C \Aabrk{e^{\pm2\pi i \, \czeta_1},u_1}_+
\Aabrk{e^{\pm2\pi i \, \czeta_2},u_2}_+ \Bbrk{q}
\quad
(\Theta_{12} \Theta_{12}^{-1} = 1) .
\eeq
The series $\Theta_{12}^{-1}$ is absolutely convergent
in the domain
\(0 < - \textit{Im} \, \czeta_{12} < \textit{Im} \, \tau\).
\item
\(\OMG{n} \! \left( \czeta_1,u_1;\, \dots;\, \czeta_k+\tau,u_k;\,
\dots;\, \czeta_n,u_n \right)\) is a symmetric formal series belonging to
\(\C \Albrk{e^{\pm2\pi i \, \czeta_1},u_1} ;\) $\dots$;
\(\Arbrk{e^{\pm2\pi i \, \czeta_n},u_n} \Bbrk{q}\)
(i.e., \(\OMG{n} (\czeta_1,u_1;\) $\dots;$ \(\czeta_n,u_n)\) $=$
\(\OMG{n} (\czeta_{\sigma_1},u_{\sigma_1};\) $\dots;$
\(\czeta_{\sigma_n},u_{\sigma_n})\)).
As a series in the conformal time differences $\czeta_{j \, j+1}$,
$\OMG{n} (\czeta_{12},$ $\dots,$ $\czeta_{n-1 \, n};$ $u_1,$ $\dots,$ $u_n)$
satisfies the property
\beqa\label{Omega}
&&
\OMG{n} \!
\left( \czeta_{12}+\lambda_1\tau,\dots,
\czeta_{n-1 \, n}+\lambda_{n-1}\tau
;\, u_1, \dots, u_n \right)
\, =
\nn && \quad \, = \,
\exp
\left\{\raisebox{14pt}{\hspace{-2pt}}\right.
- 2 \pi i
\left(\raisebox{14pt}{\hspace{-2pt}}\right.
\Su_{j,\, k \, = \, 1}^{m-1}
\lambda_j \, A_{jk}^{\left( n \right)} \, \lambda_k
\left.\raisebox{14pt}{\hspace{-2pt}}\right) \tau
\, - \,
4 \pi i
\Su_{j,\, k \, = \, 1}^{m-1}
\lambda_j \, A_{jk}^{\left( n \right)} \, \czeta_{k\, k+1}
\left.\raisebox{14pt}{\hspace{-2pt}}\right\} \times
\nn && \hspace{24pt} \times \,
\OMG{n} \! \left( \czeta_{12},\dots,\czeta_{n-1 \, n};\, u_1, \dots, u_n \right)
\eeqa
for all \(\left( \lambda_1,\dots,\lambda_{n-1} \right) \in \Z^{n-1}\),
where
\(A_{jk}^{\left( n \right)} = A_{kj}^{\left( n \right)}\)
are fixed integer constants (depending just on their indices)
and the symmetric matrix
\(\left\{\raisebox{9pt}{\hspace{-3pt}}\right.
A_{jk}^{\left( n \right)}
\left.\raisebox{9pt}{\hspace{-3pt}}\right\}\mtrx_{j,k = 1}^{n-1}\)
is positive definite.
\item
Let $R$ be a commutative complex algebra.
The space of Taylor series
$F ( \czeta_{12},$ $\dots,$ $\czeta_{n-1 \, n};$ $\tau)$
in $q^{\frac{1}{2}}$ which have polynomial coefficients belonging to
$R \Brkl{e^{\pm \pi i \, \czeta_{12}}}$, $\dots$,
$\Brkr{e^{\pm \pi i \, \czeta_{n-1 \, n}}}$
and obey
for all \(\lambda_1,\dots,\lambda_{n-1} \in \Z\)
the properties
\beqa
\label{eaB.10}
&&
\hspace{-10pt}
F \! \left( \czeta_{12} \hspace{-1pt} + \hspace{-1pt} \lambda_1, \dots,
\czeta_{n-1 \, n} \hspace{-1pt} + \hspace{-1pt} \lambda_{n-1}; \tau \right)
\, = \, \left( -1 \right)^{\Su_{k \, = \, 1}^{n-1}
\lambda_k \varepsilon_k^{\left( 1 \right)}}
F \! \left( \czeta_{12}, \dots, \czeta_{n-1 \, n}; \tau \right)
, \qquad
\\ \label{F_Omega}
&&
\hspace{-10pt}
F \!
\left( \czeta_{12} \hspace{-1pt} + \hspace{-1pt} \lambda_1\tau,\dots,
\czeta_{n-1 \, n} \hspace{-1pt} + \hspace{-1pt} \lambda_{n-1}\tau
;\tau \right)
\, = \,
\left( -1 \right)^{\Su_{k \, = \, 1}^{n-1}
\lambda_k \varepsilon_k^{\left( \tau \right)}}
\times
\nn &&
\hspace{-10pt}
\hspace{24pt} \times \,
\exp
\left\{\raisebox{14pt}{\hspace{-2pt}}\right.
- 2 \pi i N
\left(\raisebox{14pt}{\hspace{-2pt}}\right.
\Su_{j,\, k \, = \, 1}^{m-1}
\lambda_j \, A_{jk}^{\left( n \right)} \, \lambda_k
\left.\raisebox{14pt}{\hspace{-2pt}}\right) \tau
\, - \,
4 \pi i N
\Su_{j,\, k \, = \, 1}^{m-1}
\lambda_j \, A_{jk}^{\left( n \right)} \, \czeta_{k\, k+1}
\left.\raisebox{14pt}{\hspace{-2pt}}\right\} \times
\nn &&
\hspace{-10pt}
\hspace{24pt} \times \,
F \! \left( \czeta_{12},\dots,\czeta_{n-1 \, n};\tau \right)
,
\eeqa
where
\(\varepsilon_k^{\left( 1 \right)},\) \(\varepsilon_k^{\left( \tau \right)} =\)
\(0,1\) (\(k = 1,\dots, n-1\)), \(N \in \N\) and
\(\left\{\raisebox{9pt}{\hspace{-3pt}}\right.
A_{jk}^{\left( n \right)}
\left.\raisebox{9pt}{\hspace{-3pt}}\right\}\mtrx_{j,k = 1}^{n-1}\)
is integral positive matrix,
is a finitely generated module over $R \Bbrk{q^{\frac{1}{2}}}$.
In other words,
there exists a finite number of fixed (complex) series
$F^{\left( N \right)}_c ( \czeta_{12},$ $\dots,$ $\czeta_{n-1 \, n};$ $\tau)$,
\(c = 1,\dots,C_N\),
obeying the properties~(\ref{eaB.10}) and (\ref{F_Omega})
and such that
$F ( \czeta_{12},$ $\dots,$ $\czeta_{n-1 \, n};$ $\tau)$ $=$
$\mathop{\sum}\limits_{c \, = \, 1}^{C_N}$
$G_c \left( \tau \right)$
$F^{\left( N \right)}_c ( \czeta_{12},$ $\dots,$ $\czeta_{n-1 \, n};$ $\tau)$
for
\(G_c \left( \tau \right) \in R \Bbrk{q^{\frac{1}{2}}}\).
Moreover, the basic series
$F^{\left( N \right)}_c ( \czeta_{12},$ $\dots,$ $\czeta_{n-1 \, n};$ $\tau)$
can be chosen absolutely convergent to analytic functions
for \(\mathit{Im} \tau > 0\) and \(\czeta_{k \, k+1} \in \C\)
(\(k = 1,\dots,n-1\)).
If we have a multicomponent series
$F_{\aa_1\dots\aa_n} ( \czeta_{12},$ $\dots,$
$\czeta_{n-1 \, n};$ $\tau)$ of the above type which, in addition,
is $\Z_2$--symmetric (in the sense of Eq.~(\ref{eqn2.Z2}))
then we can expand it in a finite $R \Bbrk{q^{\frac{1}{2}}}$--linear
combination of $\Z_2$--symmetric basic series
$F^{\left( N \right)}_{\aa_1\dots\aa_n;c} ( \czeta_{12},$ $\dots,$
$\czeta_{n-1 \, n};$ $\tau)$.
\end{plist}
\end{proposition}

The proof of
Proposition~\ref{pr:3.4}
is straightforward. We present it
in
Appendix~B.

Let us note that taking the ratios
\beq\label{bas_ell}
E_c^{\left( N \right)} \!
\left( \czeta_{12},\dots,\czeta_{n-1 \, n};\, u_1, \dots, u_n;\, \tau \right)
\, := \,
\frac{
F^{\left( N \right)}_c \! \left( \czeta_{12},\dots,\czeta_{n-1 \, n};\, \tau \right)
}{
\OMG{n} \! \left( \czeta_{12},\dots,\czeta_{n-1 \, n};\, u_1, \dots, u_n \right)^N}
\eeq
for \(c = 1,\dots,C_N\),
using the basic systems in
Proposition~\ref{pr:3.4}~(\textit{d}),
we obtain systems of formal series (for \(N \in \N\)) belonging to
\(\C \Aabrk{e^{\pm2\pi i \, \czeta_1},u_1}_+\) $\dots$
\(\Aabrk{e^{\pm2\pi i \, \czeta_n},u_n}_+ \Bbrk{q}\)
which are absolutely convergent in the domain
\(0 < - \textit{Im} \, \czeta_{jk} < \textit{Im} \, \tau\)
for \(1 \leqslant j < k \leqslant n\)
to elliptic functions in every $\czeta_k$
(Proposition~\ref{pr:3.4}~(\textit{b}) and~(\textit{c})):
\beqa
&&
E_c^{\left( N \right)} \!
\left( \czeta_{12} \hspace{-1pt} + \hspace{-1pt} \lambda_1, \dots,
\czeta_{n-1 \, n} \hspace{-1pt} + \hspace{-1pt} \lambda_{n-1}; \tau \right)
\, = \, \left( -1 \right)^{\Su_{k \, = \, 1}^{n-1}
\lambda_k \varepsilon_k^{\left( 1 \right)}}
E_c^{\left( N \right)} \!
\left( \czeta_{12}, \dots, \czeta_{n-1 \, n}; \tau \right)
, \qquad
\nn
&&
E_c^{\left( N \right)} \!
\left( \czeta_{12} \hspace{-1pt} + \hspace{-1pt} \lambda_1\tau,\dots,
\czeta_{n-1 \, n} \hspace{-1pt} + \hspace{-1pt} \lambda_{n-1}\tau
;\tau \right)
\, = \,
\left( -1 \right)^{\Su_{k \, = \, 1}^{n-1}
\lambda_k \varepsilon_k^{\left( \tau \right)}}
E_c^{\left( N \right)} \!
\left( \czeta_{12},\dots,\czeta_{n-1 \, n};\tau \right)
.
\nn && \label{ell_prop1}
\eeqa
The finite temperature correlation functions can be
written as linear combinations of such ratios
with coefficients that are $\tau$--dependent spherical
functions (or, at least, formal Fourier series) in~$u_k$.

\medskip

We are now ready to find out the general structure of the
Gibbs (thermal) correlation functions.
When one considers the thermodynamic properties of quantum fields
additional assumptions are always needed (see footnote~1).
In our framework we impose the \textit{minimal assumption}
that the conformal Hamiltonian $H$ has
\textit{finite dimensional} eigenspaces.
This makes it possible to introduce the partition function
$Z \left( \tau \right)$, and the thermal mean values
\(\La
\phi_{\aa_1;\, \nnu_1,m_1,\sigma_1}\) $\dots$
\(\phi_{\aa_k;\, \nnu_k,m_k,\sigma_k} \Ra_q\)
(\(q=e^{2\pi i \tau}\))
of products of compact picture modes $\phi_{\aa;\, \nnu,m,\sigma}$
of fields \(\phi_{\aa} \! \left( \czeta,u \right)\) as
formal power series in $q^{\frac{1}{2}}$,
\begin{eqnarray}\label{neqq1}
& \hspace{-12pt}
Z \left( \tau \right)
\hspace{-1pt} = \hspace{-1pt}
\mathit{tr}_{\DOM{}} \left( q^H \right)
\hspace{-1pt} = \hspace{-1pt}
\text{\large \(\mathop{\sum}\limits_{j = 0}^{\infty}
\mathop{\sum}\limits_{\sigma}\)} \,
\La \! \Psi_{j\sigma} \! \Vl \!
\Psi_{j\sigma} \! \Ra q^{\frac{j}{2}}
\! , \quad
& \\ \label{neqq2} & \hspace{1pt}
\La \phi_{\aa_1;\, \nnu_1,m_1,\sigma_1}\dots
\phi_{\aa_k;\, \nnu_k,m_k,\sigma_k} \Ra_q
:=
\Txfrac{1}{Z \left( \tau \right)} \, \mathit{tr}_{\DOM{}}
\left\{ \phi_{\aa_1;\, \nnu_1,m_1,\sigma_1}\dots
\phi_{\aa_k;\, \nnu_k,m_k,\sigma_k} q^H \right\}
=
& \nonumber \\ & \hspace{-12pt}
= \,
\Txfrac{1}{Z \left( \tau \right)}
\,
\text{\large \(\mathop{\sum}\limits_{j \, = \, 0}^{\infty} \,
\mathop{\sum}\limits_{\sigma}\)} \,
\La \Psi_{j\sigma} \Vl
\phi_{\aa_1;\, \nnu_1,m_1,\sigma_1}\dots
\phi_{\aa_k;\, \nnu_k,m_k,\sigma_k} \,
\Psi_{j\sigma} \Ra q^{\frac{j}{2}}
&
\end{eqnarray}
where \(\left\{ \Psi_{j\sigma} \right\}_{j\sigma}\) is an orthonormal
basis in the Hilbert state space consisting of eigenvectors of $H$,
\(H \Psi_{j\sigma} = \txfrac{j}{2} \, \Psi_{j\sigma}\)\gvspc{9pt}.
(Note that \(\DOM = \mathit{Span}_{\C} \left\{ \Psi_{j\sigma} \right\}_{j\sigma}\)
because of
Proposition~\ref{prp:2.2}~(\textit{c});
note also that the series of $Z\left( \tau \right)$ has a leading term $1$
so that it is invertible in $\C \Bbrk{q^{\frac{1}{2}}}$--see Fact~B.1.)
The cyclic property of the traces over each (finite dimensional)
eigenspace of $H$ will imply the KMS property:
\beqa\label{KMS}
&
\La \phi_{\aa_1;\, \nnu_1,m_1,\sigma_1} \dots
\phi_{\aa_n;\, \nnu_k,m_k,\sigma_k} \Ra_q
\, = & \nn & = \,
\La \phi_{\aa_2;\, \nnu_2,m_2,\sigma_2} \dots
\phi_{\aa_n;\, \nnu_k,m_k,\sigma_k}
\, q^H \, \phi_{\aa_1;\, \nnu_1,m_1,\sigma_1} \, q^{-H}
\Ra_q
\, = & \nn & = \,
q^{-\nnu_1}
\La \phi_{\aa_2;\, \nnu_2,m_2,\sigma_2} \dots
\phi_{\aa_n;\, \nnu_k,m_k,\sigma_k}
\phi_{\aa_1;\, \nnu_1,m_1,\sigma_1}
\Ra_q
\,
\eeqa
(according to Eq.~(\ref{eqn2.22a}))
as an equality in \(\C \Bbrk{q^{\frac{1}{2}}}\).
Summing then over all triples \(\nnu_l,m_l,\sigma_l\) in
the corresponding expansions of
$\phi_{\aa_l} \left( \czeta_l,u_l \right)$ by~(\ref{eqn2.20a})
we obtain the KMS equation
\beqa\label{KMS_eq}
&
\La \phi_{\aa_1} \! \left( \czeta_1,\, u_1 \right)
\dots \phi_{\aa_n} \! \left( \czeta_n,\, u_n \right) \Ra_q
\, = & \nn & =  \,
\La
\phi_{\aa_2} \! \left( \czeta_2,\, u_2 \right)
\dots \phi_{\aa_n} \! \left( \czeta_n,\, u_n \right)
\phi_{\aa_1} \! \left( \czeta_1+\tau,\, u_1 \right)
\Ra_q
\, \
\eeqa
as an equality of formal series belonging to
\(\C \Bbrk{q^{\frac{1}{2}}}\)
\(\Aalbrk{e^{\pm2\pi i \, \czeta_1},u_1} ;\) $\dots;$
\(\Aarbrk{e^{\pm2\pi i \, \czeta_n},u_n}\).

On the other hand, one can perform the sum in the trace
\(\mathit{tr}_{\DOM{}}
\left\{\raisebox{9pt}{\hspace{-3pt}}\right.
\phi_{\aa_1} \! \left( \czeta_1,\, u_1 \right)\)
$\dots$ \(\phi_{\aa_n} \! \left( \czeta_n,\, u_n \right)\)
\(q^H \left.\raisebox{9pt}{\hspace{-3pt}}\right\}\)
first over the fields' modes and then over the energy levels
(taking the sum in the powers of~$q^{\frac{1}{2}}$),
\begin{eqnarray}\label{neqq1-a}
& \mathit{tr}_{\DOM{}}
\left\{ \phi_{\aa_1} \! \left( \czeta_1, u_1 \right)
\dots \phi_{\aa_n} \! \left( \czeta_n, u_n \right) q^H \right\}
=
& \nonumber \\ & \hspace{-12pt}
= \,
\text{\large \(\mathop{\sum}\limits_{j \, = \, 0}^{\infty} \,
\mathop{\sum}\limits_{\sigma}\)} \,
\La \Psi_{j\sigma} \Vl
\phi_{\aa_1} \! \left( \czeta_1,\, u_1 \right)
\dots \phi_{\aa_n} \! \left( \czeta_n,\, u_n \right) \,
\Psi_{j\sigma} \Ra q^{\frac{j}{2}}
; &
\end{eqnarray}
this
gives a meaning to Eq.~(\ref{KMS_eq}) as an equality in the space
\(C\Aabrk{e^{\pm2\pi i \, \czeta_1},u_1}_+\) $\dots$
\(\Aabrk{e^{\pm2\pi i \, \czeta_n},u_n}_+ \Bbrk{q^{\frac{1}{2}}}\)
(according to Eq.~(\ref{loc_def})).
Now, if we multiply both sides of~(\ref{KMS_eq}) by
$\OMG{n} (\czeta_{12},$ $\dots,$ $\czeta_{n-1 \, n};$ $u_1,$ \(\dots, u_n)^N\)
for some \(N \mgrt 0\), setting
\beq\label{F_set}
\!\!\!
F_{\aa_1\dots\aa_n} \!
\left( \czeta_{12}, \dots, \czeta_{n-1 \, n}; u_1,\dots, u_n; \tau \right)
:=
\OMG{n}^N \!
\La \phi_{\aa_1} \! \left( \czeta_1, u_1 \right)
\dots \phi_{\aa_n} \! \left( \czeta_n, u_n \right) \Ra_{\! q}
,
\eeq
we find that $F_{\aa_1\dots\aa_n}$ are \textit{symmetric}
formal Fourier series
belonging to
\(\C \Albrk{e^{\pm \pi i \, \czeta_1},u_1} ;\) $\dots$;
\(\Arbrk{e^{\pm \pi i \, \czeta_n},u_n} \Bbrk{q^{\frac{1}{2}}}\)
(i.e., with \textit{symmetric} polynomial coefficients
multiplying each power of
$q^{\frac{1}{2}}\,$)
and they also obey the properties~(\ref{eaB.10}) and~(\ref{F_Omega})
with \(\varepsilon_k^{\left( 1 \right)} =\)
\(\varepsilon_k^{\left( 2 \right)} =\)
\(\Su_{\ell \, = \, k+1}^{n} p_{\ell}\)
for \(k = 1,\dots,n-1\)
($p_{\ell}$ being the fermion parities).
The first statement is verified by using the rationality (Theorem~3.3)
and Eq.~(\ref{div}), while the second uses
Proposition~\ref{pr:3.4}~(\textit{c})
and the KMS equation~(\ref{KMS_eq})
combined with the $\Z_2$--symmetry~(\ref{eqn2.Z2}).
Thus, we can apply
Proposition~\ref{pr:3.4}~(\textit{d})
(with $R$, the algebra $\C \Brkl{u_1,$ $\dots,}$ $\Brkr{u_n}$
of harmonic polynomials in $u_1,$ $\dots,$ $u_n$ restricted to $\Sr^{D-1}$)
obtaining the expansion
\beqa\label{expansion}
& \hspace{-30pt}
\La \phi_{\aa_1} \! \left( \czeta_1, u_1 \right)
\dots \phi_{\aa_n} \! \left( \czeta_n, u_n \right) \Ra_q
\, = & \nn & \hspace{-30pt} = \,
\mathop{\sum}\limits_{c \, = \, 1}^{C_N} \,
G_{c;\hspace{1pt} \aa_1\dots\aa_n} \!
\left( u_1,\dots, u_n; \tau \right)
E_c^{\left( N \right)} \!
\left( \czeta_{12},\dots,\czeta_{n-1 \, n};\, u_1, \dots, u_n;\, \tau \right)
&
\eeqa
in the basic elliptic functions of the system~(\ref{bas_ell}).
The coefficients
\(G_{c;\hspace{1pt} \aa_1\dots\aa_n} (u_1,\) $\dots,$ $u_n;$ \(\tau)\)
are Taylor series in $q^{\frac{1}{2}}$ with symmetric
polynomial coefficients in \(u_1,\dots,u_n \in \Sr^{D-1}\)
(i.e.,
\(G_{c;\hspace{1pt} \aa_1\dots\aa_n} \in\)
\(\C \Brkl{u_1,}\) $\dots,$ \(\Brkr{u_n} \Bbrk{q^{\frac{1}{2}}}\)).

We thus end up with the following result.

\begin{theorem}
\prlabel{adprp1}
Under the assumptions of Sect.~\ref{sec.2n}
and the additional condition that the conformal Hamiltonian $H$ has
finite dimensional eigenspaces every $n$--point thermal correlation
function
\(\La \phi_{\aa_1} \! \left( \czeta_1,\, u_1 \right)\)
$\dots$ \(\phi_{\aa_n} \! \left( \czeta_n,\, u_n \right) \Ra_q\)
admits a formal series representation of type~(\ref{expansion})
where \(G_{c;\hspace{1pt} \aa_1\dots\aa_n} ( u_1,\) $\dots,$ \(u_n; \tau)\)
are Taylor series in $q^{\frac{1}{2}}$ with symmetric (harmonic)
polynomial coefficients in \(u_1,\) $\dots,$ \(u_n \in \Sr^{D-1}\)
and
\(E_c^{\left( N \right)} ( \czeta_{12},\dots,\)
\(\czeta_{n-1 \, n};\, u_1,\) $\dots,$ \(u_n; \tau )\)
(\(c = 1,\dots,C_N\))
are some fixed series
which belong to \(\C \Aabrk{e^{\pm2\pi i \, \czeta_1},u_1}_+\) $\dots$
\(\Aabrk{e^{\pm2\pi i \, \czeta_n},u_n}_+ \Bbrk{q^{\frac{1}{2}}}\)
and
are absolutely convergent in the domain
\beq\label{ell_domain}
\left\{\raisebox{9pt}{\hspace{-3pt}}\right.
\left( \czeta_1,u_1;\dots;\czeta_n,u_n;
\tau \right)
\hspace{-1pt} \in \hspace{-1pt}
\left( \C \hspace{-1pt} \times \hspace{-1pt} \Sr^{D-1} \right)^{\hspace{-1pt}n}
\!\hspace{-1pt} \times \hspace{-1pt} \hcom : \hspace{1pt}
0 \hspace{-1pt} < \hspace{-2pt}
- \textit{Im} \hspace{1pt} \czeta_{jk} \hspace{-2pt} < \hspace{-1pt}
\textit{Im} \hspace{1pt} \tau
\hspace{2pt}
(1 \hspace{-2pt} \leqslant \hspace{-2pt}
j \hspace{-2pt} < \hspace{-2pt}
k \hspace{-2pt} \leqslant \hspace{-2pt} n)
\left.\raisebox{9pt}{\hspace{-3pt}}\right\}
\eeq
to $\Z_2$--symmetric (in the sense of Eq.~(\ref{eqn2.Z2}))
meromorphic functions over
\(\left( \C \times \Sr^{D-1} \right)^n \! \times \hcom\).
Moreover,
the resulting functions
are doubly periodic (resp., antiperiodic) in \(\czeta_m\) with
periods $1$ and $\tau$ if $\phi_{\aa_m}$ is a bosonic (resp., fermionic)
field, for \(m = 1,\dots, n\).
\end{theorem}

The problem of summability of the
angular coefficients \(G_{c;\hspace{1pt} \aa_1\dots\aa_n} ( u_1,\) $\dots,$ \(u_n; \tau)\)
is still open.
Let us note first that if we exchange the order of summation in
$G_{c;\hspace{1pt} \aa_1\dots\aa_n}$ first summing in the powers of $q^{\frac{1}{2}}$
and then over the harmonic polynomials in $u_k$ we will
obtain $G_{c;\hspace{1pt} \aa_1\dots\aa_n}$ as elements of the space
\(\C \Bbrk{q^{\frac{1}{2}}}\) \(\Bbrk{u_1,\dots,u_n}\), i.e.,
the space of infinite harmonic power series in
$u_1,$ $\dots,$ $u_n$ with coefficients in $\C \Bbrk{q^{\frac{1}{2}}}$.
These coefficients can be expressed by a finite set of
series in $q^{\frac{1}{2}}$ of type (\ref{neqq1}) and~(\ref{neqq2}),
using the operations of summation and multiplication,
since all above considerations have been made in a purely algebraic setting.
Thus, if we assume that the partition function and all thermal mean
values of products of fields' modes are absolutely
convergent series for \(\left|\raisebox{9pt}{\hspace{-3pt}}\right.
q^{\frac{1}{2}} \left.\raisebox{9pt}{\hspace{-3pt}}\right| < 1\)
we obtain $G_{c;\hspace{1pt} \aa_1\dots\aa_n}$ as infinite formal Fourier series in
$(u_1,$ $\dots,$ \(u_n) \in \Sr^{\left( D-1 \right)n}\) whose
convergence should be further assumed in order to end up with
elliptic finite temperature correlation functions.
Let us conclude this discussion with the remark that in
a chiral conformal QFT
(which is, essentially, an $1$--dimensional theory)
there are no angular variables $u$ so that
Theorem~\ref{adprp1}
actually states
the existence of the finite temperature correlation functions
as elliptic functions under the assumptions of convergence of the
partition function and the thermal mean values of the
product of the fields' modes.

\begin{corollary}
\prlabel{cr:3.7}
In the assumptions of
Theorem~\ref{adprp1}
let the finite temperature correlation functions absolutely converge
in the domain~(\ref{ell_domain}) to meromorphic functions.
Then these function are elliptic of the type of $E_c^{\left( N \right)}$
appearing in the representation~(\ref{F_set}).
\end{corollary}

In the following section we shall calculate the finite temperature
correlation functions in free fields' models and will see that
they satisfy the above assumptions and are indeed elliptic functions.

\section{Free Field Models}\label{SEC:3}

\subsection{General properties of
thermal correlation functions of free fields}\label{Sec:3}

A generalized free field is defined as a Fock space representation
of the Heisenberg--Dirac algebra with generators
\(\phi_{\aa\left\{ n,\, m,\, \sigma \right\}}\) as in Eq.~(\ref{eqn2.1})
(\(\aa = 1,\, \dots,\, I\)).
It is completely determined by its $2$--point function
\begin{equation}\label{eqn4.3}
\lvac \, \phi_{\aa} \left( z \right) \,
\sphi_{\bb} \left( w \right) \, \rvac
\, = \,
\iota_{z,w} \, \Wf_{\aa\bb} \left( z,\, w \right)
\, , \quad
\Wf_{\aa\bb} \left( z,\, w \right) \, = \,
\frac{Q_{\aa\bb} \left( z-w \right)}{
\left[\raisebox{9pt}{\hspace{-3pt}}\right. \left( z-w \right)^2
\left.\raisebox{9pt}{\hspace{-3pt}}\right]^{\mu_{\aa\bb}}}
\, , \
\, \
\end{equation}
where
$Q_{\aa\bb} \left( z \right)$ are polynomials and we recall that $\iota_{z,w}$
stands for the Taylor expansion of $\Wf_{\aa\bb} \left( z,\, w \right)$
in $w$ whose coefficients are rational functions in~$z$.
Note that the $\iota_{z,w}$ operation is the $z$-picture counterpart of the
``$i0 \! \left( x^0 \! - \!y^0 \right)$'' prescription in Minkowski
space which turns,
for example, the rational function
$\Txfrac{1}{\left( x \!- \!y \right)^2}$, into
the distribution \(\Txfrac{1}{
\left( x \! - \! y \right)^2 \!\hspace{-2pt} + \! i0 \!
\left( x^0 \! - \! y^0 \right)}\).
Then the generating function of the modes' (anti)commutation relations is
\begin{eqnarray}\label{eqn4.1}
&
\phi_{\aa} \left( z \right)\, \sphi_{\bb} \left( w \right)
- \left( -1 \right)^{p_{\aa}p_{\bb}}
\sphi_{\bb} \left( w \right)\, \phi_{\aa} \left( z \right)
\, = \,
\iota_{z,w} \, \Wf_{\aa\bb} \left( z,\, w \right) -
\iota_{w,z} \, \Wf_{\aa\bb} \left( z,\, w \right)
\, = &
\nonumber \\
& = \, \iota_{z,w} \, \Wf_{\aa\bb} \left( z,\, w \right) -
\left( -1 \right)^{p_{\aa}p_{\bb}}
\iota_{w,z} \, \Wf_{\bb\aa} \left( w,\, z \right)
 \, . &
\end{eqnarray}

The annihilation operators are the modes $\phi_{\aa\left\{ n,m,\sigma \right\}}$
with \(n < 0\).
The Fock space is generated by the \textit{one particle state space}
$\DOM_1$
spanned by the vectors
\(
\phi_{\aa\left\{ n,m,\sigma \right\}} \rvac
\)
for \(
n \geqslant 0
\) and its
hermitian scalar product
is
determined by
the contributions of the Laurent modes
\(\lvac \phi_{\aa\left\{ n,m,\sigma \right\}}\)
\(\phi_{\bb\left\{ n,m,\sigma \right\}}^* \rvac\)
to the $2$--point function~(\ref{eqn4.3}).
We will not assume, in general, that the inner product in
$\DOM_1$ is positive definite.
The
rational two--point function $\Wf_{\aa\bb} \left( z,\, w \right)$
is, by assumption, conformally invariant
with respect to a cocycle $\pi_z \left( g \right){\hspace{-2pt}}_{\aa}^{\bb}$.

The partition function
\(\mathit{tr}_{\DOM{}} \left( q^H \right)\) and the other traces below
are understood as traces
taken over some (pseudo)orthonormal basis of $\DOM$
consisting of eigenvectors of $H$ (as in Sect.~\ref{msec:3}).
It is a Taylor series in $q^{\frac{1}{2}}$ which is always convergent for
\(q^{\frac{1}{2}} = e^{\hspace{1pt} i \pi\hspace{1pt}\tau}\) with
\(\textit{Im}\, \tau > 0\) (\(|q^{\frac{1}{2}}| < 1\))
since the degree of degeneracy of the conformal energy level
$\txfrac{n}{2}$ in the
1-particle state space has an upper bound of the form
\(C_1 \! \left(\hspace{-2pt}\begin{array}{c} n+C_2 \\ D-1
\end{array}\hspace{-2pt}\right)\)
with some positive constants~$C_{1,2}$.
More specifically,
due to the spin--statistics theorem (which follows from the
rationality of~(\ref{eqn4.3})), the integer conformal energy levels
$n$ in $\DOM_1$ should belong to the
bosonic 1-particle subspace while the half-integer ones,
$n-\txfrac{1}{2}$, belong to the
fermionic subspace.
Then, the partition function is determined by the
\textit{dimensions} of these energy spaces.
Let us denote these dimensions by
$d_{\bb} \left( n \right)$ and $d_f \left( n \right)$
(for the bosonic and fermionic 1-particle spaces of
energies $n$ and $n-\txfrac{1}{2}$, respectively);
then we will have
\begin{equation}\label{eqn4.4n}
Z \left( \tau \right) \, := \, \mathit{tr}_{\DOM{}} \left( q^H \right) \, = \,
\mathop{\prod}\limits_{n \, = \, 1}^{\infty} \,
\frac{
\left(\raisebox{10pt}{\hspace{-3pt}}\right.
1 + q^{n-\frac{1}{2}}
\left.\raisebox{10pt}{\hspace{-3pt}}\right)^{
d_f
\left( n \right)}}{
\left(\raisebox{10pt}{\hspace{-3pt}}\right.
1 - q^{n}
\left.\raisebox{10pt}{\hspace{-3pt}}\right)^{d_{\bb}
\left( n \right)}
}
\, . \
\end{equation}

It is also easy to see that the temperature mean value~(\ref{neqq2})
of the products of (compact picture) modes
$\phi_{\aa;\, \nnu,m} \! \left( u \right)$ (see Eq.~(\ref{eqn2.20a})),
is absolutely convergent.
Moreover,
it is expressed by Wick theorem in terms of ``1-'' and ``2-point'' Gibbs
expectation values
\(\La \hspace{-1.5pt}
\phi_{\aa; \nnu,m} \! \left( u \right)
\hspace{-2.5pt} \Ra_{\hspace{-1pt}q}\) and
\(\La \hspace{-1.5pt}
\phi_{\aa; \nnu_1,m_1} \! \left( u_1 \right)
\sphi_{\bb; \nnu_2,m_2} \! \left( u_2 \right)
\hspace{-2.5pt} \Ra_{\hspace{-1pt}q}\),
where
$\sphi_{\bb;\, \nnu,m} \! \left( u \right)$
($=$ $\Su_{\sigma} \, \sphi_{\bb;\, \nnu,m,\sigma} \, \hr{m}{\sigma} \! \left( u \right)$)
are the modes of the conjugate
field $\sphi_{\bb} \left( \czeta,\, u \right)$
(see Eq.~(\ref{eqn2.22a})).
Combining
the KMS property~(\ref{KMS})
\beq\label{eqn4.5}
\hspace{-20pt}
\La
\phi_{\aa;\, \nnu_1,m_1} \! \left( u_1 \right) \,
\sphi_{\bb;\, \nnu_2,m_2} \! \left( u_2 \right)
\Ra_{q}
\, = \,
q^{-\nnu_1}
\La
\sphi_{\bb;\, \nnu_2,m_2} \! \left( u_1 \right) \,
\phi_{\aa;\, \nnu_1,m_1} \! \left( u_2 \right)
\Ra_{q}
\, ,
\eeq
with the canonical (anti)commutation relations~(\ref{eqn4.1}) of the modes
we obtain
\begin{eqnarray}\label{eqn4.6}
&
\La
\phi_{\aa;\, \nnu_1,m_1} \! \left( u_1 \right) \,
\sphi_{\bb;\, \nnu_2,m_2} \! \left( u_2 \right)
\Ra_{q}
\, = & \nonumber \\ & = \,
\Txfrac{1}{1-\left( -1 \right)^{p_{\aa}p_{\bb}} q^{\nnu_1}}
\lvac \left[\raisebox{9pt}{\hspace{-2pt}}\right.
\phi_{\aa;\, \nnu_1,m_1} \! \left( u_1 \right),\,
\sphi_{\bb;\, \nnu_2,m_2} \! \left( u_2 \right) \,
\left.\raisebox{9pt}{\hspace{-2pt}}\right]_{-\left( -1 \right)^{p_{\aa}p_{\bb}}} \rvac
\, . &
\end{eqnarray}

\begin{theorem}
\prlabel{thr:4.1}
The series
\(\La \phi_{\aa} \left( \czeta_1, u_1 \right)
\sphi_{\bb} \left( \czeta_2, u_2 \right) \Ra_{q}\) is
absolutely convergent for
\(0 < - \mathit{Im} \, \czeta_{12} < \mathit{Im} \, \tau\)
to an elliptic function in $\czeta_{12}$.
It can be written as a series
\begin{equation}\label{eqn4.14a}
\La \phi_{\aa} \left( \czeta_1, u_1 \right)
\sphi_{\bb} \left( \czeta_2, u_2 \right) \Ra_{q}
=
\mathop{\sum}\limits_{k \, = \, -\infty}^{\infty}
\left( -1 \right)^{kp_{\aa}p_{\bb}} \,
W_{\aa\bb} \left( \czeta_{12} \! + \! k \, \tau; u_1, u_2 \right)
\, , \
\end{equation}
absolutely convergent in the same domain;
here $W_{\aa\bb} \left( \czeta_{12};\, u_1,\, u_2 \right)$
is the meromorphic vacuum correlation function
\beqa
\label{eqn4.13a}
W_{\aa\bb} \left( \czeta_{12};\, u_1,\, u_2 \right)
\, := \,
\lvac \phi_{\aa} \left( \czeta_1,\, u_1 \right)
\sphi_{\bb} \left( \czeta_2,\, u_2 \right) \rvac
\, . \
\eeqa
The functions (\ref{eqn4.14a}) are manifestly doubly periodic elliptic
functions in~$\czeta_{12}$.
\end{theorem}

\begin{proof}
First observe that \(\sphi_{\bb;\, \nnu,m} \! \left( u \right) \rvac = 0\)
if \(\nnu \geqslant 0\)
(in accord with
Proposition~\ref{prp:2.2})
and therefore,
\(\lvac \phi_{\bb;\, \nnu,m} \! \left( u \right) = 0\)
if \(\nnu \leqslant 0\).
Thus, at most one term contributes to the (anti) commutator in the
right hand side of~(\ref{eqn4.6}) and in fact:
\beqa
&&
\lvac \left[\raisebox{9pt}{\hspace{-2pt}}\right.
\phi_{\aa;\, \nnu_1,m_1} \! \left( u_1 \right),\,
\sphi_{\bb;\, \nnu_2,m_2} \! \left( u_2 \right) \,
\left.\raisebox{9pt}{\hspace{-2pt}}\right]_{-\left( -1 \right)^{p_{\aa}p_{\bb}}} \rvac
\, = \,
\nn && \ \ =
\hev_{\nnu_1} \!
\lvac
\hspace{-1pt}
\phi_{\aa; \nnu_1,m_1} \! \left( u_1 \right)
\sphi_{\bb; \nnu_2,m_2} \! \left( u_2 \right)
\hspace{-1pt}
\rvac -
\nn && \ \ \hspace{10pt}
-
\left( -1 \right)^{p_{\aa}p_{\bb}} \!
\hev_{-\nnu_1} \!
\lvac
\hspace{-1pt}
\sphi_{\bb; \nnu_2,m_2} \! \left( u_2 \right)
\phi_{\aa; \nnu_1,m_1} \! \left( u_1 \right)
\hspace{-1pt}
\rvac
\nonumber
\eeqa
where $\hev_s$ is the characteristic function of the positive
numbers (\(\hev_s := 1\) for \(s>0\) and \(\hev_s = 0\) otherwise).
Expanding for \(|q^{\frac{1}{2}}|<1\)
(\(q^{\frac{1}{2}} = e^{i \pi \tau}\)) the prefactor in
the right hand side of~(\ref{eqn4.6}), we find:
\begin{eqnarray}\label{eqn}
&& \hspace{-16pt}
\La \hspace{-2pt}
\phi_{\aa} \! \left( \czeta_1,u_1 \right)
\sphi_{\bb} \! \left( \czeta_2,u_2 \right) \hspace{-2pt} \Ra_q
\hspace{-2pt} = \hspace{-1pt}
\nonumber \\ && \hspace{-16pt}
\hspace{10pt} = \hspace{-1pt}
\mathop{\sum}\limits_{\text{\normalsize
\(\mathop{}\limits^{\nnu_1,m_1}_{\nnu_2,m_2}\)}} \hspace{-3pt}
\La \hspace{-2pt}
\phi_{\aa;\, \nnu_1,m_1} \! \left( u_1 \right)
\sphi_{\bb;\, \nnu_2,m_2} \! \left( u_2 \right) \hspace{-2pt}
\Ra_{\hspace{-1pt} q}
e^{-2\pi i \, \left( \nnu_1 \czeta_1 + \nnu_2 \czeta_2 \right)} \,
= \nonumber \\ && \hspace{-16pt}
\hspace{10pt} = \hspace{-1pt}
\mathop{\sum}\limits_{
\nnu,m_1,m_2
} \hspace{1pt}
\mathop{\sum}\limits_{k \, = \, 0}^{\infty} \hspace{-1pt}
\left(\raisebox{10pt}{\hspace{-2pt}}\right.
\hev_{\nnu} \hspace{-1pt} \left( -1 \right)^{kp_{\aa}p_{\bb}} \hspace{-1pt}
q^{k\nnu} \hspace{-1pt}
\lvac \hspace{-1pt}
\phi_{\aa;\, \nnu,m_1} \! \left( u_1 \right)
\sphi_{\bb;\, -\nnu,m_2} \! \left( u_2 \right) \hspace{-1pt}
\rvac +
\nonumber \\ && \hspace{59pt} + \,
\hev_{-\nnu} \hspace{1pt}
\left( -1 \right)^{k p_{\aa}p_{\bb}}
\hspace{1pt}
q^{-\left( k+1 \right)\nnu} \hspace{1pt}
\lvac \hspace{-1pt}
\sphi_{\bb;\, -\nnu,m_2} \! \left( u_2 \right)
\phi_{\aa;\, \nnu,m_1} \! \left( u_1 \right) \hspace{-1pt}
\rvac
\left.\raisebox{10pt}{\hspace{-2pt}}\right)
\times \hspace{-10pt} \nonumber \\ && \hspace{42pt} \hspace{16pt} \times \
e^{-2\pi i \, \nnu \, \czeta_{12}} \,
.
\raisebox{14pt}{}
\end{eqnarray}
If we first perform the sum over the indices
$\nnu$, $m_1$, $m_2$
in the right hand side of~(\ref{eqn})
we obtain (due to
Theorem~\ref{thm:2.3})
the series expansion in Eq.~(\ref{eqn4.14a}):
indeed, the first term in the sum gives
\(\mathop{\sum}\limits_{k = 0}^{\infty} \,
\left(-1 \right)^{kp_{\aa}p_{\bb}}\)
\(W_{\aa\bb} \left( \czeta_{12} \! + \! k \, \tau; u_1, u_2 \right)\)
while the second gives
\(\mathop{\sum}\limits_{k = 1}^{\infty} \,
\left( -1 \right)^{\left( -k+1 \right)p_{\aa}p_{\bb}}\)
\(W_{\bb\aa}' \left( -\czeta_{12} \! - \! k \, \tau; u_2, u_1 \right)\),
where
$$
W_{\bb\aa}' \left( \czeta_{12}; u_1, u_2 \right) \, := \,
\lvac \sphi_{\bb} \left( \czeta_1,\, u_1 \right)
\phi_{\aa} \left( \czeta_2,\, u_2 \right) \rvac,
$$
so we should further apply the symmetry property
\begin{equation}\label{eqn4.8}
W_{\aa\bb} \left( \czeta_{12}; u_1, u_2 \right) \, = \,
\left( -1 \right)^{p_{\aa}p_{\bb}} \,
W_{\bb\aa}' \left( -\czeta_{12}; u_2, u_1 \right)
\, . \
\end{equation}
The series~(\ref{eqn4.14a}) is absolutely convergent since
its terms behave as
\begin{eqnarray}
& \hspace{-12pt}
W_{\aa\bb} \! \left( \czeta_{12} \! + \! k \tau; u_1, u_2 \right)
\sim q^{kd_{\bb}}
e^{2\pi i \left( d_{\aa} \czeta_1 + d_{\bb} \czeta_2 \right)}
\Wf_{\aa\bb} \! \left(
e^{2\pi i \czeta_1}u_1, 0 \right)
\hspace{3pt} \text{for} \hspace{3pt} k \to \infty , &
\nonumber \\ & \hspace{-12pt}
W_{\aa\bb} \! \left( \czeta_{12} \! + \! k \tau; u_1, u_2 \right)
\sim q^{-kd_{\aa}}
e^{2\pi i \left( d_{\aa} \czeta_1 + d_{\bb} \czeta_2 \right)}
\Wf_{\aa\bb} \! \left(
0, e^{2\pi i \czeta_2}u_2 \right)
\hspace{3pt} \text{for} \hspace{3pt} k \to -\infty
. \nonumber &
\end{eqnarray}
The series of the finite temperature correlation function
\(\La \hspace{-2pt}
\phi_{\aa} \! \left( \czeta_1,u_1 \right)\)
\(\sphi_{\bb} \! \left( \czeta_2,u_2 \right) \hspace{-2pt} \Ra_q\)
given by the first equality in~(\ref{eqn}) is also absolutely
convergent for
\(0 < -\mathit{Im} \, \czeta_{12}\) \(< \mathit{Im} \, \tau\)
since the series of $W_{AB}$ absolutely converges in this domain.
\end{proof}

\begin{remark}
\rmlabel{rm:4.1}
Let $N$ be a hermitian operator,
commuting with the conformal Hamiltonian $H$ and such that
\begin{equation}\label{eqn4.11}
\left[ N,\, \sphi_{\aa} \left( z \right) \right] \, = \,
n_{\aa} \, \sphi_{\aa} \left( z \right)
\, , \quad
\left[ N,\, \phi_{\aa} \left( z \right) \right] \, = \,
- n_{\aa} \, \phi_{\aa} \left( z \right)
\, . \
\end{equation}
Then we can derive in the same way as above the following expression
for the {\it grand canonical} correlation functions
\begin{eqnarray}\label{eqn4.14b}
\La \hspace{-2pt} \phi_{\aa} \left( \czeta_1,u_1 \right)
\sphi_{\bb} \left( \czeta_1,u_1 \right) \hspace{-1pt} \Ra_{\hspace{-1pt} q,\,\mu}
\hspace{-2pt} := \hspace{-2pt} && \podr
\frac{\mathit{tr}_{\DOM{}} \left( \phi_{\aa} \left( \czeta_1,u_1 \right)
\sphi_{\bb} \left( \czeta_2,u_2 \right) \, q^H \,
e^{2 \pi i\hspace{1pt}\mu\hspace{1pt}N}
\right)}{
\mathit{tr}_{\DOM{}}
\left( q^H \, e^{\hspace{1pt}2\pi i\hspace{1pt}\mu\hspace{1pt}N} \right)}
\hspace{-2pt} = \hspace{-2pt}
\nonumber \\ = \hspace{-2pt} && \podr
\mathop{\sum}\limits_{k \, = \, -\infty}^{\infty}
e^{\pi i \, k \, \left( 2\mu+p_{\aa}p_{\bb} \right)} \,
W_{\aa\bb} \left( \czeta_{12} \! + \! k \, \tau; u_1, u_2 \right)
, \qquad
\end{eqnarray}
for real $\mu$.
(In the physical literature the grand canonical partition function is
written as
\(\mathit{tr} \! \left( e^{-\beta \, \left(H-\mu N\right)}\right)\)
where $\beta$ is the inverse temperature and $\mu$ is the
{\it chemical potential}.)
\end{remark}

\begin{remark}
\rmlabel{rm:4.2}
In the assumptions of
Corollary~\ref{cr:3.7}
one can state that
for the thermal $2$--point function
\(\La \phi_{\aa} \left( \czeta_1, u_1 \right)
\sphi_{\bb} \left( \czeta_2, u_2 \right) \Ra_q\),
of an arbitrary field \(\phi =\) \(\left\{ \phi_{\aa} \right\}\),
the right hand side of
Eq.~(\ref{eqn4.14a}) describes the \textit{most singular part}
in \(\czeta_{12}\) since it comes from the most singular part
of the operator product expansion of
\(\phi_{\aa} \left( \czeta_1, u_1 \right)
\sphi_{\bb} \left( \czeta_2, u_2 \right)\) (see Proposition~4.3 of~\cite{NT01}).
\end{remark}

\subsection{Free scalar fields}\label{sec:5}

The generalized free neutral scalar field
\(\phi \left( z \right) \equiv
\phi^{\left( d \right)} \left( z \right)\)
of dimension $d$ is determined by the unique conformally invariant scalar
2-point function~(\ref{eqn5.1}).

Many of the modes in the field expansion~(\ref{eqn2.1}) are zero
so that it is convenient to reduce the system of basic functions and actually,
organize the field modes in a slightly different way.

Let us denote by $\phi_{-d-n} \left( z \right)$
the homogeneous operator--valued polynomial of degree \(n \geqslant 0\)
contributing to the Taylor part of the expansion~(\ref{eqn2.1}) of
$\phi \left( z \right)$.
The polynomial $\phi_{-d-n} \left( \overline{z} \right)^*$
obtained conjugating the
coefficients of $\phi_{-d-n} \left( z \right)$ is denoted~by
\begin{equation}\label{eqn5.3}
\phi_{n+d} \left( z \right) \, = \, \phi_{-d-n} \left( \overline{z} \right)^*
\,
\end{equation}
(\(n \geqslant 0\)).
Due to
Proposition~\ref{prp:2.2},
the creation modes of the field
are exactly
\(\left\{\raisebox{8pt}{\hspace{-3pt}}\right. \phi_{-n-d} :\)
\(n \geqslant 0 \left.\raisebox{8pt}{\hspace{-3pt}}\right\}\) so that
the remaining nonzero field modes
\(\left\{\raisebox{8pt}{\hspace{-3pt}}\right. \phi_{n+d} :\)
\(n\) \(\geqslant\) \(0 \left.\raisebox{8pt}{\hspace{-3pt}}\right\}\)
annihilate the vacuum $\rvac$.
Thus the field $\phi \left( z \right)$ is expanded in the above modes as
follows:
\begin{equation}\label{eqn5.4}
\phi \left( z \right) \, = \,
\mathop{\sum}\limits_{
n \, = \, 0}^{\infty} \, \phi_{-n-d} \left( z \right)
\, + \,
\mathop{\sum}\limits_{
n \, = \, 0}^{\infty} \,
\left( z^{\, 2} \right)^{-n-d} \, \phi_{n+d} \left( z \right)
\, . \
\end{equation}
The commutation relation with the conformal Hamiltonian take the form
\begin{equation}\label{eqn5.5}
\left[ H,\, \phi_n \left( z \right) \right] \, = \, -n \, \phi_n \left( z \right)
\quad
(n \in \Z)
\, . \
\end{equation}

The vacuum matrix elements of products of field modes are derived from the
2-point function
\begin{equation}\label{eqn5.6}
\lvac
\phi \left( \overline{z} \right)^*
\phi \left( w \right) \rvac
\, = \,
\frac{1}{\left( 1 - 2 \, z \spr w + z^{\, 2} \, w^{\, 2} \right)^d}
\, = \,
\mathop{\sum}\limits_{n \, = \, 0}^{\infty} \, \widetilde{C}_n^d \left( z,\, w \right)
\, , \
\end{equation}
where $\widetilde{C}_n^d \left( z,w \right)$ are polynomials separately homogeneous
in $z$ and $w$ of equal degrees $n$ with generating function
\begin{equation}\label{eqn5.7}
\frac{1}{\left( 1 - 2 \, z \spr w \, \lambda + z^{\, 2} \, w^{\, 2} \, \lambda^2 \right)^d}
\, = \,
\mathop{\sum}\limits_{n \, = \, 0}^{\infty} \,
\widetilde{C}_n^d \left( z,\, w \right) \, \lambda^n
\, . \,
\end{equation}
Then
\begin{equation}\label{eqn5.8}
\lvac \phi_{m+d} \left( z \right) \phi_{-n-d} \left( w \right) \rvac
\, = \, \delta_{m,n} \, \widetilde{C}_n^d \left( z,\, w \right)
\, \
\end{equation}
for \(m,\, n \geqslant 0\).
Note that the polynomials $\widetilde{C}_n^d \left( z,w \right)$ are related to the
\textit{Gegenbauer polynomials} $C_n^d \left( t \right)$
(with generating function
\(\Txfrac{1}{\left( 1 - 2 \, t \, \lambda + \lambda^2 \right)^d}
=
\mathop{\sum}\limits_{n \, = \, 0}^{\infty} \, C_n^d \left( t \right) \, \lambda^n\))
by
\begin{equation}\label{eqn5.9}
\widetilde{C}_n^d \left( z,\, w \right) \, = \,
\left( z^{\, 2} w^{\, 2} \right)^{
\frac{\raisebox{0pt}{\footnotesize $n$}}{\raisebox{0pt}{\footnotesize $2$}}}
C_n^d \left(\raisebox{12pt}{\hspace{-2pt}}\right.
\frac{z \spr w}{
\left( z^{\, 2} w^{\, 2} \right)^{1\!/2}}
\left.\raisebox{12pt}{\hspace{-2pt}}\right)
\, . \
\end{equation}

In the real compact picture we set
\begin{equation}\label{eqn5.10}
\phi \left( \czeta,\, u \right) \, = \,
e^{2\pi i\hspace{1pt}d\hspace{1pt}\czeta} \hspace{1pt}
\phi
\left( e^{2\pi i\hspace{1pt}\czeta}\hspace{1pt} u \right)
\, = \,
\mathop{\sum}\limits_{
\raisebox{0pt}{\small \(\mathop{}\limits^{n \, \in \, \Z}_{|n| \geqslant d}\)}} \,
e^{-2\pi i\hspace{1pt}n\hspace{1pt}\czeta} \hspace{1pt}
\, \phi_{-n} \left( u \right)
\, \
\end{equation}
as a formal Fourier series in $\czeta$.
Taking into account the relation
$$
\lvac \phi \hspace{-1pt}
\left( e^{2\pi i\hspace{1pt}\czeta_1} u_1 \right)
\phi \hspace{-1pt}
\left( e^{2\pi i\hspace{1pt}\czeta_2} u_2 \right) \rvac
\, = \,
e^{2\pi i\hspace{1pt}d\hspace{1pt}\left( \czeta_1 + \czeta_2 \right)} \,
\lvac \phi \left( \czeta_1,\, u_1 \right) \phi \left( \czeta_2,\, u_2 \right)
\rvac
$$
we find
\begin{equation}\label{eqn5.11}
\lvac \phi \left( \czeta_1,\, u_1 \right) \phi \left( \czeta_2,\, u_2 \right) \rvac
\, = \,
\frac{\left( -1 \right)^d}{4^{d} \, \sin^d \pi \czeta_+ \, \sin^d \pi \czeta_-}
\, , \
\end{equation}
where \(\czeta_{\pm} = \czeta_{12} \pm \alpha\), \(\cos \, 2\pi\alpha = u_1 \spr u_2\).
Then Eq.~(\ref{eqn4.6}) takes the form
\begin{equation}\label{eqn5.12}
\La \phi \left( \czeta_1,\, u_1 \right) \phi \left( \czeta_2,\, u_2 \right) \Ra_q
\, = \,
\mathop{\sum}\limits_{k \, = \, -\infty}^{\infty}
\frac{\left( -1 \right)^d}{4^{d} \, \sin^d \pi \left( \czeta_+ \! + k \, \tau \right) \,
\sin^d \pi \left( \czeta_- \! + k \, \tau \right)}
\, . \
\end{equation}

For \(d=1\)
we obtain
\begin{equation}\label{eqn5.13}
\La \hspace{-1pt}
\phi \left( \czeta_1,\, u_1 \right)
\phi \left( \czeta_2,\, u_2 \right)
\hspace{-2pt}
\Ra_{\hspace{-1pt} q}
=
\frac{1}{4\,\pi\,\sin 2\pi\alpha}
\left(\raisebox{10pt}{\hspace{-2pt}}\right.
p_1 \hspace{-1pt} \left( \czeta_+,\tau \right)
-
p_1 \hspace{-1pt} \left( \czeta_-,\tau \right)
\left.\raisebox{10pt}{\hspace{-2pt}}\right)
\, , \
\end{equation}
where
\(p_k \left( \czeta,\, \tau \right)\) are written down in Appendix A
(see (\ref{fn.0})).
Eq.~(\ref{eqn5.13}) follows from the identity
\begin{equation}\label{eqn5.14a}
\frac{-1}{\sin \pi\czeta_+ \, \sin \pi \czeta_-}
\, = \,
\frac{1}{\sin 2\pi\alpha}
\left( \mathrm{cotg} \, \pi\czeta_+ - \mathrm{cotg} \, \pi\czeta_- \right)
\, \
\end{equation}
and~(\ref{eqn5.12}). Note that the differences in~(\ref{eqn5.14a})
and~(\ref{eqn5.13}) allows us to apply Eq.~(\ref{fn.0})
and ensures the ellipticity (double periodicity) in $\czeta_{12}$ of the
thermal correlation function.

\begin{remark}
\rmlabel{rm:5.1}
The Gibbs $2$--point function of the modes $\phi_n \left( u \right)$ in the latter example (\(d=1\))
\begin{equation}\label{n5.13n}
\La \hspace{-1pt}
\phi_{-m} \left( u_1 \right)
\phi_{n} \left( u_2 \right)
\Ra_{\hspace{-1pt} q} \, = \,
\delta_{mn} \, \frac{q^n}{1-q^n} \, \frac{\sin \, 2\pi n \alpha}{\sin \, 2\pi \alpha}
\end{equation}
(for \(u_1 \spr u_2 \, = \, \cos 2\pi \alpha\)), which can be derived directly
from the canonical commutation
relations and the KMS condition, yields the $q$--expansion of~(\ref{eqn5.13})
\begin{equation}\label{n5.14n}
\La \hspace{-1pt}
\phi \! \left( \czeta_1, u_1 \right)
\phi \! \left( \czeta_2, u_2 \right)
\hspace{-2pt}
\Ra_{\hspace{-1pt} q} \! = \hspace{-1pt}
\frac{-1}{4\sin \pi \czeta_+ \sin \pi \czeta_-} \hspace{1pt} + \hspace{1pt}
2 \! \mathop{\sum}\limits_{n \, = \, 1}^{\infty}
\frac{q^n}{1-q^n} \frac{\sin 2\pi n \alpha}{
\sin 2\pi\alpha} \cos 2\pi n \czeta_{12}
. \ \
\end{equation}
Comparing with~(\ref{eqn5.13}) we deduce a similar
expansion for $p_1$
\beq\label{n5.145n}
p_1 (\zeta,\tau)
\, = \, \pi \cot \pi \zeta + 4\pi \sum_{n=1}^{\infty}
\frac{q^n}{1-q^n} \, \sin 2 \pi n \zeta \, .
\eeq
\end{remark}

In the more general context of
Remark~\ref{rm:4.1},
for a complex scalar field of
dimension~$1$
(\(\lvac \phi \left( z_1 \right) \sphi \left( z_2 \right) \rvac =
\left( z_{12}^{\, 2} \right)^{-1}\)) taking
$N$ to be the \textit{charge} operator (with \(n=1\) in Eq.~(\ref{eqn4.11})),
we find
\begin{equation}\label{eqn5.13nn}
\La \hspace{-1pt} \phi \left( \czeta_1,\, u_1 \right)
\sphi \left( \czeta_2,\, u_2 \right) \hspace{-2pt}
\Ra_{\hspace{-1pt} q,\, \mu}
=
\frac{1}{4\,\pi\,\sin 2\pi\alpha}
\left(\raisebox{10pt}{\hspace{-2pt}}\right.
p_1 \hspace{-1pt} \left( \czeta_+,\tau,\mu \right)
-
p_1 \hspace{-1pt} \left( \czeta_-,\tau,\mu \right)
\left.\raisebox{10pt}{\hspace{-2pt}}\right)
\, \
\end{equation}
for the more general functions $p_1 \left( \czeta,\tau,\mu \right)$ of Appendix~A.

In order to find the mean energy (or the partition function)
we have to specify the space--time
dimension $D$ together with the field dimension~$d$.
We will consider the following two basic examples.

\subsubsection{Canonical free massless field in even space time dimension $D$}\label{Ssec.4.1}
The canonical free field is determined by the Laplace equation
\begin{equation}\label{eqn5.15}
\partial_{z}^{\, 2} \, \phi^{\left( d \right)} \left( z \right)
\, = \, 0
\quad \Leftrightarrow \quad
\partial_{z_1}^{\, 2} \frac{1}{\left( z_{12}^{\, 2} \right)^d} \, = \, 0
\quad \Leftrightarrow \quad
d \, = \, d_0 \, := \, \frac{D-2}{2}
\, . \
\end{equation}
The existence of the canonical free field as a GCI field requires
$D$ to be even and greater than~2.
Then the polynomials $\widetilde{C}_n^{d_0} \left( z,w \right)$ are
harmonic in both $z$ and $w$, and they determine a
\textit{positive definite} scalar product by Eq.~(\ref{eqn5.8}).
Thus, the canonical free fields satisfy the Hilbert space Wightman positivity.

The operator--valued polynomials
\(\phi_{-n-d} \left( z \right)\) are harmonic, i.~e.
\begin{equation}\label{eqn5.16}
\phi_{-n-d} \left( z \right) \, = \,
\mathop{\sum}\limits_{\sigma} \, \phi_{\left\{ 0,n,\sigma \right\}} \,
\hr{n}{\sigma} \left( z \right)
\, \
\end{equation}
in the notations of Sect.~\ref{ssec2.2},
so that the only nonzero modes of $\phi \left( z \right)$ are
$\phi_{\left\{ 0,n,\sigma \right\}}$ and
$\phi_{\left\{ -n-d_0,n,\sigma \right\}}$ for \(n = 0,\, 1,\, \dots\).
It then follows that the 1-particle eigenspace of conformal energy
\(n (\geqslant d_0)\) is isomorphic to the space of the harmonic
polynomials on $\C^D$ of degree $n-d_0$.
Its dimension
\(d^{\left( D \right)} \left( n \right)
( = d \left( n \right) \equiv
d_{\bb} \left( n \right) )\)
is thus
\begin{equation}\label{eqn5.17}
d^{\left( D \right)} \left( n \right) =
\frac{2n}{\left( 2d_0 \right) !}
\, \mathop{\prod}\limits_{k \, = \, 1-d_0}^{d_0-1} \,
\left( n-k \right)
\quad \mathrm{for} \quad D > 4
\quad (d^{\left( 4 \right)} \left( n \right) = n^2)
\, , \
\end{equation}
which is an even polynomial in $n$, for even~$D$, of degree $2d_0$, say
\begin{equation}\label{eqn5.18}
d^{\left( D \right)} \left( n \right) \, = \, \mathop{\sum}\limits_{k \, = \, 0}^{d_0} \,
c_k^{\left( D \right)} n^{2k}
\, (\, = \,
\frac{2n^2}{\left( 2d_0 \right)!} \, \mathop{\prod}\limits_{k \, = \, 1}^{d_0 -1}
\left( n^2 - k^2 \right)
\quad \text{for} \quad D > 4)
\, . \
\end{equation}
Note that \(d^{\left( D \right)} \left( n \right) = 0\) for
\(n = 1,\, \dots,\, d_0 -1\)
so that
the thermal energy mean value is
\begin{eqnarray}\label{eqn5.19}
\La \! H \! \Ra_{q} \, \equiv && \podr
\frac{\mathit{tr}_{\DOM{}} \left( H \, q^H \right)}{
\mathit{tr}_{\DOM{}} \left( q^H \right)} \, = \,
\frac{1}{Z \left( \tau \right)} \, q \, \frac{\partial}{\partial q} \, Z \left( \tau \right)
\, = \,
\mathop{\sum}\limits_{n \, = \, 1}^{\infty} \,
\frac{n \, d \left( n \right) \, q^n}{1-q^n}
\, = \,
\nonumber \\ = && \podr
\mathop{\sum}\limits_{k \, = \, 1}^{d_0+1}
c_{k-1}^{\left( D \right)} \,
\frac{B_{2k}}{4k}
\, + \,
\mathop{\sum}\limits_{k \, = \, 1}^{d_0+1}
c_{k-1}^{\left( D \right)}
G_{2k} \left( \tau \right)
\, , \
\end{eqnarray}
where $G_{2k} \left( \tau \right)$ are
the \textit{level 1 modular forms}~(\ref{fn_G}) (\ref{FTG}) and
$B_{2k}$ are the Bernoulli numbers (see Appendix~A).
This agrees with Eq.~(\ref{eqn4.4n}) since here \(d_f \left( n \right) = 0\)
and \(d_{\bb} \left( n \right) \equiv d \left( n \right)\).
Note that \(d^{\left( 2 \right)} \left( n \right) = c_0^{\left(  2\right)} = 2\),
while for \(D \geq 4\), \(c_{0}^{\left( D \right)} = 0\).
In particular, for \(D=4\) we find
\begin{equation}
\label{eqn5.21aa}
\La \! H + E_0 \! \Ra_q^{\left( 4 \right)} \, =
G_4 \left( \tau \right)
\, , \quad E_0 \, = \, \frac{1}{240}
\, . \
\end{equation}
If we interpret $E_0$ as a vacuum energy, i.~e. renormalize the
conformal Hamiltonian as \(\widetilde{H} = H + E_0\) then
its temperature mean value would be a modular form
of weight~$4$.

\begin{remark}
\rmlabel{rm:5.2}
Extrapolation to the case \(D=2\) of the above result contains
two chiral components each of them giving the energy distribution for
a $U(1)$ current
\begin{equation}\label{eqn5.20}
\La \! H + E_0 \! \Ra_q^{\left( 2 \right)} \, =
G_2 \left( \tau \right)
\, , \quad E_0 \, = \, - \, \frac{1}{24}
\, \
\end{equation}
which is {\it not} modular invariant.
\end{remark}

\subsubsection{Subcanonical field of dimension $d=1$ for $D=6$}\label{Ssec.4.2}
The scalar field of dimension \(d = 1\) in \(D=6\) space--time dimensions is
not harmonic but satisfies the fourth order equation
\(\left( \di_z^{\, 2} \right)^2 \phi \left( z \right) = 0\).
The harmonic polynomials on \(\C^6\) are now \(\widetilde{C}_n^2 \left( z,w \right)\).
The identity
\begin{equation}\label{eq5.21}
C_n^{1} \left( t \right) \, = \,
\frac{1}{n+1} \left( C_n^2 \left( t \right) - C_{n-2}^2 \left( t \right) \right)
\, \
\end{equation}
implies the following harmonic decomposition of the homogeneous polynomials
\linebreak
\(\widetilde{C}_n^1\left( z,w \right)\):
\begin{equation}\label{eqn5.22}
\widetilde{C}_n^2 \left( z,\, w \right) \, = \,
\frac{1}{n+1} \, \widetilde{C}_n^2 \left( z,\, w \right)
- \frac{1}{n+1} \, z^{\, 2} w^{\, 2} \,
\widetilde{C}_{n-2}^2 \left( z,\, w \right)
\, . \
\end{equation}
Thus we can decompose
\begin{equation}\label{eqn5.23}
\phi_{-n-1} \left( z \right) \, = \,
\phi_{-n-1}^1 \left( z \right) \, + \,
z^{\, 2} \, \phi_{-n-1}^2 \left( z \right)
\, , \
\end{equation}
where \(\phi_{n}^j \left( z \right)\) are now harmonic homogeneous
operator--valued polynomials of degrees \(n\) and \(n-2\),
respectively (as \(\phi_0^1 := 0\) and \(\phi_0^2 = \phi_1^2 := 0\)).
Then,
\begin{eqnarray}\label{eqn5.24}
\lvac \phi_{-n-1}^{1*} \left( z \right) \phi_{-n-1}^1 \left( w \right) \rvac
\, = && \podr
\frac{1}{n+1} \, \widetilde{C}_n^2 \left( z,\, w \right)
\, , \quad \nonumber \\
\lvac \phi_{-n-3}^{2*} \left( z \right) \phi_{-n-3}^2 \left( w \right) \rvac
\, = && \podr
\frac{-1}{n+1} \, \widetilde{C}_{n}^2 \left( z,\, w \right)
\, . \
\end{eqnarray}
Therefore, the 1-particle state space of conformal energy $n$ decomposes into
a pseudo--or\-tho\-go\-nal direct sum of two subspaces isomorphic to the spaces of harmonic
homogeneous polynomials of degrees $n-1$ and $n-3$, respectively:
the first will have positive definite while the second one, negative
definite metric.
In particular, the dimension of the full eigenspace of conformal energy $n$ is
\begin{equation}\label{eqn5.24nn}
d_{\bb} \left( n \right) \, = \,
d^{\left( 6 \right)} \left( n+1 \right)
\, + \,
d^{\left( 6 \right)} \left( n-1 \right)
\, = \,
\frac{n^2\left( n^2+5 \right)}{6}
\, \
\end{equation}
so that the thermal energy mean value and the vacuum energy are
\begin{equation}\label{eqn5.25nnn}
\La \! H + E_0 \! \Ra_q \, = \,
\frac{1}{6} \, G_6 \left( \tau \right) +
\frac{5}{6} \, G_{4} \left( \tau \right)
\, , \quad
E_0 \, = \,
-
\frac{1}{6} \, \frac{B_{6}}{12} -
\frac{5}{6} \, \frac{B_4}{8}
\, = \, \frac{19}{6048}
\, . \
\end{equation}

\subsection{The Weyl field}\label{Ssec.5.1}
Let us introduce the \(\left( 2\!\times\! 2 \right)\)--matrix representation
of the quaternionic algebra:
\begin{eqnarray}\label{eqn6.1}
&
Q_k \, = \, -i \, \sigma_k \, = \, - Q_k^+ \quad (k \, = \, 1,\, 2,\, 3)
\, , \quad
Q_4 \, = \, \ID &
\\ &
\sigma_{1} \, = \, \left( \begin{array}{cc} 0 & 1 \\ 1 & 0 \end{array}  \right)
\, , \quad
\sigma_{2} \, = \, \left( \begin{array}{cc} 0 & -i \\ i & 0 \end{array}  \right)
\, , \quad
\sigma_{3} \, = \, \left( \begin{array}{cc} 1 & 0 \\ 0 & -1 \end{array}  \right)
, & \nonumber \\ \label{eqn6.2}
&
Q_{\cmu}^+ \, Q_{\cnu} + Q_{\cnu}^+ \, Q_{\cmu} \, = \,
2 \delta_{\cmu\cnu} \, = \,
Q_{\cmu} \, Q_{\cnu}^+ + Q_{\cnu} Q_{\cmu}^+
\quad \text{for} \quad \cmu,\, \cnu = 1,\, \dots,\, 4
\, &
\end{eqnarray}
(\(\sigma_k\) being Pauli matrices).
In this section we will denote the hermitian matrix conjugation by a superscript ``$+$''.
The matrices
\begin{equation}\label{eqn6.3}
i \, \sigma_{\cmu\cnu} \, = \,
\frac{1}{2} \left( Q_{\cmu}^+ \, Q_{\cnu} - Q_{\cnu}^+ \, Q_{\cmu} \right)
\, , \quad
i \, \widetilde{\sigma}_{\cmu\cnu} \, = \,
\frac{1}{2} \left( Q_{\cmu} \, Q_{\cnu}^+ - Q_{\cnu} \, Q_{\cmu}^+ \right)
\, \
\end{equation}
are the selfdual and antiselfdual
antihermitian
\(\spin \left( 4 \right)\) Lie algebra generators.
We will denote also
\begin{equation}\label{dirz}
\dirz = \mathop{\sum}\limits_{\cmu = 1}^{4} \, z^{\cmu} \, Q_{\cmu}
, \ \
\dirz^+ = \, \mathop{\sum}\limits_{\cmu = 1}^{4} \, z^{\cmu} \, Q_{\cmu}^+
, \ \
\dirdi_z = \mathop{\sum}\limits_{\cmu = 1}^{4} \, Q_{\cmu} \, \di_{z^{\cmu}}
, \ \
\dirdi_z^+ = \mathop{\sum}\limits_{\cmu = 1}^{4} \, Q_{\cmu}^+ \, \di_{z^{\cmu}}
, \
\end{equation}
etc.
Note that in the definition of $\dirz^+$ we do not conjugate the coordinates~$z^{\cmu}$.
Then Eqs.~(\ref{eqn6.2}) are equivalent to
\begin{equation}\label{eqn6.3n}
\dirz_1^+ \, \dirz_2 + \dirz_2^+ \, \dirz_1 \, = \,
\dirz_1 \, \dirz_2^+ + \dirz_2 \, \dirz_1^+ \, = \,
2 \, z_1 \spr z_2
\quad
(\dirz^+ \, \dirz \, = \, \dirz \, \dirz^+ \, = \, z^{\, 2})
\, . \
\end{equation}

The generalized free Weyl fields of dimension
\(d = \txfrac{1}{2},\,
\txfrac{3}{2},\, \dots\) are two mutually conjugate
complex 2-component fields,
\begin{equation}\label{eqn6.5}
\chi^+ \left( z \right) \, = \,
\left( \schi_1 \left( z \right),\, \schi_2 \left( z \right) \right)
\quad \text{and} \quad \chi \left( z \right) \, = \,
\left( \begin{array}{c} \chi_1 \left( z \right) \\ \chi_2 \left( z \right) \end{array} \right)
\, , \
\end{equation}
transforming under the
elementary induced representations of \(\spin \left( 4 \right)\)
corresponding to the selfdual and antiselfdual representations~(\ref{eqn6.3}),
respectively.
In particular, the action of the Weyl reflection $j_W$~(\ref{eqn2.16}) is,
\begin{eqnarray}\label{eqn6.6}
\chi \left( z \right) \ \longmapsto \ && \podr
\frac{\dirz}{\left( z^{\, 2} \right)^{
d+\frac{\raisebox{0pt}{\footnotesize $1$}}{\raisebox{0pt}{\footnotesize $2$}}}} \ \chi \left( z \right)
\, ( \, \equiv \, \pi_z \left( j_W \right) \, \chi \left( z \right) \, )
\, , \quad \nonumber \\
\chi^+ \left( z \right) \ \longmapsto \ && \podr
\chi^+ \left( z \right) \,
\frac{\dirz}{\left( z^{\, 2} \right)^{
d+\frac{\raisebox{0pt}{\footnotesize $1$}}{\raisebox{0pt}{\footnotesize $2$}}}}
\, ( \, \equiv \, \pi_z^+ \left( j_W \right) \, \chi^+ \left( z \right) \, )
\, . \
\end{eqnarray}
The conformal invariant 2-point functions, characterizing the fields, have
the following matrix representation
\begin{equation}\label{eqn6.7}
\lvac \chi \left( z_1 \right) \chi^+ \left( z_2 \right) \rvac \, = \,
\frac{\dirz^+_{12}}{\left( z_{12}^{\, 2} \right)^{
d+\frac{\raisebox{0pt}{\footnotesize $1$}}{\raisebox{0pt}{\footnotesize $2$}}}}
\, , \quad
\end{equation}
\begin{equation}\label{eqn6.8}
\lvac \chi_{\alpha} \left( z_1 \right) \chi_{\beta} \left( z_2 \right) \rvac \, = \,
\lvac \chi_{\alpha}^+ \left( z_1 \right) \chi^+_{\beta} \left( z_2 \right) \rvac \, = \,
0
\, . \
\end{equation}
In particular, the invariance under the complex Weyl reflection $j_W$
is ensured by the equality
\begin{equation}\label{eqn6.9}
\frac{\dirz_1}{z^{\, 2}_1} \, \dirz_{12}^+ \, \frac{\dirz_2}{z^{\, 2}_2} \, = \,
\frac{\dirz_1^+}{z^{\, 2}_1} - \frac{\dirz_2^+}{z^{\, 2}_2}
\, . \
\end{equation}

The conjugation law~(\ref{eqn2.17}) reads
\begin{equation}\label{eqn6.11}
\chi^+ \left( \overline{z} \right)^+ \, = \,
\frac{\dirz}{\left( z^{\, 2} \right)^{
d+\frac{\raisebox{0pt}{\footnotesize $1$}}{\raisebox{0pt}{\footnotesize $2$}}}}
\ \,
\chi \left( \frac{z}{z^{\, 2}} \right)
\, \
\end{equation}
in other words, for any \(\Phi,\, \Psi \in \DOM\):
\begin{equation}\label{eqn6.12n}
\La \chi^+ \left( \overline{z} \right) \Phi \Vl \Psi \Ra \, = \,
\frac{\dirz}{\left( z^{\, 2} \right)^{
d+\frac{\raisebox{0pt}{\footnotesize $1$}}{\raisebox{0pt}{\footnotesize $2$}}}}
\ \,
\La \Phi \Vl \chi \left( \frac{z}{z^{\, 2}} \right) \Psi \Ra
\, . \
\end{equation}
Here one can explicitly verify the hermiticity of the 2-point scalar product
\begin{equation}\label{eqn6.12}
\La \chi^+ \left( \overline{z} \right) \Omega \Vl \chi^+ \left( w \right) \Omega \Ra
\, = \,
\left(
\La \chi^+ \left( w \right) \Omega \Vl \chi^+ \left( \overline{z} \right) \Omega \Ra
\right)^+
\, , \
\end{equation}
where \(\Omega = \rvac\) is the vacuum:
\begin{equation}\label{eqn6.13}
\frac{\dirz}{\left( z^{\, 2} \right)^{
d+\frac{\raisebox{0pt}{\footnotesize $1$}}{\raisebox{0pt}{\footnotesize $2$}}}} \,
\lvac \chi \left( \frac{z}{z^{\, 2}} \right) \chi^+ \left( w \right) \rvac =
\left(\raisebox{18pt}{\hspace{-3pt}}\right.
\frac{\overline{\dirw}}{\left( \overline{w}^{\, 2} \right)^{
d+\frac{\raisebox{0pt}{\footnotesize $1$}}{\raisebox{0pt}{\footnotesize $2$}}}} \,
\lvac \chi \left( \frac{\overline{w}}{\overline{w}^{\, 2}} \right)
\chi^+ \left( \overline{z} \right) \rvac
\left.\raisebox{18pt}{\hspace{-3pt}}\right)^{\hspace{-3pt} +}
\! .
\end{equation}

The conjugation law for
the compact picture generalized free Weyl field
\begin{equation}\label{eqn6.14}
\chi^+ \left( \czeta,\, u \right) \, = \,
e^{2\pi i \, d \, \czeta} \, \chi^+ \left( e^{2\pi i \, \czeta} \, u \right)
\, , \quad
\chi \left( \czeta,\, u \right) \, = \,
e^{2\pi i \, d \, \czeta} \, \chi \left( e^{2\pi i \, \czeta} \, u \right)
\, \
\end{equation}
becomes
\begin{equation}\label{eqn6.15}
\chi^+ \left( \czeta,\, u \right)^+ \, = \,
\diru \ \, \chi \left( \czeta,\, u \right)
\, . \
\end{equation}

The vacuum correlation function is diagonal in ``the moving frame'' representation
defined as follows.
For given non-collinear unit real vectors \(u_1,\, u_2 \in \Sr^{D-1} ( \subset \R^D)\)
such that \(u_1 \spr u_2 = \cos 2\pi \alpha\)
let $v$ and $\overline{v}$ be the unique complex vectors (in $\C^D$)
for which
\begin{equation}\label{eqn6.17n}
u_1 \, = \, e^{\pi i \, \alpha} \, v + e^{-\pi i \, \alpha} \, \overline{v}
\, , \quad
u_2 \, = \, e^{-\pi i \, \alpha} \, v + e^{\pi i \, \alpha} \, \overline{v}
\, . \
\end{equation}
It then follows that $v$ and $\overline{v}$ are mutually conjugate isotropic vectors
with scalar product: \(2 v \spr \overline{v} = 1\).
In this basis we have
\begin{equation}\label{eqn6.18n}
\lvac \!
\chi \! \left( \czeta_1, u_1 \right)
\chi^+ \!\hspace{-1pt} \left( \czeta_2, u_2 \right)
\rvac
\! = \!
\frac{1}{2\hspace{1pt}i
\left( -4 \sin \pi \czeta_+ \sin \pi \czeta_- \right)^{
d-\frac{\raisebox{0pt}{\footnotesize $1$}}{\raisebox{0pt}{\footnotesize $2$}}}}
\!
\left( \frac{\dirv^+}{\sin \pi \czeta_-}
\! + \!
\frac{\overline{\dirv}{\hspace{1pt}}^+}{\sin \pi \czeta_+} \right)
\! , \
\end{equation}
where \(\czeta_{\pm} = \czeta_{12} \pm \alpha\) (as in previous sections).
In the frame, in which
\(u_{1,2} =
\left( 0,\right.\) $0,$ $\pm\sin \pi\alpha,$ \(\left.\cos \pi\alpha \right)\)
the matrix $\dirv$ and its conjugate assume a simple form:
\begin{equation}\label{eqn6.19n}
\dirv^+ \, = \,
\left( \begin{array}{cc} 1 & 0 \\ 0 & 0 \end{array}  \right) \, = \,
\overline{\dirv}
\, , \quad
\overline{\dirv}{\hspace{1pt}}^+ \, = \,
\left( \begin{array}{cc} 0 & 0 \\ 0 & 1 \end{array}  \right) \, = \,
\dirv
\, . \
\end{equation}
Thus, in the \(d = \txfrac{1}{2}\) case
of a {\it subcanonical} Weyl field
the contribution of
$\czeta_+$ and $\czeta_-$ are separated.
The dimension \(d = \txfrac{3}{2}\) corresponds
to the {\it canonical free} Weyl field which will be denoted by
\(\psi := \chi\) (\(\psi^+ := \chi^+\)).
We find in this case
\begin{eqnarray}\label{canweyl1}
&& \hspace{-15pt}
\lvac \psi \left( \czeta_1,\, u_1 \right) \psi^+ \left( \czeta_2,\, u_2 \right) \rvac
=
\frac{i}{8\ \sin 2\pi\alpha}
\left(\raisebox{12pt}{\hspace{-2pt}}\right.
{\dirv^+}
\left(\raisebox{12pt}{\hspace{-2pt}}\right.
\frac{\cos \pi\czeta_-}{\sin^2 \pi\czeta_-}-
\frac{\cot 2\pi\alpha}{\sin \pi\czeta_-}+
\nonumber \\ && \hspace{-15pt} \hspace{10pt} +
\frac{1}{\sin 2\pi\alpha \, \sin \pi\czeta_+}
\left.\raisebox{12pt}{\hspace{-2pt}}\right) -
\overline{\dirv}^+
\left(\raisebox{12pt}{\hspace{-2pt}}\right.
\frac{\cos \pi\czeta_+}{\sin^2 \pi\czeta_+}
+
\frac{\cot 2 \pi\alpha}{\sin \pi\czeta_+}
-\frac{1}{\sin 2\pi\alpha \, \sin \pi\czeta_-}
\left.\raisebox{12pt}{\hspace{-2pt}}\right)
\left.\raisebox{12pt}{\hspace{-2pt}}\right)
\! , \ \ \ \
\end{eqnarray}
From the vacuum correlation functions ~(\ref{eqn6.18n}), (\ref{canweyl1})
and from Eq.~(\ref{eqn4.14a}) we deduce
\begin{eqnarray}\label{eqn6.20n}
&& \hspace{-15pt}
\La \hspace{-1pt} \chi \left( \czeta_1, u_1 \right) \chi^+ \left( \czeta_2, u_2 \right) \hspace{-1pt} \Ra_q
=
\frac{1}{2\pi\hspace{1pt}i}
\left(
p_1^{11} \left( \czeta_-,\, \tau \right) \, \dirv^{\hspace{1pt}+} +
p_1^{11} \left( \czeta_+,\, \tau \right) \, \overline{\dirv}{\hspace{1pt}}^+
\right)
\, , \
\\ \mgvspc{16pt}
&& \hspace{-15pt}
\label{canweyl2}
\La \hspace{-1pt} \psi \left( \czeta_1, u_1 \right) \psi^+ \left( \czeta_2, u_2 \right) \hspace{-1pt} \Ra_q
=
\frac{i}{8\pi \sin 2\pi\alpha}
\left(\raisebox{12pt}{\hspace{-2pt}}\right.
{\dirv^+} \!
\left(\raisebox{12pt}{\hspace{-2pt}}\right.
p_2^{11} \left( \czeta_-,\tau \right)
-\cot 2\pi\alpha \ p_1^{11} \left( \czeta_-,\tau \right)+
\nonumber \\ && \hspace{-15pt} \hspace{10pt}
+\frac{p_1^{11} \left( \czeta_+,\tau \right)}{\sin 2\pi\alpha}
\left.\raisebox{12pt}{\hspace{-2pt}}\right) -
\overline{\dirv}^{\, +} \!
\left(\raisebox{12pt}{\hspace{-2pt}}\right.
p_2^{11} \left( \czeta_+,\tau \right)
+\cot 2\pi\alpha \ p_1^{11} \left( \czeta_+,\tau \right)
-\frac{p_1^{11} \left( \czeta_-,\tau \right)}{\sin 2\pi\alpha}
\left.\raisebox{12pt}{\hspace{-2pt}}\right)
\left.\raisebox{12pt}{\hspace{-3pt}}\right)
. \ \
\end{eqnarray}
Here \(
p_k^{\kappa,\lambda} \left( \czeta,\, \tau \right)
=
\mathop{\sum}\limits_{m} \, \mathop{\sum}\limits_{n} \,
\left( -1 \right)^{m\kappa+n\lambda}\)
\(\left( \czeta+m\tau+n \right)^{-k}
\)
(see Appedix~A).
Under the assumptions of
Remark~\ref{rm:4.1},
for $N$ identified with the charge operator
(so that \(\left[ N \hspace{-1pt}, \chi^+ \!\! \left( z \right) \right]\)
$=$ \(\chi^+ \left( z \right)\),
\(\left[ N, \chi \! \left( z \right) \right] = - \chi \left( z \right)\)),
we find for \(d = \txfrac{1}{2}\)
\begin{equation}\label{eqn6.21n}
\La \hspace{-2pt}
\chi \hspace{-1pt} \left( \czeta_1,\, u_1 \right)
\chi^+ \hspace{-2pt} \left( \czeta_2,\, u_2 \right) \hspace{-2pt}
\Ra_{q,\mu}
=
\frac{1}{2\pi\hspace{1pt}i}
\left(
p_1^{11} \hspace{-2pt} \left( \czeta_-,\, \tau,\, \mu \right) \, \dirv^{\hspace{1pt}+}
\hspace{-2pt} +
p_1^{11} \hspace{-2pt} \left( \czeta_+,\, \tau,\, \mu \right) \, \overline{\dirv}{\hspace{1pt}}^+
\right)
\
\end{equation}
using the more general functions
$p_1^{\kappa,\lambda} \left( \czeta,\tau,\mu \right)$ of Appendix~A
(Eq.~(\ref{canweyl2}) is generalized in a similar way).

Note that the 1-particle scalar product is
\begin{equation}\label{eqn6.17}
\lvac \chi^+ \left( \overline{z} \right)^+ \chi^+ \left( w \right) \rvac
\, = \,
\frac{1 - \dirz \ \dirw^+}{\left( 1-2 \, z \spr w + z^2 \, w^{\, 2} \right)^{
d+\frac{\raisebox{0pt}{\footnotesize $1$}}{\raisebox{0pt}{\footnotesize $2$}}}}
\, . \
\end{equation}
This implies similarly to the scalar field case
that we can organize the field mode expansion (in the compact picture) as
\begin{equation}\label{eqn6.23}
\chi \! \left( \czeta, u \right) \hspace{-1pt} = \!\hspace{-1pt}
\mathop{\sum}\limits_{
\raisebox{0pt}{\small \(\mathop{}\limits^{n \, \in \, \Z}_{
|n
-\frac{\raisebox{0pt}{\scriptsize $1$}}{\raisebox{0pt}{\scriptsize $2$}}|
\, \geqslant \,
d}\)}} \!\hspace{-1pt}
\chi_{n-\frac{1}{2}} \! \left( u \right)
e^{i\pi \left( 1-2n \right) \czeta}
, \hspace{4pt}
\chi^{+} \!\hspace{-1pt} \left( \czeta, u \right) \hspace{-1pt} = \!\hspace{-1pt}
\mathop{\sum}\limits_{
\raisebox{0pt}{\small \(\mathop{}\limits^{n \, \in \, \Z}_{
|n
-\frac{\raisebox{0pt}{\scriptsize $1$}}{\raisebox{0pt}{\scriptsize $2$}}|
\, \geqslant \,
d}\)}} \!\hspace{-1pt}
\chi_{n-\frac{1}{2}}^{+} \! \left( u \right)
e^{i\pi \left( 1-2n \right) \czeta}
, \
\end{equation}
where
$\chi_{n+d}^{(+)} \! \left( u \right)$ and
$\chi_{-n-d}^{(+)} \! \left( u \right)$ for \(n = 0,1,\dots\)
are homogeneous polynomial in \(u \in \Sr^3\) of degree
$n$
with 2-component operator coefficients.
For \(n \geqslant 0\),
$\chi_{n+d}^{(+)} \! \left( u \right)$
correspond to annihilation operators while
$\chi_{-n-d}^{(+)} \! \left( u \right)$,
to the creation modes and we also have
\begin{eqnarray}\label{eqn6.23a}
&
\left( \chi_{\frac{1}{2}-k}^{+} \! \left( u \right) \right)^+
\, = \,
\diru \ \chi_{k-\frac{1}{2}} \! \left( u \right)
\, , \quad
\lvac
\chi_{-m-d}^{+} \! \left( u_1 \right)^{+} \,
\chi_{-n-d}^{+} \! \left( u_2 \right) \rvac
\, = & \nonumber \\ & = \,
\delta_{nm} \,
\left(
C_n^{
d+\frac{\raisebox{0pt}{\footnotesize $1$}}{\raisebox{0pt}{\footnotesize $2$}}}
\left( u_1 \spr u_2 \right)
-
C_{n-1}^{
d+\frac{\raisebox{0pt}{\footnotesize $1$}}{\raisebox{0pt}{\footnotesize $2$}}}
\left( u_1 \spr u_2 \right) \diru_1 \diru_2^+
\right)
\, &
\end{eqnarray}
for \(k \in \Z\), \(n,m= 0,1, \dots\) and \(u_1,u_2 \in \Sr^3\).

For the thermal energy mean values we will consider the two cases of
\(d = \txfrac{3}{2}\) and
\(d = \txfrac{1}{2}\), separately.

The $z$--picture canonical spinor field satisfies the Weyl--Dirac equation
\begin{equation}\label{eqn6.25}
\dirdi_{z} \, \psi \left( z \right) \, = \, 0
\, , \quad
\psi^+ \left( z \right) \,
\LVEC{\dirdi}{\raisebox{9pt}{\hspace{0pt}}}_z \, = \, 0
\, . \
\end{equation}
These equations are also valid for the compact picture modes
$\psi_{-n-\frac{3}{2}} \left( u \right)$ extended to \(u \in \R^4\).
The positive charge 1-particle state-space of conformal energy
\(\lvac \hspace{-1pt} H \hspace{-1pt} \rvac+\)
\(n+\txfrac{3}{2}\) (\(n=0,1,\dots\)), spanned
by \(\psi_{-n-\frac{3}{2}}^+ \left( u \right) \rvac\)
carries the irreducible representation
\(
\left(
\txfrac{n}{2}
,
\txfrac{n \! + \! 1}{2}
\right)\)
of
$\Spin \left( 4 \right)$ and therefore, has dimension
\(\left( n+2 \right) \left( n+1 \right)\).
The dimension of the full
1-particle space, including charge $-1$ states, is twice as big.
It has also a positive definite scalar product in view of~\cite{Ma77}.
Thus applying the general formula~(\ref{eqn4.4n}) and Eq.~(\ref{fn_G})
we find (cf.~\cite{DK02}):
\begin{eqnarray}\label{eqn6.26}
\La \! H + E_0 \! \Ra_{q}
\hspace{-1pt} = \hspace{-1pt} && \podr
E_0 +
\frac{1}{Z \left( \tau \right)} \, q \, \frac{\partial}{\partial q} \, Z \left( \tau \right)
=
\mathop{\sum}\limits_{n \, = \, 0}^{\infty}
\frac{2
\left(\raisebox{9pt}{\hspace{-2pt}}\right.
n \hspace{-1pt} + \hspace{-1pt}
\frac{\raisebox{0pt}{\small $3$}}{\raisebox{0pt}{\small $2$}}
\left.\raisebox{9pt}{\hspace{-2pt}}\right)
\left( n \hspace{-1pt} + \hspace{-1pt} 1 \right)
\left( n \hspace{-1pt} + \hspace{-1pt} 2 \right)
q^{n+\frac{\raisebox{0pt}{\footnotesize $3$}}{\raisebox{0pt}{\footnotesize $2$}}}
}{
1 + q^{n+\frac{\raisebox{0pt}{\footnotesize $3$}}{\raisebox{0pt}{\footnotesize $2$}}}
}
=
\nonumber \\ = \hspace{-1pt} && \podr
\frac{1}{4}
\left(\raisebox{10pt}{\hspace{-2pt}}\right.
G_4 \! \left(\raisebox{9pt}{\hspace{-2pt}}\right.
\frac{\tau \! + \! 1}{2}
\left.\raisebox{9pt}{\hspace{-2pt}}\right) - 8 \, G_4 \! \left( \tau \right)
\left.\raisebox{10pt}{\hspace{-2pt}}\right)
\, - \,
\frac{1}{4}
\left(\raisebox{10pt}{\hspace{-2pt}}\right.
G_2 \! \left(\raisebox{9pt}{\hspace{-2pt}}\right.
\frac{\tau \! + \! 1}{2}
\left.\raisebox{9pt}{\hspace{-2pt}}\right) - 2 \, G_2 \! \left( \tau \right)
\left.\raisebox{10pt}{\hspace{-2pt}}\right)
\, , \quad
\end{eqnarray}
\begin{equation}\label{eqn6.26nnn}
E_0 \, = \,
- \frac{1}{4} \, \frac{B_4}{8} \left( 1 \! - \! 2^3 \right)
+ \frac{1}{4} \, \frac{B_2}{4} \left( 1 \! - \! 2 \right)
\, = \, - \frac{17}{960}
\, . \
\end{equation}
Here we have used the equalities
\begin{eqnarray}\label{eqn6.27nnn}
&& \hspace{-20pt}
\mathop{\sum}\limits_{n \, = \, 1}^{\infty}
\frac{\left( 2n \! + \! 1 \right)^{2k-1} \! q^{2n+1}}{1+q^{2n+1}} =
G_{2k} \! \left(\raisebox{9pt}{\hspace{-2pt}}\right.
\tau \hspace{-1pt} + \hspace{-1pt} \frac{1}{2}
\left.\raisebox{9pt}{\hspace{-2pt}}\right) -
2^{2k-1} \, G_{2k} \! \left( 2\hspace{1pt}\tau \right)
+ \frac{B_{2k}}{4k} \, \left( 1 \! - \! 2^{2k-1} \right)
\, , \ \ \ \
\\ && \hspace{-20pt} \hspace{60pt} \label{eqn6.28nnn}
2 \left( n \! + \! \frac{1}{2} \right) n \left( n \! + \! 1 \right)
\, = \,
\frac{1}{4} \left( \left( 2n \! + \! 1 \right)^3 -
\left( 2n \! + \! 1 \right) \right)
\, . \
\end{eqnarray}

The subcanonical Weyl field and its conjugate satisfy
third order equations which assume the following form on the modes
\begin{equation}\label{eqn6.27}
\di^{\, 2}_u \,
\dirdi_{u} \, \chi_{n+\frac{1}{2}} \left( u \right) \, = \, 0
\, , \quad
\chi^+_{n+\frac{1}{2}} \left( u \right) \,
\LVEC{\dirdi}{\raisebox{9pt}{\hspace{0pt}}}_u
\, \LVEC{\di}{\raisebox{9pt}{\hspace{0pt}}}_u^{\, 2} \, = \, 0
\quad
(u \in \R^3)
\, . \
\end{equation}
The resulting $\Spin \left( 4 \right)$--representation in the positive charge
1-particle space of conformal energy
\(n \! + \! \txfrac{1}{2}\) (\(n = 0,1,\dots\))
is then isomorphic to a (pseudoorthogonal) direct sum of three irreducible
representations (for \(n \geqslant 3\)),
\begin{equation}\label{eqn6.28}
\left(
\frac{n}{2}
,\,
\frac{n \! + \! 1}{2}
\right)
\oplus
\left(
\frac{n}{2}
,\,
\frac{n \! - \! 1}{2}
\right)
\oplus
\left(
\frac{n \! - \! 2}{2}
,\,
\frac{n \! - \! 1}{2}
\right)
\, \
\end{equation}
each of them should posses a definite restriction of the scalar product.
In particular, the full dimension $d_f \left( n \right)$
of the space~(\ref{eqn6.28}) is
\begin{equation}\label{eqn6.29}
\frac{1}{2} \, d_f \left( n \right) \, = \, 3n^2+3n+2
, \quad
\left( n \! + \! \frac{1}{2} \right) d_f \left( n \right) \, = \,
\frac{3}{4} \left( 2n \! + \! 1 \right)^3 +
\frac{5}{4} \left( 2n \! + \! 1 \right)
,
\end{equation}
for all \(n=0,1,\dots\),
so that the thermal energy mean value and vacuum
energy are given by
\begin{equation}\label{eqn6.30}
\La \! H + E_0 \! \Ra_q \, = \,
\frac{3}{4}
\left(\raisebox{10pt}{\hspace{-2pt}}\right.
G_4 \! \left(\raisebox{9pt}{\hspace{-2pt}}\right.
\frac{\tau \! + \! 1}{2}
\left.\raisebox{9pt}{\hspace{-2pt}}\right) - 8 \, G_4 \! \left( \tau \right)
\left.\raisebox{10pt}{\hspace{-2pt}}\right)
\, + \,
\frac{5}{4}
\left(\raisebox{10pt}{\hspace{-2pt}}\right.
G_2 \! \left(\raisebox{9pt}{\hspace{-2pt}}\right.
\frac{\tau \! + \! 1}{2}
\left.\raisebox{9pt}{\hspace{-2pt}}\right) - 2 \, G_2 \! \left( \tau \right)
\left.\raisebox{10pt}{\hspace{-2pt}}\right)
\, , \quad
\end{equation}
\begin{equation}\label{eqn6.31}
E_0 \, = \,
- \frac{3}{4} \, \frac{B_4}{8} \left( 1 \! - \! 2^3 \right)
- \frac{5}{4} \, \frac{B_2}{4} \left( 1 \! - \! 2 \right)
\, = \, \frac{29}{960}
\, . \
\end{equation}
Note that although $G_2$ is not a modular form
the differences entering the right hand sides of (\ref{eqn6.23}) and (\ref{eqn6.30})
are multiples of $F(\tau)$ (\ref{ad_a3}) and are thus modular forms of weight
two and level $\Gamma_{\theta}$.

\subsection{The Maxwell free field}\label{Ssec.5.2}
The electromagnetic (or Maxwell) free field is a 6 component field
\(F_{\cmu\cnu} \left( z \right) = - F_{\cnu\cmu} \left( z \right)\)
(\(1 \leqslant \cmu < \cnu \leqslant 4\)).
It is convenient to write it as a 2-form:
\begin{equation}\label{eqn6.34}
F \left( z \right) \, = \, \frac{1}{2} \, F_{\cmu\cnu} \! \left( z \right)
\, dz^{\cmu} \!\wedge dz^{\cnu}
\, \
\end{equation}
which makes clear its transformation properties and conjugation law:
\begin{eqnarray}\label{eqn6.35}
& \hspace{-30pt}
U \left( g \right)
\left( F_{\cmu\cnu} \! \left( z \right) \, dz^{\cmu} \!\wedge dz^{\cnu} \right)
U \left( g^{-1} \right) =
F_{\cmu\cnu} \! \left( g \left( z \right) \right) \,
d \hspace{1pt}g \hspace{-1pt} \left( z \right)^{\cmu} \!\wedge
d \hspace{1pt}g \hspace{-1pt} \left( z \right)^{\cnu}
,
& \\ \label{eqn6.36} & \hspace{-30pt}
\left( F_{\cmu\cnu} \! \left( z \right) \, dz^{\cmu} \!\wedge dz^{\cnu} \right)^*
=
F_{\cmu\cnu} \! \left( z^* \right) \,
d\overline{z}^{\cmu} \!\wedge d\overline{z}^{\cnu}
\quad (z^* =
\Txfrac{\overline{z}}{\overline{z}^{\, 2}},
\ \
\left( dz^{\cmu} \right)^* = d \left( z^* \right)^{\cmu})
, &
\end{eqnarray}
where \(g = e^{t \Omega}\) for a real conformal generator
$\Omega$ as in Eq.~(\ref{eqn2.6}).
The 2-point function is
\begin{equation}\label{eqn6.37}
\lvac F_{\cmu_1\cnu_1} \! \left( z_1 \right)
F_{\cmu_2\cnu_2} \! \left( z_2 \right) \rvac \, := \,
\frac{
r_{\cmu_1\cmu_2} \! \left( z_{12} \right) r_{\cnu_1\cnu_2} \! \left( z_{12} \right) -
r_{\cmu_1\cnu_2} \! \left( z_{12} \right) r_{\cnu_1\cmu_2} \! \left( z_{12} \right)}{
\left( z_{12}^{\, 2} \right)^2} \, ,
\end{equation}
\(r_{\cmu\cnu} \left( z \right) := \delta_{\cmu\cnu} -
2 \, \Txfrac{z_{\cmu} z_{\cnu}}{z^{\, 2}}\).\gvspc{-7pt}
It is verified to satisfy the \textit{Maxwell equations}
\begin{equation}\label{eqn6.38}
d F \left( z \right) \, = \, 0
\, , \quad
d * \! \left( F \right) \left( z \right) \, = \, 0
\, , \
\end{equation}
$*$ being \textit{Hodge} star:
\(* \! \left( F \right)_{\cmu\cnu} \! \left( z \right) \, := \,
\varepsilon_{\cmu\cnu\rho\sigma} \, F^{\rho\sigma} \! \left( z \right)\).

To compute the (compact picture) finite temperature correlation functions
\(\La F_{\cmu_1\cnu_1} \left( \czeta_1, u_1 \right)\)
\(F_{\cmu_2\cnu_2} \left( \czeta_2, u_2 \right) \Ra_q\)
we again use the diagonal frame
in which, $2v$ $=$ $(0,$ $0,$ $-i,$ $1)$,
\(u_{1,2} = \left( 0,\right.\)
$0,$ $\pm\sin \pi\alpha,$ \(\left.\cos \pi\alpha \right)\); then there
exist
linear combinations of the field components
\begin{equation}\label{eqn6.40new}
\sqrt{2} \, F_1^{\pm} \! = \! F_{23} \hspace{-1pt} \pm \hspace{-1pt} F_{14}
, \hspace{2pt}
\sqrt{2} \, F_2^{\pm} \! = \! F_{31} \hspace{-1pt} \pm \hspace{-1pt} F_{24}
, \hspace{2pt}
\sqrt{2} \, F_3^{\pm} \! = \! F_{12} \hspace{-1pt} \pm \hspace{-1pt} F_{34}
, \hspace{2pt}
\sqrt{2} \, F_{\pm}^{\varepsilon} \! = \!
F_1^{\varepsilon} \hspace{-1pt} \pm \hspace{-1pt} iF_2^{\varepsilon}
\hspace{4pt}
\end{equation}
(\(\varepsilon = \pm\))
such that
\begin{eqnarray}\label{eqn6.39}
&
\lvac F_+^+ \! \left( \czeta_1, u_1 \right)
F_-^- \! \left( \czeta_2, u_2 \right) \rvac
=: \Wf_0 \left( \czeta_{12}, \alpha \right) =
& \nonumber \\ & =
\frac{\raisebox{1pt}{\(
1
\)}}{\raisebox{-4pt}{\(
4 \sin^3 \! 2 \pi \alpha
\)}}
\left(
\mathrm{cotg} \hspace{1pt} \pi \czeta_-
\hspace{-1pt} - \hspace{-1pt}
\mathrm{cotg} \hspace{1pt} \pi \czeta_+
\right)
-
\frac{\raisebox{1pt}{\(
1
\)}}{\raisebox{-4pt}{\(
4 \sin 2 \pi \alpha
\)}}
\left(
\frac{\raisebox{1pt}{\(
\cos \hspace{1pt} \pi\czeta_+
\)}}{\raisebox{-4pt}{\(
\sin^3 \hspace{-1pt} \pi\czeta_+
\)}}
\hspace{-1pt} - \hspace{-1pt}
\frac{\raisebox{1pt}{\(
\mathrm{cotg} \hspace{1pt} 2\pi\alpha
\)}}{\raisebox{-4pt}{\(
\sin^2 \hspace{-1pt} \pi\czeta_+
\)}}
\right)
;
& \nonumber \\ &
\lvac F_-^+ \! \left( \czeta_1, u_1 \right)
F_+^- \! \left( \czeta_2, u_2 \right) \rvac
= \Wf_0 \left( \czeta_{12}, -\alpha \right);
\hspace{9pt} \mgvspc{12pt}
\lvac F_3^+ \! \left( \czeta_1, u_1 \right)
F_3^- \! \left( \czeta_2, u_2 \right) \rvac =
& \hspace{-45pt} \nonumber \\ & \hspace{-5pt} =
\frac{\raisebox{1pt}{\(
1
\)}}{\raisebox{-4pt}{\(
4 \sin^2 \! 2 \pi \alpha
\)}}
\left(
\frac{\raisebox{1pt}{\(
1
\)}}{\raisebox{-4pt}{\(
\sin^2 \hspace{-1pt} \pi\czeta_+
\)}}
\hspace{-1pt} + \hspace{-1pt}
\frac{\raisebox{1pt}{\(
1
\)}}{\raisebox{-4pt}{\(
\sin^2 \hspace{-1pt} \pi\czeta_-
\)}}
\hspace{-1pt} + \hspace{-1pt}
2 \mathrm{cotg} \hspace{1pt} 2 \pi \alpha
\left(
\mathrm{cotg} \hspace{1pt} \pi \czeta_+
\hspace{-1pt} - \hspace{-1pt}
\mathrm{cotg} \hspace{1pt} \pi \czeta_-
\right)
\hspace{-1pt}\right)
& \hspace{5pt}
\end{eqnarray}
(\(\czeta_{\pm} = \czeta_{12} \pm \alpha\)).
The corresponding finite temperature correlation functions read:
\begin{eqnarray}\label{eqn6.40}
&
\La F_+^+ \! \left( \czeta_1, u_1 \right)
F_-^- \! \left( \czeta_2, u_2 \right) \Ra_q
\! =: \! \Wf_q \left( \czeta_{12}, \alpha \right) \! = \!
\frac{\raisebox{1pt}{\(
1
\)}}{\raisebox{-4pt}{\(
4 \sin^3 \! 2 \pi \alpha
\)}}
\!
\left( p_1 \left( \czeta_-, \tau \right)
\hspace{-1pt} - \hspace{-1pt}
p_1 \left( \czeta_+, \tau \right) \right)
\! - \!\!
& \nonumber \\ & -
\frac{\raisebox{1pt}{\(
1
\)}}{\raisebox{-4pt}{\(
4 \sin 2 \pi \alpha
\)}}
\left(
\frac{\raisebox{1pt}{\(
1
\)}}{\raisebox{-4pt}{\(
2\pi
\)}}
p_3 \left( \czeta_+, \tau \right)
\hspace{-1pt} - \hspace{-1pt}
\mathrm{cotg} \, 2\pi \alpha \, p_2 \left( \czeta_+, \tau \right)
\right)
;
& \nonumber \\ &
\La F_-^+ \! \left( \czeta_1, u_1 \right)
F_+^- \! \left( \czeta_2, u_2 \right) \Ra_q
= \Wf_q \left( \czeta_{12}, -\alpha \right);
\hspace{9pt} \mgvspc{12pt}
\La F_3^+ \! \left( \czeta_1, u_1 \right)
F_3^- \! \left( \czeta_2, u_2 \right) \Ra_q =
& \nonumber \\ & \!\! = \!
\frac{\raisebox{1pt}{\(
1
\)}}{\raisebox{-4pt}{\(
4 \sin^2 \! 2 \pi \alpha
\)}}
\left(
p_2 \left( \czeta_+, \tau \right)
\hspace{-2pt} + \hspace{-1pt}
p_2 \left( \czeta_-, \tau \right)
\hspace{-2pt} + \hspace{-1pt}
2 \mathrm{cotg} \hspace{1pt} 2 \pi \alpha
\left(
p_1 \left( \czeta_+, \tau \right)
\hspace{-2pt} - \hspace{-1pt}
p_1 \left( \czeta_-, \tau \right)
\right)
\right)
\! .
& \hspace{15pt}
\end{eqnarray}

In order to find the thermal energy mean value for the Maxwell field
we have to compute the dimension \(d(n)(\equiv d_{\bb} \left( n \right))\)
of the 1-particle state space of conformal energy $n$,
spanned by \(F_{\cmu\cnu;\, -n} \left( z \right)\rvac\) where the mode
\(F_{\cmu\cnu;\, -n} \left( z \right)\) is a homogeneous (harmonic) polynomial of
degree \(n-2\), satisfying the Maxwell equations.
To this end we display the $SO(4)$ representation content of the modes
satisfying the Maxwell equations.
Decomposing the antisymmetric tensor $F_{\cmu\cnu}$ into selfdual and antiselfdual
parts, \((1,0) \oplus (0,1)\), we see that the full space of homogeneous
skewsymmetric--tensor valued polynomials in $z$ of degree \(n-2\) generically
splits into a direct sum of three conjugate pairs of \(SU(2) \times SU(2)\)
representations; for instance,
\(
\left( 1,0 \right) \otimes
\left(\txfrac{n \hspace{-1pt} - \hspace{-1pt} 2}{2},\,
\txfrac{n \hspace{-1pt} - \hspace{-1pt} 2}{2}\right)
=
\left(\txfrac{n}{2},\,
\txfrac{n \hspace{-1pt} - \hspace{-1pt} 2}{2}\right)
\oplus
\left(\txfrac{n \hspace{-1pt} - \hspace{-1pt} 2}{2},\,
\txfrac{n \hspace{-1pt} - \hspace{-1pt} 2}{2}\right)
\oplus
\left(\txfrac{n \hspace{-1pt} - \hspace{-1pt} 4}{2},\,
\txfrac{n \hspace{-1pt} - \hspace{-1pt} 2}{2}\right)
\)
(for \(n>3\)).
Maxwell equations imply that only two of the resulting six
representations, those with maximal weights, appear in
the energy $n$ 1-particle space:
\(\left(\txfrac{n}{2},\,
\txfrac{n \hspace{-1pt} - \hspace{-1pt} 2}{2}\right)
\oplus
\left(\txfrac{n \hspace{-1pt} - \hspace{-1pt} 2}{2},\,
\txfrac{n}{2}\right)\). Thus,
\begin{equation}\label{eqHFM1}
d \left( n \right) = 2 \left( n^2-1 \right)
\end{equation}
and using (\ref{eqn5.19}) we then find
\begin{equation}\label{eqHFM2}
\La \hspace{-1pt} H + E_0 \hspace{-1pt} \Ra_q \, =
2 G_4(\tau)-2G_2(\tau)
\, , \quad
E_0 \, = \,
-2\,\frac{B_4}{8}+2\,\frac{B_2}{4} = \frac{11}{120}
\, . \
\end{equation}

\begin{remark}
\rmlabel{rm:5n}
Let us consider a generalized free vector field
$\lgf_{\cmu} \! \left( z \right)$
independent of $F_{\cmu\cnu} \! \left( z \right)$ (i.~e., commuting with it)
and having two point function
\begin{equation}\label{7.37}
\lvac \lgf_\cmu \left( z_1 \right) \lgf_\cnu \left( z_2 \right)\rvac \, = \, C \,
\frac{r_{\cmu\cnu} \left( z_{12} \right)}{z_{12}^{\, 2}}
\, , \quad
\di_{z^{\cmu}} \, \lgf_{\cnu} \left( z \right) \, = \,
\di_{z^{\cnu}} \, \lgf_{\cmu} \left( z \right)
\, \
\end{equation}
(the last equality means that $\lgf_{\cmu} \left( z \right)$ is a
``{\it longitudinal}'' field but we note that there is no GCI scalar field
$s \left( z \right)$ such that \(\lgf_{\cmu} \left( z \right) =
\di_{z^{\cmu}} \, s \left( z \right)\)).
The field $\lgf_{\cmu} \left( z \right)$ satisfies the third order equation
\begin{equation}\label{7.38}
\di_z^{\, 2} \ \di_z \spr \lgf \left( z \right) \, = \, 0
\, \
\end{equation}
and it then follows that the conformal energy $n$ state space has dimension
\begin{equation}\label{7.39}
d_{\lgf} \left( n \right) \, = \,
\left(\hspace{-2pt} \begin{array}{c} n+3 \\ 3 \end{array} \hspace{-2pt}\right)
-
\left(\hspace{-2pt} \begin{array}{c} n-1 \\ 3 \end{array} \hspace{-2pt}\right)
\, = \,
\left( n+1 \right)^2 + \left( n-1 \right)^2
\, = \, 2 \left( n^2+1 \right)
\, . \
\end{equation}
Thus the thermal energy mean value in the state space of
$F_{\cmu\cnu} \left( z \right)$ and $\lgf_{\cmu} \left( z \right)$ will be
\begin{equation}\label{7.40}
\La \hspace{-1pt} H_F+H_{\lgf} + E_0 \hspace{-1pt} \Ra_q \, =
4 \, G_4 \left( \tau \right)
\, , \quad
E_0 \, = \, \frac{1}{60}
\, \
\end{equation}
($H_F$ and $H_{\lgf}$ being the conformal Hamiltonians of the
corresponding subsystems).
We can interpret the state space of $\lgf_{\cmu} \left( z \right)$
as the space of \textit{pure gauge transformations} and then
the full state space of
$F_{\cmu\cnu} \left( z \right)$ and $\lgf_{\cmu} \left( z \right)$
has the meaning of the space of all gauge potentials.
\end{remark}

\section{The Thermodynamic Limit}\label{Sect.7n}

\subsection{Compactified Minkowski space as a ``finite box'' approximation}\label{SSect.7.1}

We shall now substitute $z$ in Eqs.~(\ref{eq1.2}) and (\ref{t2.2})
by $\Txfrac{z}{R}$ thus treating $\Sr^{D-1}$ and $\Sr^{1}$ in
the definition of $\M$ as a sphere and a circle of radius \(R \, (> 0)\).
Performing further the Minkowski space dilation
\(\left( 2R \right)^{\Omega_{-1D}} :\)
\(x^{\mu} \mapsto \Txfrac{x^{\mu}}{2R}\), \(\mu = 0,\dots,D-1\)
(see Eq.~(\ref{e2.3n}))
on the (real) variable \((\mzeta =) \, x\) in (\ref{t2.2}) we find
\(z \! \left( x;R \right) \hspace{-2pt} = \hspace{-1pt}
R \, z \! \left( \Txfrac{x}{2R} \right)\) or
\beq\label{e7.1}
\mbf{z} \! \left( x;R \right) \hspace{-2pt} = \hspace{-1pt}
\frac{\mbf{x}}{
2 \omega \! \left( \hspace{-1pt}\Txfrac{x}{2R}\hspace{-1pt} \right) \mgvspc{13pt}}
, \ \,
z_D \! \left( x;R \right) - R
\hspace{-2pt} = \hspace{-2pt}
\frac{i x^0 \! - \! \Txfrac{x^{\, 2}}{2R \mgvspc{-2pt}}}{
2 \omega \! \left( \hspace{-1pt}\Txfrac{x}{2R}\hspace{-1pt} \right) \mgvspc{13pt}}
, \ \,
2 \omega \! \left( \hspace{-1pt}\frac{x}{2R}\hspace{-1pt} \right)
\hspace{-2pt} = \hspace{-1pt}
1 + \frac{x^{\, 2}}{4R^2} - i \hspace{1pt} \frac{x^{0}}{R}
. \ \
\eeq
The stability subgroup of \(z \left( x;R \right) = 0 \, ( \in T_+)\)
in $\confgr$ is conjugate to the maximal compact subgroup
\(\compgr \subset \confgr\):
\beq\label{e7.2}
\compgr \! \left( 2R \right) \, = \,
\left( 2R \right)^{\Omega_{-1D}} \compgr \, \left( 2R \right)^{-\Omega_{-1D}}
\, , \quad
\compgr \, \equiv \, \compgr \!
\left(\raisebox{9pt}{\hspace{-2pt}}\right.
1
\left.\raisebox{9pt}{\hspace{-2pt}}\right)
\, \cong \,
\nfrc{U \left( 1 \right) \times \Spin \left( D \right)}{\Z_2}
\, . \
\eeq
In particular, the hermitian $U \left( 1 \right)$--generator $H \left( 2R \right)$,
which acts in the $z$--coordinates~(\ref{e7.1})
as the Euler vector field $z \spr \Txfrac{\di}{\di z}$, is
mutually conjugate to \(H \equiv H \! \left( 1 \right)\),
\beq\label{e7.3}
H \left( 2R \right) \, = \,
\left( 2R \right)^{\Omega_{-1D}} H \,
\left( 2R \right)^{-\Omega_{-1D}}
\, , \quad
H \, \equiv \,
H \!
\left(\raisebox{9pt}{\hspace{-2pt}}\right.
1
\left.\raisebox{9pt}{\hspace{-2pt}}\right)
\, . \
\eeq

For large $R$ and finite $x$ the variables \(\left( \mbf{z},z_D-R \right)\)
approach the (Wick rotated) Minkowski space coordinates
\(\left( \mbf{x},ix^0 \right)\).
In particular, for \(x^0 = 0 \, ( \, =\czeta)\),
the real $\left( D-1 \right)$--sphere \(z^{\, 2} = R^2\)
can be viewed as a $\mathit{SO} \! \left( D \right)$--invariant
``box'' approaching for \(R \to \infty\) the flat
space $\R^{D-1}$.
Thus the conformal compactification of Minkowski space also
plays the role of a convenient tool for studying the thermodynamic
limit of thermal expectation values.
This interpretation is justified in view of the following:

\begin{proposition}
\prlabel{pr:7.1}
The asymptotic behaviour of \(z \left( x; R \right) -R e_D\)
(\(e_D = \left( \Mbf{0},1 \right)\)) for large $R$ is:
\beqa\label{e7.4}
&
\mbf{z} \left( x;R \right) \, = \, \mbf{x} \, + \,
O \left(\raisebox{10pt}{\hspace{-2pt}}\right. \Txfrac{\|x\|^2}{R}
\left.\raisebox{10pt}{\hspace{-2pt}}\right)
\, , \quad
z_D \left( x;R \right) - R \, = \, i \hspace{1pt} x^0 \, + \,
O \left(\raisebox{10pt}{\hspace{-2pt}}\right. \Txfrac{\|x\|^2}{R}
\left.\raisebox{10pt}{\hspace{-2pt}}\right)
, &
\\ \label{e7.5} &
H_R \, := \, \Txfrac{H \left( 2R \right)}{R} \, = \,
P_0 \, + \, \Txfrac{1}{4R^2} \, K_0
\ ( \, = \, P_0 +
O \left(\raisebox{10pt}{\hspace{-2pt}}\right. \Txfrac{1}{R}
\left.\raisebox{10pt}{\hspace{-2pt}}\right)
\ \in i \hspace{1pt} \cnfalg \, )
\, , &
\eeqa
where
\(\left\| x \right\| := \sqrt{\left( x^0 \right)^2 + \left| \mbf{x} \right|^2}\)
for \(x = \left( x^0, \mbf{x} \right) \in M\) and
$i P_0$ is the real
conformal algebra generator of the Minkowski time ($x^0$) translation
(see Sect.~\ref{ssec2.1}).
The operator $H_R$ is the physical conformal Hamiltonian
(of dimension inverse length).
\end{proposition}

\begin{proof}
Eq.~(\ref{e7.4}) is obtained by a straightforward computation.
It follows from Eqs.~(\ref{e2.3n}) and (\ref{conf_ham}) that
\beq\label{e7.7}
H \, = \, \frac{1}{2} \left( P_0 + K_0 \right)
\, . \
\eeq
To derive Eq.~(\ref{e7.5}) one should then use (\ref{e7.7})
and the equations
\(\lambda^{\Omega_{-1D}} P_0 \lambda^{-\Omega_{-1D}} =\) \(\lambda \hspace{1pt} P_0\),
\(\lambda^{\Omega_{-1D}} K_0 \lambda^{-\Omega_{-1D}} =\) \(\lambda^{-1} K_0\);
hence,
\(
H \left( 2R \right) =\) \(\left( 2R \right)^{\Omega_{-1D}}\) $H$
\(\left( 2R \right)^{-\Omega_{-1D}} =\)
\(R P_0 +\) \(\Txfrac{1}{4R} \, K_0\).
\end{proof}

\begin{remark}
\rmlabel{rm:7.1}
The observation that the universal cover of $\M$,
the Einstein universe
\(\widetilde{M} =\) $\R \times \Sr^{D-1}$
(for \(D=4\)), which admits a
\textit{globally causal structure}, is locally undistiguishable
from $M$ for large $R$ has been emphasized over 30 year ago
by Irving Segal (for a concise expos\'e and further
references~--~see~\cite{S71}).
For a fixed choice, $\Omega_{-1D}$, of the dilation generator
in~(\ref{e7.2}) he identifies the Minkowski energy $P_0$
with the scale covariant component of $H_R$.
With this choice $M$ is osculating $\M$ (and hence $\widetilde{M}$)
at the north pole \(\left( \mbf{z},z_D \right)\) $=$
\(\left( \Mbf{0},R \right)\) (respectively, \(\czeta = 0\),
\(u = e_D\)), identified with the origin \(x=0\) in $M$.
(The vector fields associated with $H_R$ and $P_0$ coincide
at this point.)
\end{remark}

Using the Lie algebra limit \(\mathop{\lim}\limits_{R \to \infty} \, H_R = P_0\)
implied by~(\ref{e7.5}),
one can approximate the Minkowski energy operator $P_0$
for large $R$ by the physical conformal Hamiltonian $H_R$.
As we shall see below, the fact that in all considered free field models
in dimension \(D=4\) the conformal mean energy
is a linear combination of modular forms $G_{2k} \left( \tau \right)$
with highest weight \(2k=4\),
has a remarkable corollary: the \textit{density} $\MED$
of the physical mean energy has a limit reproducing the
\textit{Stefan--Boltzmann} law
\beq\label{e7.7n}
\MED \hspace{-1pt} \left( \beta \right) \, := \,
\mathop{\lim}\limits_{R \to \infty} \,
\frac{\La \hspace{-2pt} H_R \hspace{-2pt} \Ra_{q_{\beta}}}{\VOL{R}}
\, = \, \frac{C}{\beta^4}
\quad \text{for} \quad q_{\beta} \, := \, e^{-\beta}
\, \
\eeq
where $C$ is some constant,
\(\beta = \Txfrac{1}{k T}\) is the inverse temperature $T$
multiplied by the Boltzmann constant $k$) and
\(\VOL{R} := 2\pi^2 R^3\) is the volume of the $3$--sphere
of radius~$R$ at a fixed time (say \(x^0 = 0 = \czeta\)).
We will calculate this limit for two cases: the model of a free
scalar filed in \(D=4\) (see Sect.~\ref{Ssec.4.1})
which we will further denote by $\varphi$ and
the Maxwell free field model introduced in Sect.~\ref{Ssec.5.2}.

\begin{proposition}
\prlabel{pr:7.2}
For the free scalar field $\varphi$ in dimension \(D=4\)
we have the following behaviour
of the mean energy density
for \(\Txfrac{R}{\beta} \mgrt 1\)
\beq\label{e7.8}
\MED_R^{\left( \hspace{-1pt} \varphi \hspace{-1pt} \right)}
\! \left( \beta \right) \, := \,
\frac{1}{\VOL{R}} \, \frac{\mathit{tr}_{\DOM{}} \,
H_R \hspace{2pt} e^{-\beta H_R}}{
\mathit{tr}_{\DOM{}} \, e^{-\beta H_R}}
\, = \, \left( \frac{\pi^2}{30} -
\frac{1}{480 \hspace{1PT} \pi^2} \, \frac{\beta^4}{R^4} +
O \! \left(\raisebox{10pt}{\hspace{-2pt}}\right.
e^{-4\pi^2 \frac{R}{\beta}}
\left.\raisebox{10pt}{\hspace{-2pt}}\right) \right)
\frac{1}{\beta^4}
\, .
\eeq
The corresponding result for of the Maxwell free field
$F_{\mu\nu}$ is
\beq\label{e7.9}
\MED_R^{\left( \hspace{-1pt} F \hspace{-1pt} \right)}
\! \left( \beta \right) \, = \,
\left( \frac{\pi^2}{15} -
\frac{1}{6} \, \frac{\beta^2}{R^2} +
\frac{1}{
4 \pi^3
} \, \frac{\beta^3}{R^3} -
\frac{11}{240 \hspace{1pt}\pi^2} \, \frac{\beta^4}{R^4} +
O \! \left(\raisebox{10pt}{\hspace{-2pt}}\right.
e^{-4\pi^2 \frac{R}{\beta}}
\left.\raisebox{10pt}{\hspace{-2pt}}\right) \right)
\frac{1}{\beta^4}
\, .
\eeq
\end{proposition}

\begin{proof}
The hermitian operators $H$ and $H \left( 2R \right)$ are
unitarily equivalent due to Eq.~(\ref{e7.3}).
This leads to the fact that
\(\mathit{tr}_{\DOM{}} \,q^{H \left( 2R \right)}\) and
\(\mathit{tr}_{\DOM{}} \, H \left( 2R \right) q^{H \left( 2R \right)}\)
do not depend on $R$.
Then Eqs.~(\ref{eqn5.17}) and (\ref{eqHFM2}) imply that in the two
models under consideration we have
\beq\label{e7.11n}
\MED_R^{\left( \hspace{-1pt} \varphi \hspace{-1pt} \right)}
\! \left( \beta \right)
=
\frac{
G_4 \!
\left(\raisebox{10pt}{\hspace{-3pt}}\right.
\Txfrac{i \beta}{2 \pi R}
\left.\raisebox{10pt}{\hspace{-3pt}}\right)
\hspace{-1pt} -
\Txfrac{1}{240} \mgvspc{-10pt}
}{R \VOL{R}}
\, , \hspace{6pt}
\MED_R^{\left( \hspace{-1pt} F \hspace{-1pt} \right)}
\! \left( \beta \right)
=
\frac{2 G_4 \!
\left(\raisebox{10pt}{\hspace{-3pt}}\right.
\Txfrac{i \beta}{2 \pi R}
\left.\raisebox{10pt}{\hspace{-3pt}}\right)
\hspace{-1pt} -
2 G_2 \!
\left(\raisebox{10pt}{\hspace{-3pt}}\right.
\Txfrac{i \beta}{2 \pi R}
\left.\raisebox{10pt}{\hspace{-3pt}}\right)
\hspace{-1pt} -
\Txfrac{11}{120} \mgvspc{-10pt}
}{R \VOL{R}}
\, . \hspace{2pt}
\eeq
Using further the relations
\beq\label{e7.10}
G_2 \! \left( \tau \right) \, = \, \frac{1}{\tau^2}
\, G_2 \! \left(\raisebox{10pt}{\hspace{-2pt}}\right.
\frac{-1}{\tau}
\left.\raisebox{10pt}{\hspace{-2pt}}\right)
-
\frac{i}{4 \pi \tau}
\, , \quad
G_4 \! \left( \tau \right) \, = \, \frac{1}{\tau^4}
\, G_4 \! \left(\raisebox{10pt}{\hspace{-2pt}}\right.
\frac{-1}{\tau}
\left.\raisebox{10pt}{\hspace{-2pt}}\right)
\, \
\eeq
(which are special cases of~(\ref{ad_a1}) and (\ref{ad_a2}))
we find
\beqa\label{e7.11}
& \hspace{-10pt}
\MED_R^{\left( \hspace{-1pt} \varphi \hspace{-1pt} \right)}
\! \left( \beta \right)
\, = \, \Txfrac{1}{\beta^4}
\left(\raisebox{10pt}{\hspace{-2pt}}\right.
8 \pi^2 G_4 \!
\left(\raisebox{10pt}{\hspace{-2pt}}\right.
\Txfrac{2 \pi i \hspace{1pt} R}{\beta}
\left.\raisebox{10pt}{\hspace{-2pt}}\right)
-
\Txfrac{\beta^4}{480 \pi^2 R^4}
\left.\raisebox{10pt}{\hspace{-2pt}}\right)
\, , & \\ \label{e7.12} & \hspace{-10pt}
\MED_R^{\left( \hspace{-1pt} F \hspace{-1pt} \right)}
\! \left( \beta \right)
= \Txfrac{1}{\beta^4}
\left(\raisebox{10pt}{\hspace{-2pt}}\right.
16 \pi^2
G_4 \!
\left(\raisebox{10pt}{\hspace{-2pt}}\right.
\Txfrac{2 \pi i \hspace{1pt} R}{\beta}
\left.\raisebox{10pt}{\hspace{-2pt}}\right)
+
\Txfrac{4 \beta^2}{R^2}
G_2 \!
\left(\raisebox{10pt}{\hspace{-2pt}}\right.
\Txfrac{2 \pi i \hspace{1pt} R}{\beta}
\left.\raisebox{10pt}{\hspace{-2pt}}\right)
+
\Txfrac{\beta^3}{
4 \pi^3
R^3}
-
\Txfrac{11 \beta^4}{240 \hspace{1pt} \pi^2 R^4}
\left.\raisebox{10pt}{\hspace{-2pt}}\right)
\! .
& \qquad
\eeqa
Finally, to obtain Eqs.~(\ref{e7.8}) and (\ref{e7.9})
one should apply the expansion~(\ref{FTG})
implying that
\beqa
G_2 \! \left(\raisebox{10pt}{\hspace{-2pt}}\right.
\frac{2 \pi i \hspace{1pt} R}{\beta}
\left.\raisebox{10pt}{\hspace{-2pt}}\right)
= - \frac{1}{24} +
O \! \left(\raisebox{10pt}{\hspace{-2pt}}\right.
e^{-4\pi^2 \frac{R}{\beta}}
\left.\raisebox{10pt}{\hspace{-2pt}}\right)
, \quad
G_4 \! \left(\raisebox{10pt}{\hspace{-2pt}}\right.
\frac{2 \pi i \hspace{1pt} R}{\beta}
\left.\raisebox{10pt}{\hspace{-2pt}}\right)
= \frac{1}{240} +
O \! \left(\raisebox{10pt}{\hspace{-2pt}}\right.
e^{-4\pi^2 \frac{R}{\beta}}
\left.\raisebox{10pt}{\hspace{-2pt}}\right)
. \
\nonumber
\eeqa
\end{proof}

\begin{remark}
\rmlabel{rm:7.2}
In order to make comparison with the familiar expression
for the black body radiation it is instructive to restore
the dimensional constants $h$ and $c$ setting
\(H_R = \Txfrac{hc}{R} \, H \hspace{-1pt} \left( 2R \right)\)
(instead of (\ref{e7.4})).
The counterpart of~(\ref{e7.11n}) and~(\ref{FTG}) then reads
\beq\label{bl_b}
\La H_R \Ra_q =
\frac{h c}{R}
\left(\raisebox{10pt}{\hspace{-3pt}}\right.
G_4 \!
\left(\raisebox{10pt}{\hspace{-3pt}}\right.
\frac{i \hspace{1pt} h \hspace{0.5pt} c \hspace{0.5pt} \beta}{R}
\left.\raisebox{10pt}{\hspace{-3pt}}\right)
\hspace{-2pt} - \hspace{-1pt} E_0
\left.\raisebox{10pt}{\hspace{-3pt}}\right)
=
\frac{hc}{R} \,
\mathop{\sum}\limits_{n \, = \, 1}^{\infty} \hspace{3pt}
\frac{n^3
e^{- n \text{\footnotesize \(\Txfrac{h \hspace{0.5pt} c \hspace{0.5pt} \beta}{R}\)}}
\mgvspc{-8pt}}{\mgvspc{19pt}
1 - e^{- n \text{\footnotesize \(\Txfrac{h \hspace{0.5pt} c \hspace{0.5pt} \beta}{R}\)}}}
\, . \
\eeq
Each term in the infinite sum in the right hand side is a constant
multiple of Plank's black body radiation formula for frequency
\beq\label{freq}
\nu = n \frac{c}{R} .
\eeq
Thus, for finite $R$, there is a minimal frequency, $\Txfrac{c}{R}$.
Using the expansion in~(\ref{bl_b}) one can also find an alternative
integral derivation of the limit mean energy density
\(\MED_R^{\left( \hspace{-1pt} \varphi \hspace{-1pt} \right)}
\! \left( \beta \right)\)~(\ref{e7.11n}):
\beq\label{e7.16}
\MED_R^{\left( \hspace{-1pt} \varphi \hspace{-1pt} \right)}
\! \left( \beta \right)
\, = \,
\frac{1}{2 \pi^2 \hspace{-0.3pt} h^3 \hspace{-0.3pt} c^3 \hspace{-0.3pt} \beta^4}
\mathop{\sum}\limits_{n \, = \, 1}^{\infty} \hspace{-1pt}
\frac{\left( n
\Txfrac{h \hspace{0.5pt} c \hspace{0.5pt} \beta}{R} \right)^{\hspace{-2pt}3}
e^{- n \text{\footnotesize \(\Txfrac{h \hspace{0.5pt} c \hspace{0.5pt} \beta}{R}\)}}
\mgvspc{-8pt}}{\mgvspc{19pt}
1 - e^{- n \text{\footnotesize \(\Txfrac{h \hspace{0.5pt} c \hspace{0.5pt} \beta}{R}\)}}}
\, \frac{h \hspace{0.5pt} c \hspace{0.5pt} \beta}{R}
\ \mathop{\longrightarrow}\limits_{R \to \infty} \
\frac{\pi^2}{30 \hspace{-0.3pt} h^3 \hspace{-0.3pt} c^3 \hspace{-0.3pt} \beta^4}
\, \
\eeq
since the sum in the right hand side goes to the integral
\(\mathop{\int}\limits_{0}^{\infty}
\Txfrac{t^3 e^{-t}}{1 - e^{-t} \mgvspc{9pt}} \, d t
= \Txfrac{\pi^4}{15}\).
\end{remark}

\begin{remark}
\rmlabel{rm:7.3}
We observe that the constant $C$ in~(\ref{e7.7n}) in both
considered models
is equal to $\Txfrac{c_1}{30 \hspace{1pt} \pi^2}$, where
$c_1$ is the coefficient to the $G_4$--modular form in
$\La \hspace{-1pt} H \hspace{-1pt} \Ra_q$ (see Eq.~(\ref{eqn5.16})).
If we use in the definition~(\ref{e7.5}) of $H_R$
the Hamiltonian \(H \left( 2R \right) + E_0'\) instead of
$H \left( 2R \right)$,
\(\widetilde{H}_R := \Txfrac{H \left( 2R \right) + E_0'}{R}\),
then this will only reflect on the (non--leading)
terms $c_4 \Txfrac{\beta^4}{R^4}$
in (\ref{e7.8}) (\ref{e7.9}) replacing them
by $\Txfrac{E_0' - E_0}{2 \pi^2} \Txfrac{\beta^4}{R^4}$,
where $E_0$ is the ``vacuum energy'' for the corresponding models
(i.e., $E_0$ is $\Txfrac{1}{240}$ and $\Txfrac{11}{120}$
for the fields $\varphi$ and $F_{\mu\nu}$, respectively).
\end{remark}

\subsection{Infinite volume limit of the thermal correlation functions}\label{SSect.7.2}

We shall study the \(R \to \infty\) limit on the example of
a free scalar field $\varphi$ in four dimensions.

Denote by
$\varphi^{\MINK} \! \left( x \right)$
(the canonically normalized) \(D=4\) free massless scalar field
with $2$--point function
\beq\label{ea5.17}
\lvac \varphi^{\MINK} \! \left( x_1 \right)
\varphi^{\MINK} \! \left( x_2 \right) \rvac \, = \,
\left( 2\pi \right)^{-2} \left( x_{12}^{\, 2} + i \, 0 \, x_{12}^0 \right)^{-1}
\quad
\eeq
(\(x_{12} = x_1 - x_2\),
\(x_{12}^{\, 2} = \mbf{x}_{12}^{\, 2} - \left( x_{12}^0 \right)^2\)).
We define, in accord with
Proposition~\ref{pr:7.1},
a finite volume approximation of its thermal correlation function by
\beq\label{e7.18n}
\La \hspace{-1pt}
\varphi^{\MINK} \! \left( x_1 \right) \varphi^{\MINK} \! \left( x_2 \right)
\hspace{-1pt} \Ra_{\hspace{-1pt} \beta, R}
\, := \,
\frac{\mathit{tr}_{\DOM{}} \
\varphi^{\MINK} \! \left( x_1 \right) \varphi^{\MINK} \! \left( x_2 \right) \,
e^{-\beta H_{R}}}{
\mathit{tr}_{\DOM{}} \
e^{-\beta H_{R}}}
\, \
\eeq
and will be interested in the thermodynamic limit,
\beq\label{e7.19n}
\La \hspace{-1pt}
\varphi^{\MINK} \! \left( x_1 \right) \varphi^{\MINK} \! \left( x_2 \right)
\hspace{-1pt} \Ra_{\hspace{-1pt} \beta, \infty} :=
\mathop{\lim}\limits_{R \to \infty}
\La \hspace{-1pt}
\varphi^{\MINK} \! \left( x_1 \right) \varphi^{\MINK} \! \left( x_2 \right)
\hspace{-1pt} \Ra_{\hspace{-1pt} \beta, R}
\, . \
\eeq

{\samepage
\begin{proposition}
\prlabel{pr:7.3}
The limit (\ref{e7.19n}) (viewed as a meromorphic function) is given by
\beq\label{e7.20n}
\La \hspace{-1pt}
\varphi^{\MINK} \! \left( x_1 \right) \varphi^{\MINK} \! \left( x_2 \right)
\hspace{-1pt} \Ra_{\hspace{-1pt} \beta, \infty}
=
\Txfrac{
\sinh
\hspace{1pt} 2 \pi
\Txfrac{\left| \mbf{x}_{12} \right| \mgvspc{-4pt}}{\mgvspc{9pt} \beta}
}{
8 \pi \beta \left| \mbf{x}_{12} \right|}
\left(\raisebox{16pt}{\hspace{-3pt}}\right.
\cosh
\hspace{1pt} 2 \pi
\Txfrac{\left| \mbf{x}_{12} \right|}{\beta}
-
\cosh
\hspace{1pt} 2 \pi
\Txfrac{x_{12}^0}{\beta}
\left.\raisebox{16pt}{\hspace{-3pt}}\right)^{\hspace{-1pt}-1}
\!\! , \
\eeq
(\(\left| \mbf{x}_{12} \right| := \sqrt{\mbf{x}_{12}^{\, 2}} \equiv
\sqrt{\left( x^1_{12} \right)^2 + \left( x^2_{12} \right)^2 +
\left( x^3_{12} \right)^2}\)).
\end{proposition}
}

We shall \textit{prove} this statement by relating
$\varphi^{\MINK} \! \left( x \right)$ to the compact picture field
$\varphi \left( \czeta, u \right)$
\((\equiv \phi^{\left( 1 \right)} \left( \czeta,u \right))\)
whose thermal $2$--point function was computed in Sect.~\ref{sec:5}.

First, we use Eq.~(\ref{WF}) to express
$\varphi^{\MINK} \! \left( x \right)$ in terms of the $z$--picture
field (corresponding to the $R$--depending chart~(\ref{e7.1}))
\beq\label{e7.17}
2 \pi \,\varphi^{\MINK} \! \left( x \right) \, = \,
\frac{1}{2\omega \!
\left(\raisebox{10pt}{\hspace{-3pt}}\right.
\Txfrac{x}{2R}
\left.\raisebox{10pt}{\hspace{-3pt}}\right)} \
\varphi_R \! \left( z \left( x; R \right) \right)
\, \
\eeq
(since \(dz^{\, 2} = \omega \!
\left(\raisebox{10pt}{\hspace{-3pt}}\right.
\Txfrac{x}{2R}
\left.\raisebox{10pt}{\hspace{-3pt}}\right)^{-\hspace{-1pt} 2}
\Txfrac{dx^{\, 2}}{4}\), cp.~(\ref{t2.3})).
The factor $2\pi$ in front of $\varphi^{\MINK}$ accounts for
the different normalization conventions for the $x$--
and $z$--picture fields (we have
\(\lvac \varphi \! \left( z_1 \right)\)
\(\varphi \! \left( z_2 \right) \rvac =\) $\left( z_{12}^{\, 2} \right)^{-1}$
instead of~(\ref{ea5.17})).

As a second step we express
$\varphi_R \! \left( z \right)$~--~and
thus $\varphi^{\MINK} \! \left( x \right)$~--~in
terms of the compact picture field
$\varphi_R \! \left( \czeta,u \right)$:
\beqa\label{e7.21n}
&
\varphi_R \! \left( \czeta, u \right) \, := \,
R \hspace{1pt} e^{2 \pi i \czeta} \varphi \!
\left(\raisebox{9pt}{\hspace{-3pt}}\right.
R \hspace{1pt} e^{2 \pi i \czeta} u
\left.\raisebox{9pt}{\hspace{-3pt}}\right)
, & \nn &
2 \pi \hspace{1pt}
\varphi^{\MINK} \left( x \right) =
\Txfrac{1}{2R \left|\raisebox{10pt}{\hspace{-2pt}}\right.
\omega \!
\left(\raisebox{10pt}{\hspace{-3pt}}\right.
\Txfrac{x}{2R}
\left.\raisebox{10pt}{\hspace{-3pt}}\right)
\left.\raisebox{10pt}{\hspace{-3pt}}\right|
\mgvspc{14pt}}
\hspace{1pt} \varphi_R \!
\left(\raisebox{10pt}{\hspace{-3pt}}\right.
\czeta \! \left(\raisebox{10pt}{\hspace{-3pt}}\right.
\Txfrac{x}{2R}
\left.\raisebox{10pt}{\hspace{-3pt}}\right)
,
u \! \left(\raisebox{10pt}{\hspace{-3pt}}\right.
\Txfrac{x}{2R}
\left.\raisebox{10pt}{\hspace{-3pt}}\right)
\left.\raisebox{10pt}{\hspace{-3pt}}\right)
. & \qquad
\eeqa
Here $\czeta$ and $u$ are determined as functions of
$\Txfrac{x}{2R}$ from \(e^{2 \pi i \czeta} u
= \Txfrac{z \left( x; R \right)}{R} =
z \! \left(\raisebox{10pt}{\hspace{-3pt}}\right.
\Txfrac{x}{2R}
\left.\raisebox{10pt}{\hspace{-3pt}}\right)\)
($z \! \left( x \right)$ is the transformation~(\ref{t2.2}));
in deriving the second equation in (\ref{e7.21n})
we have used the relation
\(
e^{4 \pi i \czeta} =
\Txfrac{z \left( x; R \right)^{2}}{R^2} =
\overline{\omega \!
\left(\raisebox{10pt}{\hspace{-3pt}}\right.
\Txfrac{x}{2R}
\left.\raisebox{10pt}{\hspace{-3pt}}\right)}
\,
\omega \!
\left(\raisebox{10pt}{\hspace{-3pt}}\right.
\Txfrac{x}{2R}
\left.\raisebox{10pt}{\hspace{-3pt}}\right)^{-1}
\).

Next we observe that
$\varphi_R \hspace{-1pt} \left( \czeta, u \right)$
are mutually conjugate (for different $R$) just as
$H \left( 2R \right)$ in Eq.~(\ref{e7.3}).
(To see this one can use an intermediate
``dimensionless'' coordinates
\(\widetilde{z} \hspace{-1pt} \left( x; R \right)
= \Txfrac{z}{R} =
z \! \left(\raisebox{10pt}{\hspace{-3pt}}\right.
\Txfrac{x}{2R}
\left.\raisebox{10pt}{\hspace{-3pt}}\right)\),
which differs from (\ref{t2.2}) just by the dilation $\left( 2R \right)^{\Omega_{-1D}}$.)
It follows that its vacuum and thermal $2$--point function
with respect to the Hamiltonian $H \! \left( 2R \right)$
do not depend on $R$ and coincide with~(\ref{eqn5.11})
(for \(d=1\)) and~(\ref{eqn5.13}).
Thus
\beq\label{e7.22}
4 \pi^2
\La \hspace{-2pt}
\varphi^{\MINK} \!\hspace{-1pt} \left( x_1 \right)
\hspace{-1pt}
\varphi^{\MINK} \!\hspace{-1pt} \left( x_2 \right)
\hspace{-2pt}
\Ra_{\hspace{-2pt} \beta,R}
\hspace{1pt} = \hspace{1pt}
\Txfrac{
p_1 \! \hspace{-1pt} \left( \czeta_{12} \hspace{-2pt} + \hspace{-2pt} \alpha
, \hspace{-1pt}
\tau_R \hspace{-1pt} \right)
\hspace{-1pt} - \hspace{-1pt}
p_1 \! \hspace{-1pt} \left( \czeta_{12} \hspace{-2pt} - \hspace{-2pt} \alpha
, \hspace{-1pt}
\tau_R \hspace{-1pt} \right)
}{16 \hspace{-0.5pt} \pi \hspace{-0.5pt} R^2 \hspace{-1pt}
\left| \omega_1 \omega_2 \right| \sin 2\pi\alpha}
\hspace{-1pt}
\eeq
for
\(\omega_k =
\omega \! \left(\raisebox{10pt}{\hspace{-3pt}}\right.
\Txfrac{x_k}{2R}
\left.\raisebox{10pt}{\hspace{-3pt}}\right)\),
\(\czeta_{12} =
\czeta \! \left(\raisebox{10pt}{\hspace{-3pt}}\right.
\Txfrac{x_1}{2R}
\left.\raisebox{10pt}{\hspace{-3pt}}\right)
-
\czeta \! \left(\raisebox{10pt}{\hspace{-3pt}}\right.
\Txfrac{x_2}{2R}
\left.\raisebox{10pt}{\hspace{-3pt}}\right)\),
\(\cos \hspace{1pt} 2\pi \alpha =
u \! \left(\raisebox{10pt}{\hspace{-3pt}}\right.
\Txfrac{x_1}{2R}
\left.\raisebox{10pt}{\hspace{-3pt}}\right)
\spr
u \! \left(\raisebox{10pt}{\hspace{-3pt}}\right.
\Txfrac{x_2}{2R}
\left.\raisebox{10pt}{\hspace{-3pt}}\right)\),
\(\tau_R = \Txfrac{i \beta}{2 \pi R}\).
In order to perform the \(R \to \infty\) limit we derive
the large $R$ behaviour of
$\left| \omega_k \right|$, $\czeta_{12}$ and $\alpha$:
\beqa\label{e7.23}
&
2 \pi \czeta_{12} \hspace{-1pt} = \hspace{-1pt}
\Txfrac{x_{12}^0 \mgvspc{-4pt}}{\mgvspc{9pt} R}
\hspace{-1pt}
\left(\raisebox{10pt}{\hspace{-3pt}}\right.
1 \hspace{-1pt} + \hspace{-1pt}
O \! \left(\raisebox{10pt}{\hspace{-3pt}}\right.
\Txfrac{\left\| x_1 \right\|^2 \hspace{-2pt} +
\hspace{-1pt} \left\| x_2 \right\|^2 \mgvspc{-4pt}}{\mgvspc{9pt} R^2}
\left.\raisebox{10pt}{\hspace{-3pt}}\right)
\left.\raisebox{10pt}{\hspace{-4pt}}\right)
\hspace{-1pt} , \quad \hspace{-1pt}
2 \pi \alpha =
\Txfrac{\left| \mbf{x}_{12} \right| \mgvspc{-4pt}}{\mgvspc{9pt} R}
\left(\raisebox{10pt}{\hspace{-3pt}}\right.
1 \hspace{-1pt} + \hspace{-1pt}
O \! \left(\raisebox{10pt}{\hspace{-3pt}}\right.
\Txfrac{\left\| x_1 \right\|^2 \hspace{-2pt} +
\hspace{-1pt} \left\| x_2 \right\|^2 \mgvspc{-4pt}}{\mgvspc{9pt} R^2}
\left.\raisebox{10pt}{\hspace{-3pt}}\right)
\left.\raisebox{10pt}{\hspace{-4pt}}\right)
\hspace{-1pt} ,
& \nn &
4 \left| \omega_k \right|^2 = 1 +
O \! \left(\raisebox{10pt}{\hspace{-2pt}}\right.
\Txfrac{\left\| x_k \right\|^2 \mgvspc{-4pt}}{\mgvspc{9pt} R^2}
\left.\raisebox{10pt}{\hspace{-2pt}}\right)
, &
\eeqa
(\(\left\| x \right\| := \sqrt{\left( x^0 \right)^2 + \left| \mbf{x} \right|^2}\))
following from
\beqa\label{nnnnn}
&
\cos 2 \pi \czeta_k =
\Txfrac{1 \hspace{-1pt} + \hspace{-1pt}
\left(\raisebox{12pt}{\hspace{-3pt}}\right.
\Txfrac{x_k \mgvspc{-3pt}}{\mgvspc{8pt} 2R}
\left.\raisebox{12pt}{\hspace{-3pt}}\right)^2
\mgvspc{-9pt}}{\mgvspc{9pt} 2
\left| \omega_k \right|}
, \quad
\sin 2 \pi \czeta_k =
\Txfrac{x^0_k \mgvspc{-4pt}}{\mgvspc{9pt} 2 R \left| \omega_k \right|}
, \quad
\mbf{u} =
\Txfrac{\mbf{x}_k \mgvspc{-4pt}}{\mgvspc{9pt} 2 R \left| \omega_k \right|}
, \quad
u_4 =
\Txfrac{1 \hspace{-1pt} - \hspace{-1pt}
\left(\raisebox{12pt}{\hspace{-3pt}}\right.
\Txfrac{x_k \mgvspc{-3pt}}{\mgvspc{8pt} 2R}
\left.\raisebox{12pt}{\hspace{-3pt}}\right)^2
\mgvspc{-9pt}}{\mgvspc{9pt} 2 R \left| \omega_k \right|}
, & \nn &
4 \sin^2 \pi \alpha =
\left( u_1 - u_2 \right)^2 =
\Txfrac{\left| \mbf{x}_{12} \right|^2 \mgvspc{-4pt}}{\mgvspc{9pt} R^2}
\left(\raisebox{10pt}{\hspace{-2pt}}\right.
1 +
O \! \left(\raisebox{10pt}{\hspace{-2pt}}\right.
\Txfrac{\left\| x_1 \right\|^2 \hspace{-2pt} +
\hspace{-1pt} \left\| x_2 \right\|^2 \mgvspc{-4pt}}{\mgvspc{9pt} R^2}
\left.\raisebox{10pt}{\hspace{-2pt}}\right)
\left.\raisebox{10pt}{\hspace{-2pt}}\right)
. & \nonumber
\eeqa
To evaluate the small $\tau_R$ (large $R$) limit of
the difference of $p_1$--functions in~(\ref{e7.22})
we use (\ref{eqnA.19n}), (\ref{nA.11}) and (\ref{ad_a2})
to deduce
\beq\label{e7.24}
p_1 \left( \zeta, \tau \right) = \frac{1}{\tau}
\left(\raisebox{10pt}{\hspace{-2pt}}\right.
p_1
\left(\raisebox{10pt}{\hspace{-2pt}}\right.
\frac{\zeta}{\tau}, \frac{-1}{\tau}
\left.\raisebox{10pt}{\hspace{-2pt}}\right)
- 2 \pi i \zeta
\left.\raisebox{10pt}{\hspace{-2pt}}\right)
\, . \
\eeq
Eq.~(\ref{e7.24}) implies, on the other hand, that
\beq\label{e7.25}
p_1 \!
\left(\raisebox{10pt}{\hspace{-2pt}}\right.
\frac{\czeta_{12} \hspace{-1pt} \pm \hspace{-1pt} \alpha}{\tau_R},
\frac{-1}{\tau_R}
\left.\raisebox{10pt}{\hspace{-2pt}}\right)
\mathop{\approx}\limits_{R \to \infty}
p_1 \!
\left(\raisebox{10pt}{\hspace{-2pt}}\right.
\frac{x_{12}^0 \hspace{-1pt} \pm \hspace{-1pt}
\left| x_{12} \right|}{i \beta},
\frac{i 2 \pi R}{\beta}
\left.\raisebox{10pt}{\hspace{-2pt}}\right)
\mathop{\longrightarrow}\limits_{R \to \infty}
\pi i \,
\coth \!
\left(\raisebox{10pt}{\hspace{-3pt}}\right.
\pi \,
\frac{x_{12}^0 \hspace{-1pt} \pm \hspace{-1pt}
\left| x_{12} \right|}{\beta}
\left.\raisebox{10pt}{\hspace{-2pt}}\right)
.
\eeq
Inserting~(\ref{e7.23})--(\ref{e7.25}) into (\ref{e7.22})
we complete the proof of~(\ref{e7.20n}) and hence of Proposition~\ref{pr:7.3}.

\begin{remark}
\rmlabel{rm:7.4}
The physical thermal correlation functions should be, in fact,
defined as distributions which amounts to giving
integration rules around the poles.
To do this one should view~(\ref{e7.20n}) as a boundary
value of an analytic function in $x_{12}$ for
\(x_{12}^0 \to x_{12}^0 - i \varepsilon\), \(\varepsilon > 0\),
\(\varepsilon \to 0\) (cf.~(\ref{ea5.17})).
It is not difficult to demonstrate that the limit \(\varepsilon \to +0\)
and \(R \to \infty\) in (\ref{e7.19n}) commute.
Using (\ref{nnnnn}) we can also compute the $\Txfrac{1}{R\beta}$
correction to~(\ref{e7.20n}):
\beq\label{e7.27}
\La \hspace{-1pt}
\varphi^{\MINK} \! \left( x_1 \right) \varphi^{\MINK} \! \left( x_2 \right)
\hspace{-1pt} \Ra_{\hspace{-1pt} \beta, R}
\ \, \mathop{\approx}\limits_{R \mgrt \beta} \ \,
\La \hspace{-1pt}
\varphi^{\MINK} \! \left( x_1 \right) \varphi^{\MINK} \! \left( x_2 \right)
\hspace{-1pt} \Ra_{\hspace{-1pt} \beta, \infty}
- \,
\frac{1}{4 \pi^2 \beta R}
.
\eeq
\end{remark}

To obtain the Fourier expansion of the result
we combine Eqs.~(\ref{e7.22}) (\ref{e7.23}) with the
$q$--series (\ref{n5.145n}) and set
(as in Remark~\ref{rm:7.2})
\beq\label{e7.28}
\frac{n}{R} \, = \, p
\, , \quad
\mathop{\sum}\limits_{n \, = \, 1}^{\infty} \,
\frac{1}{R} \, f \!
\left(\raisebox{10pt}{\hspace{-2pt}}\right.
\frac{n}{R} ;\, x, \beta
\left.\raisebox{10pt}{\hspace{-2pt}}\right)
\ \mathop{\longrightarrow}\limits_{R \to \infty} \
\mathop{\int}\limits_{\hspace{-7pt} 0}^{\hspace{7pt} \infty} \!
f \! \left( p; x, \beta \right) \, dp
. \
\eeq
The result is
\beqa\label{e7.29}
&
\left( 2\pi \right)^2
\La \hspace{-1pt}
\varphi^{\MINK} \! \left( x_1 \right) \varphi^{\MINK} \! \left( x_2 \right)
\hspace{-1pt} \Ra_{\hspace{-1pt} \beta, \infty}
\, = \,
\Txfrac{1 \mgvspc{-3pt}}{\mgvspc{9pt}
x_{12}^{\, 2} + i \hspace{1pt} 0 x_{12}^0} \, + \,
& \nn & \, + \,
\Txfrac{2 \mgvspc{-3pt}}{\mgvspc{10pt} \left| \mbf{x}_{12} \right|} \,
\mathop{\text{\LARGE $\int$}}\limits_{\hspace{-7pt} 0}^{\hspace{7pt} \infty} \!
\Txfrac{e^{-\beta p} \mgvspc{-3pt}}{\mgvspc{10pt} 1 - e^{-\beta p}} \,
\cos \left( p x_{12}^0 \right)
\sin \left( p \left| \mbf{x}_{12} \right| \right)
\, dp
\, . \
&
\eeqa

\medskip

To conclude: the conformal compactification $\M$ of Minkowski space $M$
plays a dual role.

On one hand, it can serve as a \textit{symmetric finite box} approximation
to $M$ in the study of finite temperature equilibrium states.
In fact, any finite inverse temperature $\beta$ actually fixes a Lorentz
frame (cf.~\cite{Bu03}) so that the symmetry of a Gibbs state is described
by the $7$--parameter ``Aristotelian group'' of ($3$--dimensional) Euclidean
motions and time translations.
In the passage from $M$ to $\M$ the Euclidean group is deformed to the
(stable) compact group of $4$--dimensional rotations while the group
of time translations is compactified to $U \left( 1 \right)$.
Working throughout with the maximal ($7$--parameter) symmetry allows to
write down simple explicit formulae for both finite $R$ and the
``thermodynamic limit''.

On the other hand, taking $\M$ (and its universal cover,
\(\widetilde{M} = \R \times \Sr^{3}\)) not as an auxiliary
finite volume approximation
but as a model of a static space--time, we can view $R$
as a (large but) finite quantity and use the above discussion as
a basis for studding finite $R$ corrections to the
Minkowski space formulae.
It is a challenge from this second point of view to study the conformal
symmetry breaking by considering massive fields in $\widetilde{M}$.

\section{Concluding Remarks}\label{sec:7}

Periodicity of (observable) GCI fields in the conformal time variable
$\czeta$ suggests that their Gibbs (finite temperature) correlation functions
should be (doubly periodic) elliptic functions in the conformal time differences
with second period proportional to the (complexified) inverse  absolute
temperature.
We give arguments
(Theorem~\ref{adprp1}, Corollary~\ref{cr:3.7})
that this
is indeed the case in a GCI Wightman theory.
Explicit constructions are presented of elliptic 2-point functions of
free fields in an even number of space-time dimensions.

If a field $\psi \left( \czeta,u \right)$
of dimension $d$ and its conjugate satisfy the strong locality property
(\ref{eqn2.2}), i.e., if
\begin{equation}\label{n7.1}
\left( \cos 2\pi\czeta_{12}
- u_1 \spr u_2 \right)^N \!
\left[
\psi \! \left( \czeta_1,u_1 \right)
\psi^* \!\hspace{-1pt} \left( \czeta_2,u_2 \right)
\hspace{-1pt} - \hspace{-1pt} (-1)^{2d}
\psi^* \!\hspace{-1pt} \left( \czeta_2,u_2 \right)
\psi \! \left( \czeta_1,u_1 \right) \right]
\hspace{-1pt} = \hspace{-1pt} 0
\hspace{4pt}
\end{equation}
for \(N \geqslant N_{\psi}\)
then the Gibbs 2-point function
$\La \psi \left( \czeta_1,u_1 \right) \psi^*\left( \czeta_2,u_2 \right)\Ra_q$
has exactly two poles in a fundamental domain, centred around the origin
of the $\czeta_{12}$ plane, of leading order $N_{\psi}$ at the points
\begin{equation}\label{n7.2}
\czeta_{12} \, = \, \pm \alpha
\quad \text{for} \quad u_1 \spr u_2 \, = \,
\cos \, 2\pi\alpha
\, , \quad
0 \leqslant \alpha < \frac{1}{2}
.
\end{equation}
For a rank $\ell$ symmetric tensor field $\psi$ of dimension $d$
the integer $N_{\psi}$ coincides with \(d+\ell\);
for an irreducible spin-tensor field in \(D=4\), of
\(S(U(2)\times U(2))\)--weight
\(\left( d;\, j_1,j_2 \right)\), we have \(N_{\psi} = d+j_1+j_2\).

The conformal energy mean value in an equilibrium Gibbs state (with
suitably shif\-t\-ed vacuum energy) appears as a superposition of modular
forms of different weights. Postulating this property for the photon
energy (associated with the Maxwell stress tensor $F$) requires including
(non-physical) gauge degrees of freedom
(otherwise, the non--modular term $-2G_2\left( \tau \right)$
contributes to~(\ref{eqHFM2})).
The result is then a modular
form of weight 4 (Sect.~\ref{Ssec.5.2}, Eq.~(\ref{7.40})).
The same is true for the free
massless scalar field for \(D=4\), while the energy mean of a
\(d=\Txfrac{3}{2}\) Weyl field is a superposition
of modular forms (\ref{eqn6.26}) of weight 4 and 2 (and level
$\Gamma_{\theta}$~--~see Appendix~A, Eq.~(\ref{ad_a3}) and the text following it).
The question arises whether relaxing the condition of Wightman positivity one
cannot find an (indefinite metric) interacting Weyl field model whose
energy mean value is a (homogeneous) modular form of weight four (as
suggested by the study of chiral conformal models in \(1+1\) space--time
dimension). More generally, the role of modular invariance in higher
dimensional conformal field models still awaits its full
understanding.

\section*{Acknowledgments}

The authors' interest in the modular properties of energy
distributions in higher dimensional conformal field theory was
stimulated
by an early suggestion of Maxim Kontsevitch.
Discussions with Petko Nikolov are also gratefully acknowledged.
We thank Seif Randjbar-Daemi and the Abdus Salam International
Centre for Theoretical Physics in Trieste for invitation and support during
a late stage of this work.
Discussions with Detlev Buchholz in G\"ottingen led to including
the addition of the present Sect.~\ref{Sect.7n}.
We acknowledge partial support of the Alexander von Humboldt Foundation
and the hospitality of the Institut f\"ur Theoretische Physik der Universit\"at
G\"ottingen
during the final stage of this work.
Our work is supported in part by the Research Training Network within the
Framework Programme 5 of the European Commission under contract HPRN-CT-2002-00325
and by the Bulgarian National Council for Scientific Research under contract PH-1406.

\appendix

\addtocounter{section}{1}
\renewcommand{\thesection}{\Alph{section}}
\section*{Appendix A. \ Basic Elliptic Functions}\label{ap.A}
\addcontentsline{toc}{section}{Appendix A. \ Basic Elliptic Functions}

In this Appendix we define the basic elliptic functions
used in the paper and list their properties and relations
with the conventional functions.

Recall that an elliptic function $f \! \left( \zeta \right)$ is
a meromorpic function on \(\C \, ( \ni \zeta)\) which is doubly periodic.
Its periods can be chosen (after rescalling by a nonzero complex constant)
to be $1$ and $\tau$ with
\(
\tau \in \hcom
\, ( \, :=
\left\{ \tau' \in \C \, : \, \mathit{Im} \, \tau' > 0 \right\} )
\).
Thus, \(f \! \left( \zeta \right) = f \! \left( \zeta + m + n\tau \right)\)
for \(m,\, n \in \Z\) and hence, $f$ is completely determined by its values
in the \textit{fundamental domain}
$\fundom$ $:=$
\(\left\{\raisebox{9pt}{\hspace{-3pt}}\right. \zeta \in \C :\)
\(\zeta =\) \(\lambda +\) \(\mu \tau,\)
$0$ $\leqslant$ \(\lambda,\mu\) $<$
\(1 \left.\raisebox{9pt}{\hspace{-3pt}}\right\}\).
By the Liouville's theorem, $f \! \left( \zeta \right)$ should have
at least one pole in $\fundom$ if it is nonconstant:
otherwise it will be bounded nonconstant entire function in $\zeta$
which is not possible.
Integrating $f \! \left( \zeta \right)$ and
$\Txfrac{f' \! \left( \zeta \right)}{f \! \left( \zeta \right)}$
over the boundary \(\di \fundom\)
(or, over the shifted \(\di \fundom + c\), if necessary)
we conclude in addition (by the Cauchy theorem on one hand,
and the double periodicity, on the other) that:
(\textit{i}) the sum of the residues of the simple poles
of $f$ lying in $\fundom$ is zero and,
(\textit{ii}) the sum of multiplicities of all zeros minus
the sum of multiplicities of all poles of $f$ in $\fundom$
is also zero.
In particular, $f$ cannot have just one simple pole in $\fundom$.
Therefore, if the singular part of $f$ in $\fundom$ has the form:
\beq\label{f_bas0}
\Su_{k \, = \, 1}^{K} \Su_{s \, = \, 1}^{S} \, N_{k,s} \
\frac{1}{\left( \zeta-\zeta_s \right)^k}
\, \
\eeq
for some \(K,S \in \N\), \(N,N_{s,k} \in \C\), \(\zeta_s \in \fundom\)
(\(k = 1,\dots, K\), \(s=1,\dots,S\)),
then $f$ can be represented in a finite sum:
\beq\label{f_bas}
f \left( \zeta \right) = N \, + \,
\Su_{k \, = \, 1}^{K} \Su_{s \, = \, 1}^{S} \, N_{k,s} \
p_k \! \left( \zeta-\zeta_s, \tau \right)
\, \
\eeq
where $p_k \! \left( \zeta,\tau \right)$ are, roughly speaking, equal to:
\beq\label{p_rough}
p_k \left( \zeta, \tau \right) \, := \,
\Su_{m,n \in \Z} \frac{1}{\left( \zeta+m+n\tau \right)^k}
\, . \
\eeq
The series~(\ref{p_rough}) are absolutely convergent for \(k \geqslant 3\)
and
\beq\label{p_k}
p_{k+1} \! \left( \zeta,\tau \right) \, = \, - \frac{1}{k}
\left( \di_{\zeta} p_k \right) \left( \zeta,\tau \right)
\, . \
\eeq
For \(k = 1,2\) one should specialize the order of summation or, alternatively,
add regularizing terms.
In such a way we arrive at the standard Weierstrass functions~\cite{FNTP}:
\beqa
\label{fn_W2}
\hspace{-5pt}
\zfun \left( \zeta,\, \tau \right)
\, = && \podr
\frac{1}{\zeta} +
\mathop{\sum}\limits_{
\mathop{}\limits^{\left( m,n \right) \, \in}_{
\in \, \Z^2 \backslash \left\{ \left( 0,\, 0 \right) \right\}}}
\left[ \frac{1}{\zeta + m \tau + n} -
\frac{1}{m \tau + n} +
\frac{\zeta}{\left( m \tau + n \right)^2}
\right]
\,
,
\\
\label{fn_W1}
\hspace{-5pt}
\pfun \left( \zeta,\, \tau \right)
\, = && \podr
\frac{1}{\zeta^2} +
\mathop{\sum}\limits_{
\mathop{}\limits^{\left( m,n \right) \, \in}_{
\in \, \Z^2 \backslash \left\{ \left( 0,\, 0 \right) \right\}}}
\left[ \frac{1}{\left( \zeta + m \tau + n \right)^2} -
\frac{1}{\left( m \tau + n \right)^2} \right]
.
\eeqa
Thus $\zfun \! \left( \zeta,\tau \right)$ and $\pfun \! \left( \zeta,\tau \right)$
are odd and even meromorphic functions in $\zeta$, respectively, and
\begin{equation}\label{fn_cn1}
\left( \partial_{\zeta} \zfun \right) \left( \zeta,\, \tau \right) \, = \,
-
\pfun \left( \zeta,\, \tau \right)
\, , \quad
\left( \partial_{\zeta} \pfun \right) \left( \zeta,\tau \right) \, = \, - 2 \, p_3 \! \left( \zeta,\tau \right)
\, . \
\end{equation}
Since $p_3 \left( \zeta,\tau \right)$ is elliptic it then follows that
$\pfun \! \left( \zeta,\tau \right)$ is also elliptic.
The function $\zfun \! \left( \zeta,\tau \right)$ cannot be elliptic
(by the property (\textit{i}) above) and, in fact,
\begin{eqnarray}\label{fn_cn2}
\zfun \left( \zeta+1,\, \tau \right)
\, = && \podr
\zfun \left( \zeta,\, \tau \right)
- 8 \hspace{1pt} \pi^2 \hspace{1pt} G_2 \! \left( \tau \right)
\, , \quad
\\ \label{fn_cn3}
\zfun \left( \zeta+\tau,\, \tau \right)
\, = && \podr
\zfun \left( \zeta,\, \tau \right)
- 8 \hspace{1pt} \pi^2 \hspace{1pt} G_2 \! \left( \tau \right) \hspace{1pt} \tau
- 2 \pi i
\,
\end{eqnarray}
where
\beq\label{fn_G}
G_{2k} \left( \tau \right)
\, = \,
\frac{\left( 2k-1 \right)!}{2\left( 2\pi i \right)^{2k}} \,
\left\{\raisebox{14pt}{\hspace{-2pt}}\right.
\mathop{\sum}\limits_{n \, \in \, \Z \backslash \left\{ 0 \right\}}
\frac{1}{n^{2k}} \, +
\mathop{\sum}\limits_{m \, \in \, \Z \backslash \left\{ 0 \right\}} \,
\mathop{\sum}\limits_{n \, \in \, \Z}
\frac{1}{\left( m \tau + n \right)^{2k}}
\left.\raisebox{14pt}{\hspace{-2pt}}\right\}
\,
\eeq
(\(k = 1,2,\dots\))
are the $G$--modular functions also playing central role in this work.
Hence, $\zfun \! \left( \zeta,\tau \right)$ and $\pfun \! \left( \zeta,\tau \right)$ are
possible candidates for $p_1$ and $p_2$ and they are indeed used as basic functions
in~\cite{Zh96}.
As we have explained in the introduction we prefer to work with (anti)periodic function in $\zeta$
with period $1$ and on the other hand, to preserve the relation~(\ref{p_k})
for all \(k \in \N\)
so that this naturally fixes
\beqa\label{eqnA.19n}
& p_1 \hspace{-1pt} \left( \zeta, \tau \right) & := \,
\zfun \left( \zeta, \tau \right)
+
8 \pi^2 G_2 \! \left( \tau \right) \zeta
, \ \
\\ \label{eqnA.21a}
& p_2 \hspace{-1pt} \left( \zeta,\, \tau \right) & := \,
\pfun \left( \zeta,\, \tau \right)
-
8 \pi^2 G_2 \! \left( \tau \right)
\, . \
\eeqa

For \(k>1\), the above introduced $G_{2k} \left( \tau \right)$ are
\textit{modular forms of weight} $2k$ (and level 1):
\begin{equation}\label{ad_a1}
\frac{1}{\left( c\tau \hspace{-1pt} + \hspace{-1pt} d \right)^{2k}} \,
G_{2k} \left(\raisebox{10pt}{\hspace{-2pt}}\right.
\frac{a\tau+b}{c\tau+d}
\left.\raisebox{10pt}{\hspace{-2pt}}\right) \, = \,
G_{2k} \left( \tau \right)
\quad \text{for} \quad
\left(\hspace{-1pt}\begin{array}{cc}
a & b \\
c & d
\end{array}\hspace{-1pt}\right) \in \mathit{SL} \left( 2,\, \Z \right)
\, , \
\end{equation}
while for \(k=1\) we have instead
\begin{equation}\label{ad_a2}
\frac{1}{\left( c\tau \hspace{-1pt} + \hspace{-1pt} d \right)^2} \,
G_{2} \left(\raisebox{10pt}{\hspace{-2pt}}\right.
\frac{a\tau+b}{c\tau+d}
\left.\raisebox{10pt}{\hspace{-2pt}}\right) \, = \,
G_{2} \left( \tau \right)
\, + \,
\frac{i \hspace{1pt} c}{
4\hspace{1pt}\pi\left( c\tau \hspace{-1pt} + \hspace{-1pt} d \right)}
\, . \
\end{equation}
In applications to CFT there appear more general modular forms like
\begin{equation}\label{ad_a3}
F(\tau ) \, := \, 2 \, G_{2}(\tau )-G_{2}
\left(\raisebox{10pt}{\hspace{-2pt}}\right.
\frac{\tau+1}{2}
\left.\raisebox{10pt}{\hspace{-2pt}}\right)
\, \
\end{equation}
which is invariant under the index $2$ subgroup $\Gamma_{\theta}$ of
$\mathit{SL} \left( 2,\, \Z \right)$ generated by $S$ and $T^{2}$
where
$S$ is given by (\ref{eq1.1}) and
\(
T=
\left(\hspace{-1pt}\begin{array}{cc}
1 & 1 \\
0 & 1
\end{array}\hspace{-1pt}\right)
\).
We note that the normalization factor in the definition of the modular forms
$G_{2k} \left( \tau \right)$~(\ref{fn_G}) is chosen so that
the coefficient to $q$ in their Fourier expansion is 1.
Then one finds that all Fourier coefficients (except the constant term)
are positive integers:
\begin{equation}\label{FTG}
G_{2k} \left( \tau \right)
\, = \,
-\frac{B_{2k}}{4k} +
\mathop{\sum}\limits_{n \, = \, 1}^{\infty} \,
\frac{n^{2k-1}}{1-q^{n}} \, q^{n}
\, = \,
\frac{1}{2} \, \zeta \left( 1 \! - \! 2k \right)
+ \mathop{\sum}\limits_{n \, = \, 1}^{\infty} \,
\sigma_{2k-1} \! \left( n \right) \, q^{n}
\, \
\end{equation}
where \(\sigma _{l} \! \left( n \right)
=\mathop{\sum}\limits_{r \mid \hspace{1pt} n}r^{l}\)
(sum over all positive divisors $r$ of $n$),
$B_{l}$ are the Bernoulli numbers,
and \(\zeta \left( s \right)\) is the Riemann $\zeta$--function.

Let us mention also the modular transformation properties of
the Weierstrass functions~(\ref{fn_W1}) and~(\ref{fn_W2})
\begin{eqnarray}\label{nA.11}
\frac{1}{c\tau \hspace{-1pt} + \hspace{-1pt} d} \,
\zfun \left(\raisebox{10pt}{\hspace{-2pt}}\right.
\frac{\zeta}{c\tau+d},\, \frac{a\tau+b}{c\tau+d}
\left.\raisebox{10pt}{\hspace{-2pt}}\right) \, = && \podr
\zfun \left( \zeta, \tau \right)
\, , \quad
\\ \label{nA.12}
\frac{1}{\left( c\tau \hspace{-1pt} + \hspace{-1pt} d \right)^2} \,
\pfun \left(\raisebox{10pt}{\hspace{-2pt}}\right.
\frac{\zeta}{c\tau+d},\, \frac{a\tau+b}{c\tau+d}
\left.\raisebox{10pt}{\hspace{-2pt}}\right) \, = && \podr
\pfun \left( \zeta, \tau \right)
\, . \
\end{eqnarray}
Thus, our $p_1 \! \left( \zeta,\tau \right)$~(\ref{nA.11}) and
$p_2 \! \left( \zeta,\tau \right)$~(\ref{nA.12}) will
obey inhomogeneous modular transformation laws (as in the
example of Eq.~(\ref{e7.24})). (This is the price for
preserving the periodicity property for \(\zeta \mapsto \zeta+1\).)

We will use also the Jacobi \(\vartheta\)--functions
\cite{M83} \cite{Y97}:
\begin{eqnarray}
\label{fn.3}
\vartheta \left( \zeta,\, \tau \right) \, := && \podr
\mathop{\sum}\limits_{n \, = \, -\infty}^{\infty} \hspace{2pt}
e^{\hspace{1pt}
\pi \hspace{1pt} i \hspace{0.5pt}
\hspace{1pt} \left( n^2 \hspace{1pt} \tau + 2 \hspace{1pt} n \hspace{1pt} \zeta \right)
}
\, \equiv \,
\vartheta_{00} \left( \zeta,\,\tau \right)
\, , \
\\
\label{fn.2}
\vartheta_{\lambda\kappa} \left( \zeta,\,\tau \right) \, := && \podr
e^{\hspace{1pt}
\pi \hspace{0.5pt} i \hspace{0.5pt} \tau
\frac{\raisebox{0pt}{\small \(\lambda^2\)}}{\raisebox{0pt}{\small \(4\)}}
+
\pi \hspace{0.5pt} i \hspace{0.5pt} \lambda
\left( \zeta +
\frac{\raisebox{0pt}{\small \(\kappa\)}}{\raisebox{0pt}{\small \(2\)}} \right)
}\,
\vartheta \left(\raisebox{10pt}{\hspace{-2pt}}\right.
\zeta + \frac{\lambda \hspace{1pt} \tau \hspace{-1pt} + \hspace{-1pt} \kappa}{2}
,\, \tau
\left.\raisebox{10pt}{\hspace{-2pt}}\right)
\end{eqnarray}
for \(\kappa,\, \lambda = 0,\, 1\),
which have the following properties
(for \(\kappa,\, \lambda = 0,\, 1\)):
\begin{eqnarray}\label{fn.4}
\vartheta_{\lambda\kappa} \left( \zeta + m \tau + n,\, \tau \right)
\, = && \podr
\left( -1 \right)^{m \hspace{0.5pt} \kappa + n \hspace{0.5pt} \lambda}
e^{\hspace{1pt} - \hspace{1pt} \pi \hspace{1pt} i \hspace{1pt}
\left(
m^2 \hspace{1pt} \tau \hspace{1pt} + 2 \hspace{1pt} m \hspace{1pt} \zeta
\right)}
\,
\vartheta_{\lambda\kappa} \left( \zeta,\, \tau \right)
\, , \quad
\\ \label{fn.5}
\vartheta_{\lambda\kappa} \left( -\zeta,\, \tau \right)
\, = && \podr
\left( -1 \right)^{\lambda\kappa} \,
\vartheta_{\lambda\kappa} \left( \zeta,\, \tau \right)
\, , \quad
\\ \label{fn.6}
\vartheta_{\lambda\kappa} \left( \zeta,\, \tau \right)
\, = && \podr 0
\quad \Leftrightarrow \quad
\zeta \, \in \,
\left(\raisebox{10pt}{\hspace{-2pt}}\right.
\Z + \frac{1 \hspace{-1pt} - \hspace{-1pt} \lambda}{2}
\left.\raisebox{10pt}{\hspace{-2pt}}\right)
\hspace{0.5pt} \tau +
\Z + \frac{1 \hspace{-1pt} - \hspace{-1pt} \kappa}{2}
\, \quad
\end{eqnarray}
(Eqs.~(\ref{fn.4}) and~(\ref{fn.5}) are first proven for the series~(\ref{fn.3})
and then for the other functions~(\ref{fn.2}); Eq.~(\ref{fn.6})
follows from Eqs.~(\ref{fn.5}) and~(\ref{fn.2}) since the first one means that
\(\vartheta_{11} \left( \zeta,\, \tau \right)\) is odd in $\zeta$.)
We are using in Sect.~\ref{mssec:3.2p} the fact that
the odd $\vartheta$--function, $\vartheta_{11}$, can be written in the form
\beq\label{eqn_new}
\vartheta_{11} \! \left( \zeta,\tau \right) \, = \,
-2 \, \Su_{n \, = \, 0}^{\infty} \,
\left( -1 \right)^n e^{i \pi \tau \, \left( n + \smtxfrac{1}{2} \right)^2} \,
\sin \left( 2n \! + \! 1 \right) \!\pi \zeta
\, . \
\eeq

Returning to our set
$\left\{ p_k \! \left( \zeta,\tau \right) \right\}$
of basic elliptic functions
we can rewrite (the qausielliptic)
$p_1 \! \left( \zeta, \tau \right)$ as
\beqa\label{motive1}
p_1 \! \left( \zeta,\tau \right) = && \podr \! \!
\mathop{\lim}\limits_{N \, \to \, \infty}
\mathop{\sum}\limits_{n \, = \, -N}^N
\pi \,
\mathrm{cotg} \left[\raisebox{9pt}{\hspace{-2pt}}\right. \pi \hspace{-1pt}
\left(\raisebox{9pt}{\hspace{-3pt}}\right.
\zeta \hspace{-2pt} + \hspace{-1pt} n \hspace{1pt} \tau
\left.\raisebox{9pt}{\hspace{-2pt}}\right)
\left.\raisebox{9pt}{\hspace{-3pt}}\right]
= \!
\nn = && \podr \! \!
\mathop{\sum}\limits_{n \, = \, -\infty}^{\infty}
\left\{
\pi \,
\mathrm{cotg} \left[\raisebox{9pt}{\hspace{-2pt}}\right. \pi \hspace{-1pt}
\left(\raisebox{9pt}{\hspace{-3pt}}\right.
\zeta \hspace{-2pt} + \hspace{-1pt} n \hspace{1pt} \tau
\left.\raisebox{9pt}{\hspace{-2pt}}\right)
\left.\raisebox{9pt}{\hspace{-3pt}}\right]
\hspace{-1pt} + \hspace{-1pt} i \pi \
\mathrm{sgn} \hspace{-1pt} \left( n \right)
\right\} \! =
\\ \label{motive2}
= && \podr \! \!
\mathop{\lim}\limits_{M \to \infty} \mathop{\lim}\limits_{N \to \infty} \,
\mathop{\sum}\limits_{m = -M}^{M} \, \mathop{\sum}\limits_{n = -N}^{N} \,
\frac{1}{\zeta+m+n\tau}
,
\eeqa
where \(\mathrm{sgn} \left( n \right) :=
\Txfrac{|n|}{n}\)
for \(n \neq 0\) and \(\mathrm{sgn} \left( 0 \right) := 0\).
Indeed,
first note that the second sum in Eq.~(\ref{motive1}) is absolutely convergent
since the absolute value of the summand
has a behaviour as
\(
e^{\hspace{1pt} - \pi
\left| n \right|
\hspace{1pt} \mathit{Im} \, \tau
}\).
Then Eq.~(\ref{motive2}) follows from Euler's identity:
\beq\label{Eull}
\frac{\pi
\cos^{\hspace{1pt}1-\lambda} \hspace{1pt} \pi \zeta}{
\sin \hspace{1pt} \pi \zeta}
\, = \,
\mathop{\lim}\limits_{N \, \to \, \infty} \
\mathop{\sum}\limits_{n \, = \, -N}^N \,
\frac{\left( -1 \right)^{n\hspace{0.5pt}\lambda}}{\zeta + n}
\quad (\lambda = 0,\, 1)
\, . \
\eeq
Finally, to obtain the first Eq.~(\ref{motive1}) we take the difference
between both sides and observe that it is an elliptic function in $\zeta$,
in accord to Eqs.~(\ref{fn_cn3}) and~(\ref{eqnA.19n}).
On the other hand, this difference is regular in $\zeta$
in the fundamental domain $\fundom$,
because of Eqs.~(\ref{fn_W2}) and~(\ref{motive2}),
so that it is a constant which is actually zero since it is obviously
an odd function in $\zeta$.

Eq.~(\ref{motive1}) is closely related to the general form of the elliptic correlation functions
arising in the free field GCI models according to
Theorem~\ref{thr:4.1}
(see Eq.~(\ref{eqn4.14a})).
In view of the more general situation of the ``grand canonical'' corelation functions
in
Remark~\ref{rm:4.1}
(Eq.~(\ref{eqn4.14b})) we are led to introduce for
\(\kappa,\, \lambda = 0,\, 1\), \(\tau \in \hcom\),
\(\zeta \in \C \,\backslash\! \left( \Z \tau + \Z \right)\) and \(\mu \in \R\)
\begin{equation}\label{fn.0}
p_1^{\kappa, \lambda} \hspace{-2pt} \left( \zeta,\tau,\mu \right)
\hspace{-1pt} = \hspace{-2pt}
\mathop{\sum}\limits_{n \hspace{1pt} = \hspace{1pt} -\infty}^{\infty}
\hspace{-1pt}
\left[ \hspace{0pt}
\frac{\pi
\cos^{\hspace{1pt}1-\lambda} \hspace{-2pt}
\left[\raisebox{9pt}{\hspace{-2pt}}\right. \pi \hspace{-1pt}
\left(\raisebox{9pt}{\hspace{-3pt}}\right.
\zeta \hspace{-2pt} + \hspace{-1pt} n \hspace{1pt} \tau
\left.\raisebox{9pt}{\hspace{-2pt}}\right)
\left.\raisebox{9pt}{\hspace{-3pt}}\right]
}{
\sin \left[\raisebox{9pt}{\hspace{-2pt}}\right. \pi \hspace{-1pt}
\left(\raisebox{9pt}{\hspace{-3pt}}\right.
\zeta \hspace{-2pt} + \hspace{-1pt} n \hspace{1pt} \tau
\left.\raisebox{9pt}{\hspace{-2pt}}\right)
\left.\raisebox{9pt}{\hspace{-3pt}}\right]
}
\hspace{-1pt} + \hspace{-1pt} i \pi
\left( 1 \hspace{-2pt} - \hspace{-2pt} \lambda \right)
\mathrm{sgn} \hspace{-1pt} \left( n
\right)
\hspace{0pt} \right] \hspace{-1pt}
e^{\hspace{1pt} \pi i \hspace{1pt} n \hspace{0.5pt}
\left( 2 \mu + \kappa \right)}
\! . \hspace{4pt}
\end{equation}
For \(\left| n \right| \mgrt 0\) the absolute value of the summand in
the above series will have a behaviour as
\(
e^{\hspace{1pt} - \pi
\left| n \right|
\hspace{1pt} \mathit{Im} \, \tau
}\)\gvspc{-6pt}
and therefore, the series is convergent
for every \(\zeta \in \C \,\backslash\! \left( \Z \tau + \Z \right)\),
\(\mu \in \R\) and
\(\tau \in \hcom\).
It then follows that
\begin{eqnarray}\label{fn_14}
p_1^{\kappa,\lambda} \left( \zeta+\tau,\, \tau,\, \mu \right)
\, = && \podr
\left( -1 \right)^{\kappa} \,
e^{\hspace{1pt} - 2 \hspace{0.5pt} \pi \hspace{0.5pt} i \hspace{1pt} \mu} \,
p_1^{\kappa,\lambda} \left( \zeta,\, \tau,\, \mu \right)
-
\pi i
\left( 1 \hspace{-1pt} - \hspace{-1pt} \lambda \right)
\hspace{-1pt} \left(\raisebox{9pt}{\hspace{-3pt}}\right.
1 \! + e^{- \pi i \hspace{0.5pt}
\left( 2 \mu + \kappa \right)} \left.\raisebox{9pt}{\hspace{-3pt}}\right)
,
\nonumber \\
p_1^{\kappa,\lambda} \left( \zeta+1,\, \tau,\, \mu \right)
\, = && \podr
\left( -1 \right)^{\lambda}
p_1^{\kappa,\lambda} \left( \zeta,\, \tau,\, \mu \right)
\, . \quad
\end{eqnarray}
In the case of \(\kappa=\lambda=0\) we will simplify the notation setting
\begin{equation}\label{eqnA.15nn}
p_1 \left( \zeta,\, \tau,\, \mu \right) \, := \,
p_1^{00} \left( \zeta,\, \tau,\, \mu \right)
\, . \
\end{equation}

\begin{proposition}
\prlabel{pr:A.1}
The functions $p_1^{\kappa,\lambda} \left( \zeta, \tau, \mu \right)$~(\ref{fn.0})
(\(\kappa,\lambda = 0,1\))
have an analytic extension to me\-ro\-mor\-phic functions in
\(\left( \zeta,\, \tau,\, \mu \right) \in \C \times \hcom \times \C\)
given for \(\mu + \txfrac{\kappa}{2} \in \R \, \backslash \, \Z\), by
\begin{equation}\label{eqnA.18n}
p_1^{\kappa,\lambda}\hspace{-1pt} \left( \zeta,\,\tau,\,\mu \right)
\hspace{-1pt} = \hspace{-1pt}
\frac{\left( \partial_{\zeta} \, \vartheta_{11} \right) \! \left( 0, \tau \right)}{
\vartheta_{1-\lambda \hspace{1pt} 1-\kappa} \hspace{-2pt} \left( \mu, \tau \right)}
\,
\frac{\vartheta_{1-\lambda \hspace{1pt} 1-\kappa} \hspace{-2pt}
\left( \zeta\! +\! \mu, \tau \right)}{
\vartheta_{11} \hspace{-2pt} \left( \zeta, \tau \right)}
\hspace{0pt} - \hspace{0pt}
\left( 1\! -\! \lambda \right) \pi \hspace{2pt}
\mathrm{cotg} \, \pi \hspace{-2pt}
\left(\raisebox{10pt}{\hspace{-3pt}}\right.
\mu \! + \hspace{-2pt} \frac{\kappa}{2}
\left.\raisebox{10pt}{\hspace{-2pt}}\right)
. \hspace{4pt}
\end{equation}
They are regular for all \(\mu \in \R\) and
\begin{equation}\label{eqnA.19nn}
p_1 \hspace{-1pt} \left( \zeta, \tau, 0 \right)
= \frac{\left( \partial_{\zeta} \, \vartheta_{11} \right)
\! \left( \zeta, \tau \right)}{
\vartheta_{11} \! \left( \zeta, \tau \right)} \equiv
p_1 \hspace{-1pt} \left( \zeta, \tau \right)
,
\end{equation}
$p_1 \! \left( \zeta,\tau \right)$ being defined by Eq.~(\ref{eqnA.19n}).
\end{proposition}

\begin{proof}
Let \(\mu \in \R\) and take the difference
\(\Delta \left( \zeta,\, \tau,\, \mu \right)\) between the left and right hand sides
of Eq.~(\ref{eqnA.18n}).
From the properties~(\ref{fn.4})--(\ref{fn.6}) and~(\ref{fn_14}) we find that
\begin{equation}\label{fn.7}
\Delta \left( \zeta + m \tau + n,\, \tau,\, \mu \right)
=
\left( -1 \right)^{m \hspace{0.5pt} \kappa + n \hspace{0.5pt} \lambda} \,
e^{\hspace{1pt} - 2 \hspace{0.5pt} \pi \hspace{0.5pt} i \hspace{1pt}
m \hspace{0.5pt} \mu} \,
\Delta \left( \zeta,\, \tau,\, \mu \right)
\, . \
\end{equation}
(Note that the second ratio in Eq.~(\ref{eqnA.18n}) is chosen to obey the
the quasiperiodicity property~(\ref{fn.7})
and its pole coefficient
at \(\zeta = 0\) is canceled by the first ratio.)
On the other hand, \(\Delta \left( \zeta,\tau,\mu \right)\)
is analytic in $\zeta$, for fixed $\tau$ and $\mu$, outside
the lattice \(\Z \tau + \Z \subset \C\)
and since it is also regular at the origin \(\zeta=0\), Eq.~(\ref{fn.7})
then implies that \(\Delta \left( \zeta,\tau,\mu \right)\)
is an entire bounded function in $\zeta$.
By the Liouville's theorem we conclude that
\(\Delta \left( \zeta,\tau,\mu \right)\) does not depend on $\zeta$ and
it is actually zero, again by Eq.~(\ref{fn.7}).

Eq.~(\ref{eqnA.19nn}) follows in the same way
from Eqs.~(\ref{fn_cn2}), (\ref{fn_cn3}) and (\ref{fn_14}).
(The constant here is fixed by the behaviour for \(\zeta \to 0\).)
\end{proof}

For \(k = 1, 2, \dots\) we set
\begin{eqnarray}\label{eqnA.20add}
&& \hspace{-24pt}
p_{k+1}^{\kappa,\lambda} \! \left( \zeta, \tau, \mu \right) =
-\frac{1}{k} \, \di_{\zeta} \, p_{k}^{\kappa,\lambda} \!
\left( \zeta, \tau, \mu \right)
, \quad
p_k \left( \zeta, \tau, \mu \right) := p_k^{00} \left( \zeta, \tau, \mu \right)
,
\\ \label{eqnA.20a} && \hspace{-24pt}
p_{k+1}^{\kappa,\lambda} \! \left( \zeta, \tau \right) =
p_k^{\kappa,\lambda} \left( \zeta,\, \tau,\, 0 \right) ( =
-\frac{1}{k} \, \di_{\zeta} \, p_{k}^{\kappa,\lambda} \!
\left( \zeta, \tau \right) )
, \quad
p_k \left( \zeta, \tau \right) := p_k^{00} \left( \zeta, \tau \right)
. \quad
\end{eqnarray}

\begin{proposition}
\prlabel{prp:A.2}
Every function
$p_k^{\kappa,\lambda} \left( \zeta, \tau, \mu \right)$,
for \(k=2,3,\dots\), is
uniquely characterized by the conditions:
\begin{plist}
\item[{\rm (}\textit{a}{\rm )}\hspace{2pt}]
\(p_k^{\kappa,\lambda} \left( \zeta, \tau, \mu \right)\) is a meromorphic function
in \(\left( \zeta, \tau, \mu \right) \in \C \times \hcom \times \C\)
and for real $\mu$, and for all \(\tau \in \hcom\), \(k = 1,2,\dots\),
\(\kappa,\lambda = 0,1\),
it has exactly one pole in $\zeta$ at $0$
of order $k$ and residue $1$ in the domain
\(\left\{\raisebox{9pt}{\hspace{-3pt}}\right.
\alpha\tau+\beta :\) \(\alpha, \beta \in\)
\(\left[ \hspace{1pt} 0,\, 1 \right)
\left.\raisebox{9pt}{\hspace{-3pt}}\right\} \subset \C\);\gvspc{10pt}
\item[{\rm (}\textit{b}{\rm )}\hspace{2pt}]
\(p_k^{\kappa,\lambda} \left( \zeta+1, \tau, \mu \right) =
\left( -1 \right)^{\lambda}
p_k^{\kappa,\lambda} \left( \zeta, \tau, \mu \right)\);
\item[{\rm (}\textit{c}{\rm )}\hspace{2pt}]
\(p_k^{\kappa,\lambda} \left( \zeta+\tau, \tau, \mu \right) =
e^{-\pi i \hspace{1pt} \left( 2 \hspace{1pt}\mu+\kappa \right)} \,
p_k^{\kappa,\lambda} \left( \zeta, \tau, \mu \right)\).\gvspc{10pt}
\end{plist}
It also obeys the property
\begin{plist}
\item[{\rm (}\textit{d}{\rm )}\hspace{2pt}]
\(p_k^{\kappa,\lambda} \left( -\zeta, \tau, \mu \right)
= \left( -1 \right)^k \,
p_k^{\kappa,\lambda} \left( \zeta, \tau, \mu \right)\).\gvspc{10pt}
\end{plist}
The function $p_1^{\kappa,\lambda} \left( \zeta,\tau,\mu \right)$
can be fixed by the condition (d) and the relation (\ref{eqnA.20add})
connecting it with the function
\(p_2^{\kappa,\lambda} \left( \zeta,\tau,\mu \right)\).

For \(\mu \in \R\) we have the series representation
\begin{equation}\label{eqnA.22a}
p_k^{\kappa,\lambda} \left( \zeta,\tau,\mu \right)
\, = \,
\mathop{\sum}\limits_{m,\, n \, \in \, \Z}
\,
\frac{e^{\pi i \hspace{1pt} m \hspace{1pt}
	\left( 2 \hspace{1pt}\mu+\kappa \right)} \,
e^{\pi i \hspace{1pt} n \hspace{1pt} \lambda}}{
\left( \zeta+m\hspace{1pt}\tau+n \right)^k}
\, , \
\end{equation}
which is absolutely convergent for \(k \geqslant 3\) and
\(\zeta \in \C \,\backslash\! \left( \Z \tau + \Z \right)\),
and for \(k=1,2\) the sum should be taken
first in $n$ for \(|n| \leqslant N\) as \(N \to \infty\)
and then in $m$ for \(|m| \leqslant M \to \infty\).
\end{proposition}

\begin{proof}
Clearly the of functions defined by Eqs.~(\ref{eqnA.18n}) and~(\ref{eqnA.20add})
satisfy the conditions (\textit{a})--(\textit{d})
except the case of \(k=1\) in (\textit{c}).
By the argument used in the proof of
Proposition~\ref{pr:A.1}
it follows that
(\textit{a})--(\textit{c}) uniquely determine the functions
\(p_k^{\kappa,\lambda} \left( \zeta,\tau,\mu \right)\) for \(k\geqslant 2\).
The relation~(\ref{eqnA.20add}) fixes the function
\(p_1^{\kappa,\lambda} \left( \zeta,\tau,\mu \right)\) up to
an additive constant which is determined by the condition~(\textit{d}).

The derivation of Eq.~(\ref{eqnA.22a}) is based on (\ref{fn.0}) and (\ref{Eull}).
\end{proof}

Eq.~(\ref{eqnA.22a}) implies that:
\begin{eqnarray}\label{eqnA.23a}
p_k^{01} \hspace{-1pt} \left( \zeta, \tau \right)
\, = && \podr
2^{1-k} \, p_k \hspace{-1pt} \left(\raisebox{10pt}{\hspace{-3pt}}\right.
\frac{\zeta }{2}, \frac{\tau}{2}
\left.\raisebox{10pt}{\hspace{-3pt}}\right)
-
p_k \hspace{-1pt} \left( \zeta, \tau  \right)
. \ \
\nonumber \\
p_k^{10} \hspace{-1pt} \left( \zeta, \tau \right)
\, = && \podr
p_k \hspace{-1pt} \left( \zeta, 2\tau \right)
-
p_k \hspace{-1pt} \left( \zeta,\tau  \right)
\hspace{-1pt} , \ \
\nonumber \\
p_k^{11} \hspace{-1pt} \left( \zeta, \tau \right)
\, = && \podr
2^{1-k} \, p_k \hspace{-1pt} \left(\raisebox{10pt}{\hspace{-3pt}}\right.
\frac{\zeta }{2}, \frac{\tau +1}{2}
\left.\raisebox{10pt}{\hspace{-3pt}}\right)
-
p_k \hspace{-1pt} \left( \zeta, \tau \right)
. \ \
\end{eqnarray}

\addtocounter{section}{1}
\renewcommand{\thesection}{\Alph{section}}
\section*{Appendix B. \ Proof of Proposition~\ref{pr:3.4}}\label{app:B}
\addcontentsline{toc}{section}{Appendix B. \ Proof of Proposition~\ref{pr:3.4}}
\setcounter{equation}{0}

We begin by recalling a basic fact of the theory
of formal power series

\medskip

\noindent
\textbf{Fact B.1.} \quad
\textit{Let $R$ be a commutative ring with unit and
\(a \! \left( q \right)\) \(= 1\) $+$
\(\Su_{n \, = \, 1}^{\infty} a_n q^n\)
\(\in R \Bbrk{q}\) be an infinite formal
power series in a single variable $q$.
Then $a \! \left( q \right)$ is invertible
in \(R \Bbrk{q}\), i.e., there exists unique
\(b \! \left( q \right) =\)
\(\Su_{n \, = \, 0}^{\infty} b_n q^n\)
\(\in R \Bbrk{q}\) such that
\(a \! \left( q \right) b \! \left( q \right) = 1\).
Moreover, \(b_0 = 1\) and if \(a \! \left( q \right)\)
is a complex series that is absolutely convergent
and nonzero for \(\left| q \right| < \lambda\)
then \(b \! \left( q \right)\) is absolutely
convergent for \(\left| q \right| < \lambda^{-1}\).}

\begin{proof}
Noting that
\(\left(\raisebox{10pt}{\hspace{-2pt}}\right.
1 + \Su_{n \, = \, 1}^{\infty} a_n q^n
\left.\raisebox{10pt}{\hspace{-2pt}}\right)
\left(\raisebox{10pt}{\hspace{-2pt}}\right.
\Su_{n \, = \, 0}^{\infty} b_n q^n
\left.\raisebox{10pt}{\hspace{-2pt}}\right)
=
b_0 + \Su_{n \, = \, 1}^{\infty}
\left(\raisebox{10pt}{\hspace{-2pt}}\right.
b_n +
\Su_{k \, = \, 0}^{n-1} a_{n-k} b_k
\left.\raisebox{10pt}{\hspace{-2pt}}\right)
q^n\)
one can inductively determine $b_n$ starting with
\(b_0 = 1\).
If \(a \! \left( q \right)\) is absolutely convergent
and nonzero for
\(\left| q \right| < \lambda\)
then \(b \! \left( q \right)\)
will be the Taylor series of an analytic
function for
\(\left| q \right| < \lambda^{-1}\)
so that it will be absolutely convergent there.
\end{proof}

Continuing with the \textit{proof} of the
statement~(\textit{a}) of
Proposition~\ref{pr:3.4}
we note
first that $\Theta_{12}$ is obtained from~(\ref{Th})
(see also (\ref{eqn?}) (\ref{eq1.6}))
as a formal power series in $q$ with coefficients
that are polynomials, say \(\Theta_{12}^{\left( n \right)}\),
in $e^{\pm \pi i \czeta_{12}}$ and $e^{\pm \pi i \alpha}$.
Thus the coefficient in $\Theta_{12}^{\left( n \right)}$
to $e^{m \pi i \czeta_{12}}$
(for \(\left| m \right| \leqslant n\)) will be an even
trigonometric polynomial in $\alpha$ with period~$1$
(since $\Theta_{12}$, as an analytic function, is even
and periodic with period~$1$ in
$\czeta_{12}$ as well as in $\alpha$,
according to Eqs.~(\ref{fn.4}) and (\ref{fn.5}))
and hence, $\Theta_{12}^{\left( n \right)}$ is a polynomial in
\(\cos 2 \pi \alpha =\) $u_1 \spr u_2$
\(\in \C \Brk{u_1,u_2}\).
Then considering $\Theta_{12}^{\left( n \right)}$
as a polynomial in $\cos 2 \pi \alpha$ we find in the
same way that its coefficients are polynomials in
\(\cos 2 \pi \czeta_{12}\).
Summarizing, we have $\Theta_{12}$
\(\in \C \Brkl{e^{\pm 2\pi i \czeta_1},u_1;}\)
\(\Brkr{e^{\pm 2\pi i \czeta_2},u_2}\).
To prove next Eq.~(\ref{div}) we observe that
\(\Theta_{12}^{\left( n \right)}\)
is a polynomial in \(\cos 2\pi \czeta_{12}\)
(with polynomial coefficients in \(\cos 2 \pi \alpha\))
which is zero for
\(\cos 2\pi \czeta_{12} = \cos 2\pi \alpha\)
(since \(\Theta_{12} = 0\) for \(\czeta_{12} = \pm \alpha\)).
It then follows that
\(\Txfrac{\Theta_{12}^{\left( n \right)}}{
4\sin \pi \czeta_+ \sin \pi \czeta_-} \equiv
\Txfrac{\Theta_{12}^{\left( n \right)}}{
2 \left( \cos 2\pi\alpha - \cos 2\pi\czeta_{12} \right)}\)
is again a polynomial in
\(\cos 2\pi \czeta_{12}\) and \(\cos 2\pi \alpha\).
This and the second equality in~(\ref{Th}) prove
Eq.~(\ref{div}).
Since $\Theta_{12}$ is even in both, $\czeta_{12}$ and $\alpha$,
we have the symmetry \(\Theta_{12} = \Theta_{21}\).

Now the proof of the first part of
Proposition~\ref{pr:3.4}~(\textit{b})
follows from
Fact B.1, Eq.~(\ref{div}) and the existence in
\(\C \Bbrk{e^{\pm2\pi i \, \czeta_1},u_1}_+
\Bbrk{e^{\pm2\pi i \, \czeta_2},u_2}_+\)
of the inverse:
\beqa\label{inverse}
\Txfrac{1}{\sin \pi \czeta_+ \sin \pi \czeta_-}
\, = && \podr
\Txfrac{4 \, e^{-2 \pi i \czeta_{12}}}{
1 \! - \!
2 \cos \left( 2 \pi \alpha \right) e^{-2 \pi i \czeta_{12}}
\! + \! e^{-4 \pi i \czeta_{12}}}
\, = \, \\ \, = && \podr
4 \, \Su_{n \, = \, 0}^{\infty} \,
C_n^1 \left( \cos 2 \pi \alpha \right)
e^{-\left( n+1 \right) \pi i \czeta_{12}}
\eeqa
where $C_n^k \left( t \right)$ are the
Gegenbauer polynomials already used in Sect.~\ref{sec:5}.

Continuing with the proof of
Proposition~\ref{pr:3.4}~(\textit{c})
we note first that the symmetry of $\OMG{n}$ follows from
that of $\Theta_{jk}$.
To obtain Eq.~(\ref{Omega}) one first derives for \(m \in \Z\):
\beq\label{th_tr}
\Theta \! \left( \czeta_{12}+m\,\tau;\, u_1,u_2 \right) \, = \,
e^{-2\pi i \hspace{1pt} \left( m^2\,\tau + 2 \, m \, \czeta_{12} \right)} \,
\Theta \! \left( \czeta_{12};\, u_1,u_2 \right)
\eeq
using Eqs.~(\ref{Th}) and (\ref{eqnA.21a}).
Then we have for \(\lambda_1,\dots,\lambda_{n-1} \in \Z\):
\beqa\label{calc}
&&
\OMG{n} \!
\left(
\czeta_{12} + \lambda_1 \tau
,\dots,
\czeta_{n-1 \,n} + \lambda_{n-1} \tau
;\,
u_1,\dots,u_n;\, \tau \right)
\, = \,
\nn && \quad \, = \,
\mathop{\prod}\limits_{1 \, \leqslant \, l \, < \, m \, \leqslant \, n}
\Theta \! \left(\raisebox{10pt}{\hspace{-2pt}}\right.
\Su_{j \, = \, l}^{m-1}
\left( \czeta_{j\, j+1} + \lambda_j \tau \right)
;\, u_l,u_m
\left.\raisebox{10pt}{\hspace{-2pt}}\right)
\, = \,
\nn && \quad \, = \,
\exp \!
\left\{\raisebox{14pt}{\hspace{-2pt}}\right.
-2\pi i \!
\Su_{1 \, \leqslant \, l \, \leqslant \, m \, \leqslant \, n-1}
\left[\raisebox{14pt}{\hspace{-2pt}}\right.
\left(\raisebox{14pt}{\hspace{-2pt}}\right.
\Su_{j \, = \, l}^{m} \lambda_j
\left.\raisebox{14pt}{\hspace{-2pt}}\right)^{\! 2}
\tau +
2
\left(\raisebox{14pt}{\hspace{-2pt}}\right.
\Su_{j \, = \, l}^{m} \lambda_j
\left.\raisebox{14pt}{\hspace{-2pt}}\right)
\!
\left(\raisebox{14pt}{\hspace{-2pt}}\right.
\Su_{j \, = \, l}^{m} \czeta_{j \, j+1}
\left.\raisebox{14pt}{\hspace{-2pt}}\right)
\left.\raisebox{14pt}{\hspace{-2pt}}\right]
\left.\raisebox{14pt}{\hspace{-2pt}}\right\}
\times
\nn && \hspace{24pt}
\times \
\OMG{n} \!
\left(
\czeta_{12},\dots,\czeta_{n-1 \,n}
;\,
u_1,\dots,u_n;\, \tau \right)
\eeqa
so that expanding the sums in the latter exponent we arrive at
Eq.~(\ref{Omega}) with a positive definite integral matrix
\(\left\{\raisebox{9pt}{\hspace{-3pt}}\right.
A_{jk}^{\left( n \right)}
\left.\raisebox{9pt}{\hspace{-3pt}}\right\}\mtrx_{j,k = 1}^{n-1}\).

To prove the Proposition~\ref{pr:3.4}~(\textit{d})
let us write (following~\cite{M83}):
\beq\label{eB.5}
F \! \left( \czeta_{12},\dots,\czeta_{n-1 \, n}; \tau \right)
\, = \,
\Su_{\left( m_1,\dots,m_{n-1} \right) \, \in \, \Z^{n-1}}
F'_{m_1\dots m_{n-1}} \! \left( q \right) \,
e^{i \pi \Su_{k=1}^{n-1} m_k \czeta_{k \, k+1}}
\eeq
where $F'_{m_1\dots m_{n-1}} \! \left( q \right)$ are infinite
formal power series in $q^{\frac{1}{2}}$ (with coefficients in the algebra $R$).
Then the properties~(\ref{Omega})
combined with the expansion~(\ref{eB.5}) implies
that $F'_{m_1\dots m_{n-1}} \! \left( q \right)$ are nonzero if
\(m_k + \varepsilon^{\left( 1 \right)}_k = 0 \, \mathit{mod} \, 2\).
Therefore, we can rewrite the expansion~(\ref{eB.5}) in the form:
\beqa\label{eB.5n}
F \! \left( \czeta_{12},\dots,\czeta_{n-1 \, n}; \tau \right)
\, =
\Su_{\nnu_1 \, \in \, \Z +  \frac{\varepsilon^{\left( 1 \right)}_{1}}{2}}
\! \dots \!
\Su_{\nnu_{n-1} \, \in \, \Z +  \frac{\varepsilon^{\left( 1 \right)}_{n-1}}{2}}
F_{\nnu_1\dots \nnu_{n-1}} \! \left( q \right) \,
e^{2 \pi i \Su_{k=1}^{n-1} \nnu_k \czeta_{k \, k+1}}
\nn
\eeqa
(\(F_{\nnu_1\dots \nnu_{n-1}} \! \left( q \right) \in
R \Bbrk{q^{\frac{1}{2}}}\)).
Now combining the expansion~(\ref{eB.5}) with the property (\ref{eaB.10})
we obtain:
\beq\label{eB.8}
F_{\underline{\nnu} \, + \,
2 N A^{\left( n \right)} \left( \underline{\lambda} \right)}
\! \left( q \right) \, = \,
\left( -1 \right)^{\underline{\lambda} \hspace{1pt}\spr\hspace{1pt}
\underline{\varepsilon}^{\left( 2 \right)}} \,
q^{
\hspace{1.5pt} N \hspace{2.5pt} \underline{\lambda} \hspace{1pt}\spr\hspace{1pt}
A^{\left( n \right)} \! \left( \underline{\lambda} \right)
\, - \, \underline{\nnu} \hspace{1pt}\spr\hspace{1pt} \underline{\lambda}
}
\ \, F_{\underline{\nnu}} \! \left( q \right)
\eeq
where \(\underline{\nnu} :=\) \(( \nnu_1,\) $\dots,$ \(\nnu_{n-1})\),
\(\underline{\lambda} :=\) \(( \lambda_1,\) $\dots,$ \(\lambda_{n-1})\),
\(\underline{\varepsilon}^{\left( k \right)} :=\)
\(( \varepsilon^{\left( k \right)}_1,\) $\dots,$
\(\varepsilon^{\left( k \right)}_{n-1})\),
\(A^{\left( n \right)} (\underline{\lambda}) :=\)
\(\left(\raisebox{9pt}{\hspace{-2pt}}\right.
\Su_{k \, = \, 1}^{n-1} A^{\left( n \right)}_{jk} \lambda_k
\left.\raisebox{9pt}{\hspace{-2pt}}\right)_{j = 1}^{n-1}\)
and
\(\underline{\nnu} \spr \underline{\lambda} :=\)
\(\Su_{k \, = \, 1}^{n-1} \, \nnu_k \, \lambda_k\).
Thus, we can find all
the series $F_{\underline{\nnu}} \! \left( q \right)$
if we know them
for all
\(\underline{\nnu} - \Txfrac{\underline{\varepsilon}^{\left( 1 \right)}}{2}\)
belonging to a finite subset \(M \subset \Z^{n-1}\)
given by the intersection of the lattice $\Z^{n-1}$ with a
fundamental domain of its sublattice
\(2 N A^{\left( n \right)} \left(\raisebox{9pt}{\hspace{-2pt}}\right.
\Z^{n-1}
\left.\raisebox{9pt}{\hspace{-2pt}}\right)\)
(here we use the fact that $A^{\left( n \right)}$ is a nondegenerate
integral matrix).
In fact,
if we split the sum in~(\ref{eB.5n}) into two sums,
the first, over the fundamental domain
\(L + \Txfrac{\underline{\varepsilon}^{\left( 1 \right)}}{2}\)
and the second, over the its translates
\(2 N A^{\left( n \right)} \left( \underline{\lambda} \right)\)~--~we
find (using~(\ref{eB.8})):
\beq\label{exp_F}
F \! \left( \czeta_{12},\dots,\czeta_{n-1 \, n}; \tau \right)
\ =
\Su_{\underline{\nnu} \, \in \,
L + \frac{\underline{\varepsilon}^{\left( 1 \right)}_{}}{2}}
F_{\underline{\nnu}} \! \left(  q\right)
\, F^{\left( N \right)}_{\underline{\nnu}} \!
\left( \czeta_{12},\dots,\czeta_{n-1 \, n}; \tau \right)
\eeq
where the series
\beqa\label{G-ser}
&&
F^{\left( N \right)}_{\underline{\nnu}} \!
\left( \czeta_{12},\dots,\czeta_{n-1 \, n};\, \tau \right)
\, := \,
\nn
&&
\quad \, := \,
\Su_{\underline{\lambda} \, \in \, \Z^{n-1}}
\left( -1 \right)^{\underline{\lambda} \hspace{1pt}\spr\hspace{1pt}
\underline{\varepsilon}^{\left( 2 \right)}}
q^{
\hspace{1.5pt} N \hspace{2.5pt} \underline{\lambda} \hspace{1pt}\spr\hspace{1pt}
A^{\left( n \right)} \! \left( \underline{\lambda} \right)
\, - \, \underline{\nnu} \hspace{1pt}\spr\hspace{1pt} \underline{\lambda}
}
\ \,
e^{2 \pi i \,
\left( \underline{\nnu} \, + \,
2 N A^{\left( n \right)} \left( \underline{\lambda} \right) \right)
\hspace{1pt}\spr\hspace{1pt} \underline{\czeta}}
\,
\eeqa
(\(\underline{\czeta} := \left( \czeta_{12},\dots,\czeta_{n-1 \, n} \right)\))
are absolutely convergent to analytic functions in
\(\underline{\czeta}\) and $\tau$ with \(\mathit{Im} \, \tau > 0\)
according to the general theory of the theta-series.
If $F$ is symmetric as a series in $\czeta_1,$ $\dots,$ $\czeta_n$
then the above basic $F^{\left( N \right)}$ series can be further
symmetrized in $\czeta_k$ (\(k = 1,\dots, n\)).

This completes the proof of
Proposition~\ref{pr:3.4}.

\end{document}